\documentclass[amsmath]{aastex701}
\usepackage{amssymb}
\usepackage{booktabs}

\begin{document}

\title{Lunar Reflective Interferometry}

\author[orcid=0000-0002-0923-9363]{Peter W. Gorham}
\affiliation{{Dept. of Physics \& Astronomy, Univ. of Hawaii at Manoa}, {2505 Correa Rd.},{Honolulu}, {HI}, {96822}, {USA}}
\email{gorham@hawaii.edu}

\author[orcid=0009-0003-8984-388X]{T.~Joseph~W.~Lazio}
\affiliation{{Department of Climate and Space Sciences and Engineering, Univ.\ of Michigan}, {Ann Arbor}, {MI}, {48109}, {USA}}
\email{}

\author{Andrew Romero-Wolf}
\affiliation{Physical Insights, LLC., Honolulu, Hawaii}
\email{}

\author[orcid=0000-0001-8141-2653]{Patrick~S.~Allison}
\affiliation{{Dept. of Physics, Center for Cosmology and AstroParticle Physics},{Columbus}, {OH}, {43210}, {USA}}
\email{}

\author[orcid=0000-0003-2238-2698]{Marin M.~Anderson}
\affiliation{{Jet Propulsion Laboratory, California Institute of Technology}, {4800 Oak Grove Dr}, {Pasadena}, {CA}, {91109}, {USA}
}
\email{}

\author{Ali M.~Bramson}
\affiliation{{Dept.~of Earth, Atmospheric, and Planetary Sciences, Purdue University}, {550 Stadium Mall Drive}, {West Lafayette}, {IN}, {47907}, {USA}
}
\email{}

\author{Roman Burridge}
\affiliation{{Dept. of Physics \& Astronomy, Univ. of Hawaii at Manoa}, {2505 Correa Rd.},{Honolulu}, {HI}, {96822}, {USA}
}
\email{}

\author{Amy Connolly}
\affiliation{{Dept. of Physics, Center for Cosmology and AstroParticle Physics},{Columbus}, {OH}, {43210}, {USA}}
\email{}

\author{Emily S.~Costello}
\affiliation{{Hawaii Institute of Geophysics and Planetology}, {1680 East-West Road}  {Honolulu}, {HI}, {96822}, {USA}
}
\email{}

\author[orcid=0000-0002-4953-6397]{Cosmin Deaconu}
\affiliation{{Dept.~of Astronomy \& Astrophysics, Enrico Fermi Institute, Kavli Institute for Cosmological Physics, University of Chicago}, {5640 S.~Ellis Ave}, {Chicago}, {IL}, {60637}, {USA}
}
\email{}

\author{Susan Lepri}
\affiliation{{Department of Climate and Space Sciences and Engineering}, {Ann Arbor}, {MI}, {48109}, {USA}}
\email{}

\author{Christian Miki}
\affiliation{{Dept. of Physics \& Astronomy, Univ. of Hawaii at Manoa}, {2505 Correa Rd.},{Honolulu}, {HI}, {96822}, {USA}}
\email{miki@hawaii.edu}

\author[orcid=0000-0001-8344-7999]{Eric Oberla}
\affiliation{{Dept.~of Astronomy \& Astrophysics, Enrico Fermi Institute, Kavli Institute for Cosmological Physics, University of Chicago}, {5640 S.~Ellis Ave}, {Chicago}, {IL}, {60637}, {USA}
}
\email{}

\author[orcid=0000-0003-2527-8271]{Sean Peters}
\affiliation{{University of Colorado Boulder},{3775 Discovery Dr.}, {Boulder}, {CO}, {80303}, {USA}
}
\email{}

\author[orcid=0000-0001-6360-6972]{Kathryn Plant}
\affiliation{{National Radio Astronomy Observatory}, {1011 Lopezville Rd}, {Socorro}, {NM}, {87801}, {USA}
}
\email{}

\author[orcid=0000-0001-9384-6037]{Julie Rolla}
\affiliation{{Jet Propulsion Laboratory, California Institute of Technology}, {4800 Oak Grove Dr}, {Pasadena}, {CA}, {91109}, {USA}
}
\email{}

\author{Matthew Siegler}
\affiliation{{Hawaii Institute of Geophysics and Planetology}, {1680 East-West Road}  {Honolulu}, {HI}, {96822}, {USA}
}
\email{}


\begin{abstract}

We present the method of lunar reflection interferometry (\hbox{LRI}) in which a virtual interferometer can be formed by using a spacecraft-borne antenna in lunar orbit that receives both direct rays and those reflected from the lunar surface.
The technique exploits the method of images and is akin to the classic ``sea cliff'' interferometer, with the Moon's surface, primarily the lunar Maria, replacing the ocean surface.
We describe the method in detail, demonstrate that significant portions of the Moon's surface are sufficiently smooth at low radio frequencies ($\nu \lesssim 10\,\mathrm{MHz}$) for the technique to work, and outline a spacecraft instrument implementation.
We describe potential systematic errors and how they could be mitigated. We present several astrophysics applications of the method.

\end{abstract}




\section{Introduction} 
\label{sec:intro}
Astronomy has achieved high-resolution mapping of the sky across nearly the entire accessible electromagnetic spectrum, from gamma rays to meter-wavelength radio emission. At radio frequencies above $\sim10~\mathrm{MHz}$, ground-based interferometric arrays have produced images with angular resolutions reaching sub-arcsecond and even milliarcsecond scales, enabling detailed studies of compact objects, relativistic jets, and the structure of the interstellar medium~\citep{TMS2017}. 

In striking contrast, the sky below $10~\mathrm{MHz}$ has been mapped only at relatively low resolution. The terrestrial ionosphere becomes opaque at these frequencies, precluding ground-based observations except in rare cases of low ionospheric activity over select regions. Early space-based measurements, most notably those of the \textit{Radio Astronomy Explorer 2} (RAE-2), provided the first all-sky maps in this regime, revealing diffuse nonthermal Galactic emission and large-scale structure, but at angular resolutions of tens of degrees \citep{novacoBrown1978}. These observations demonstrated both the scientific richness of the low-frequency sky and the severe limitations imposed by the lack of interferometric capability in space. Many other observations have been made of the low-frequency radio sky; these are summarized in appendix~\ref{appendix1}, but in all cases the resolution has been on the scale of several degrees to tens of degrees.

Achieving sub-degree-scale resolution over the 0.1--10 MHz band requires interferometric baselines of $\sim60$ wavelengths or more, equivalent to projected antenna separations of 2--100 kilometer. Traditional concepts for space-based low-frequency interferometry have therefore relied on constellations of multiple spacecraft operating as a distributed array~\citep{jones1999alfa, vanvugt2017frequency, wijnholds2018olfar} or lunar surface arrays~\citep{jones2006low,2009SPIE.7436E..0IL,burns2019farside, burns2020transformative}. While technically feasible, such architectures have historically been considered prohibitively complex and expensive, placing them beyond the scope of cost-constrained astrophysics missions. 

A recent exception is the NASA Heliophysics Division's Sun Radio Interferometer Space Experiment~\citep[\hbox{SunRISE},][]{SunRISE2024}, which is expected to launch later this year, deploying a six-spacecraft formation-flying cubesat array to perform high-resolution imaging of solar type II and type III radio burst sources. SunRISE will also perform some astrophysical interferometric imaging of the low-frequency radio sky as a secondary science goal, subject to constraints from the solar emission and terrestrial radio-frequency interference (RFI). While SunRISE is poised to achieve important but limited high-resolution coverage, the practical barriers to low-frequency all-sky mapping remain, leaving a largely unexplored discovery space for comprehensive high-resolution imaging of the sky below the ionospheric cutoff. In this work, we present a concept that addresses this challenge through a fundamentally different approach to interferometry, enabling access to this spectral regime with a single spacecraft and a cost-effective architecture.

\begin{figure}
\centerline{~~\includegraphics[width=0.60\textwidth, trim=3mm 5mm 3mm 2mm, clip]{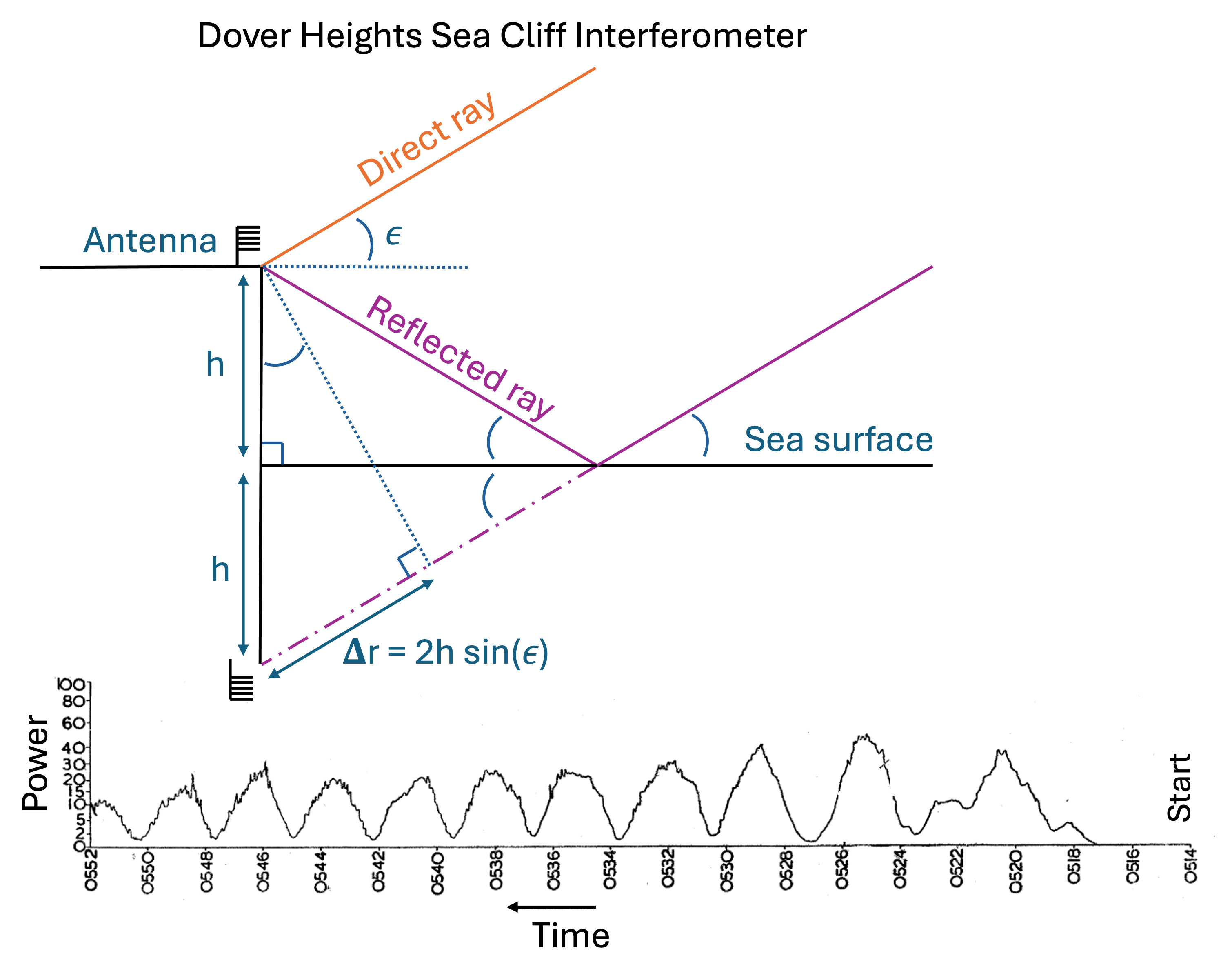}}
\vspace{-3mm}
\caption{\textit{top}~Geometry of the original sea cliff interferometer, which observed the interference of direct and reflected rays from the rising Sun at elevation~$\epsilon$.  \textit{bottom}~Power vs.\ time as measured in the original experiment~\citep{mcready1947seacliff} as the rising Sun moved through the fringe pattern of the interferometer. 
\label{fig:seacliff} 
}
\end{figure}

It is important to stress that low-radio-frequency interferometric imaging does have inherent limitations to its angular resolution, imposed by angular broadening due to scattering by turbulent plasmas in both the solar-wind-dominated interplanetary medium (IPM) and the interstellar medium (ISM). These will be discussed in more detail below, but it is sufficient to say for now that scattering in both the IPM and ISM ensures that the full angular resolution of a 100~km baseline can only be achieved at frequencies above about~3\,MHz.
Below that, angular resolution will be broadened, though not catastrophically, as we discuss in more detail below.

In addition to the degradation due to angular broadening, free-free absorption due to the warm ionized medium in the Galactic plane rapidly increases the opacity of sightlines through the Galactic plane below 10~MHz, and by around 2--3\,MHz, the Galactic plane radiation becomes attenuated to the point that the background radiation in Galactic polar directions exceeds that of the plane. At frequencies of~1\,MHz and below, observing distances are generally limited to a few hundred pc, except perhaps in some regions near the Galactic poles. Radio astronomy thus rapidly evolves with decreasing frequency within this sub-10 MHz window, due to processes that are intimately tied to the structure, composition, and dynamics of the IPM and \hbox{ISM}.
Sky mapping within this last spectral frontier thus has the potential to establish final brackets on both Galactic and some extragalactic source structure and spectral behavior, as well as crucial details on the propagation media through which we observe them.

In one of the first applications of interferometry in radio astronomy, observations with an antenna on a cliff overlooking the ocean (Figure~\ref{fig:seacliff}) were used to discover that sunspots emit strong nonthermal radio emission~\citep{mcready1947seacliff}.
This sea cliff interferometer was constructed by observing the interference of rays propagating directly to the antenna and rays reflected off the ocean, which constrained the size and location of the emission region. 
This technique made use of the ``method of images''~\citep{thomson1848mathematical} in which the reflected ray experiences an additional path length equivalent to the path to a virtual image antenna at the cliff edge below the sea surface.
For a wavelength~$\lambda$, a cliff of height~$h$, and source elevation angle~$\epsilon$, the path length difference $\Delta r = 2h\,\sin\epsilon$ introduces a phase difference $\Delta \phi = (2h\,\sin\epsilon)/\lambda$ between the direct and reflected rays, in addition to the phase shift of~$\pi$ incurred by reflection off the sea surface. 
As the Sun rose, observed peaks and minima in power (an interferometric fringe pattern) indicated that the source size was significantly smaller than the change in $\epsilon$ between constructive and destructive interference.
Subsequently, this sea cliff interferometry technique was used to measure or constrain the sizes of discrete radio sources, notably including \hbox{Cyg~A} \citep{1953AuJPh...6..420B}.


While sea cliff interferometry provided a natural economy of resources, the constraints on the baseline, which was fixed by the cliff geometry, limited the general usefulness of the approach, and it was soon supplanted in the radio astronomy context by two-element interferometers~\citep{ryle1946solar,ryle1948investigation}, in which one of the antennas was transportable, thereby providing access to a larger range of fringe spacings or spatial frequencies.
However, it has been revisited as a passive radio-sounding method in a geophysical context \citep{PassiveSounding}, for scenarios in which multiple antennas are not feasible, and for future remote sensing in the Jupiter system \citep{11783057}.

In this paper, we describe how the sea cliff interferometer technique or method of images could be extended to form an interferometer by a {\em single} lunar-orbiting spacecraft.  The technique requires low radio frequencies, $\nu \lesssim 10\,\mathrm{MHz}$, or long wavelengths, $\lambda \gtrsim 30\,\mathrm{m}$, where portions of the Moon's Fresnel-reflective surface can be considered sufficiently flat to provide specular reflections.  The value of this approach is that a single spacecraft can provide a relatively low-cost system for high-resolution imaging at these frequencies, in contrast to concepts involving several spacecraft with complex formation-flying and phase synchronization control.

The structure of this paper is as follows.
In Section~\ref{sec:tech}, we describe the technique and illustrate potential results; 
in Section~\ref{sec:systematic}, we discuss potential systematics;
in Section~\ref{sec:science}, we illustrate several examples of science use cases; and 
in Section~\ref{sec:conclude}, we present our conclusions.

\section{Methodology for Lunar Reflection Interferometry}\label{sec:tech}

Lunar Reflective Interferometry (LRI) effectively includes the lunar surface as part of the complete interferometer instrument. As such, intimate knowledge of the lunar surface is required, to fractions of a wavelength of the radio signals that will be utilized for the interference measurements. Fortunately our knowledge of the lunar surface topography has been advanced to a remarkable level in the last two decades by the Lunar Orbiting Laser Altimeter (LOLA), an instrument aboard the Lunar Reconnaissance Orbiter (LRO), combined with radar mapping from the Lunar Radar Sounder (LRS) on the Kaguya/Selene orbiter~\citep{Smith2010LOLA, Ono2010LRS}. The Digital Elevation Model (DEM) that has been derived from these missions is now at a level of resolution and accuracy that exceeds our knowledge of the Earth's surface in some aspects, and provides exactly the kind of data required to enable accurate estimates of the reflectivity of the lunar surface at virtually any location on the Moon. We will return to this in more detail below.

\begin{figure}[tb]
\centerline{~~\includegraphics[width=0.5\textwidth, trim=0mm 0mm 0mm 0mm, clip]{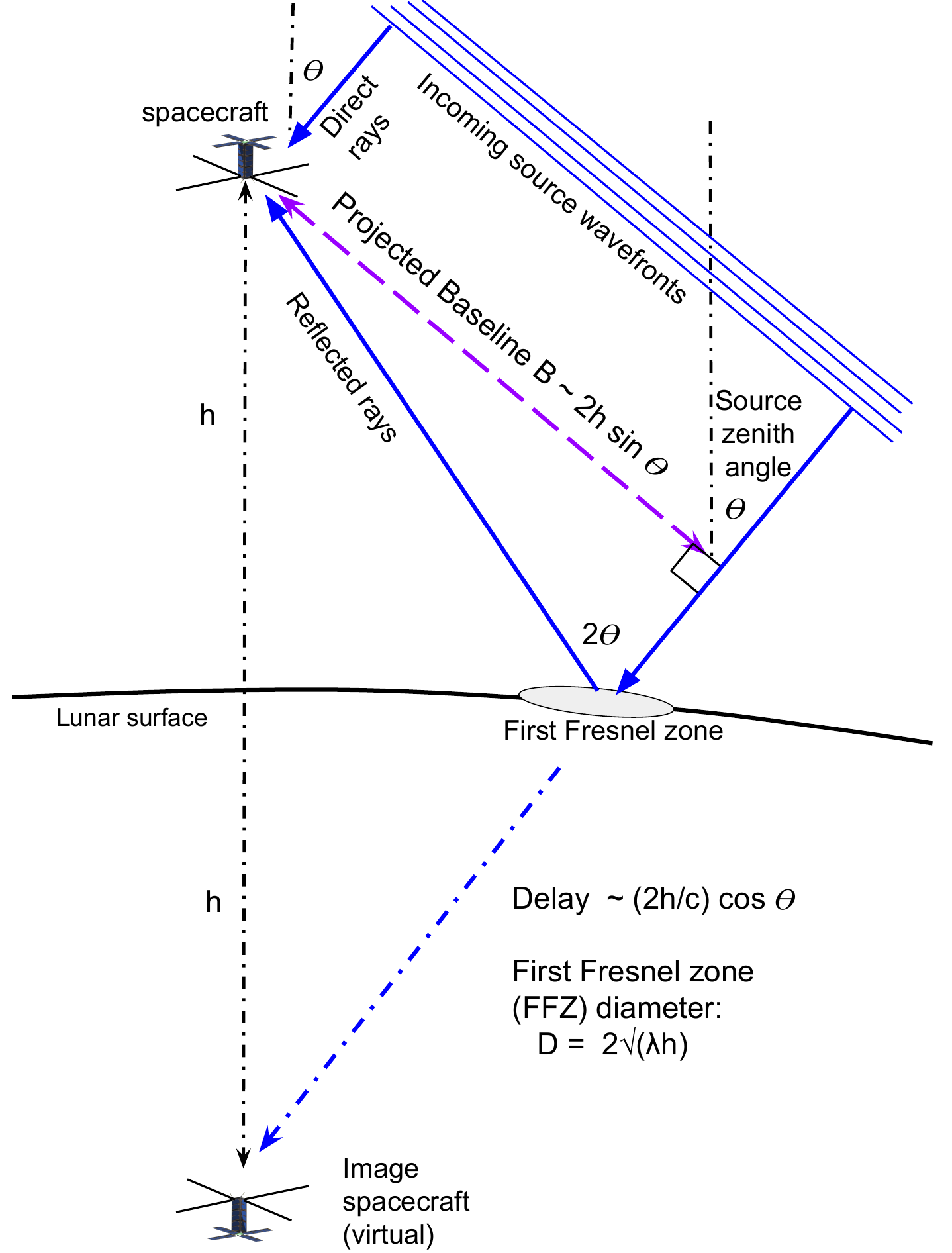}}
\vspace{-1mm}
\caption{The LRI technique forms an effective interferometer by using the interference between direct rays and those reflected from the Moon's surface.
Note that we have re-used the angle $\theta$ as in the sea cliff case, but it now indicates zenith angle rather than elevation angle.
\label{fig:LRIgeom} }
\end{figure}

Figure~\ref{fig:LRIgeom} presents the geometry for \hbox{LRI}, which is directly analogous to that of the sea cliff interferometer, with the Moon's surface replacing the ocean.
The method of images applies equally well here as it did for the sea cliff case. 
For clarity, the system is described in plane geometry but in practice at longer baselines, spherical geometry is preferred.

The sea cliff interferometer observed the modulation of the total received power as a function of time as the source rose over the horizon. 
For the \hbox{LRI} technique, the combination of wide-field mapping and polarization effects from the Brewster angle ($\sim56^{\circ}$ for the lunar regolith) tends to restrict mapping to regions closer to the zenith, although reflected signals from all angles will be preserved in the collected data.
For this reason, the geometry of Figure~\ref{fig:LRIgeom} is referenced to local zenith angle, although it is important to note that the sky position of the local zenith changes continuously as the spacecraft orbits the Moon.

LRI arises functionally from a two-path interferometer in which the second element is provided by the specular lunar surface reflection.
The antenna receives the superposition of a direct sky field and a delayed, reflected copy, $V_{\mathrm{tot}}(t) = V(t) + \alpha V(t-\tau)$, where $\tau$ is the geometric delay set by spacecraft altitude and zenith angle, and $\alpha$ is the complex reflection coefficient (including the sign for phase inversion, if applicable).  In a classical sea-cliff interferometer, the summed voltage was square-law detected and integrated, producing a proportional power output:
\begin{equation}
\label{acf-eq}
P_{\mathrm{tot}} \propto \langle V_{\mathrm{tot}}^2 \rangle = \langle V_{\mathrm{tot}} V^*_{\mathrm{tot}} \rangle = 
(1 + |\alpha^2|)\langle V(t)V^*(t) \rangle + 2{\rm Re} \left [ \alpha^* \langle V(t) V^*(t-\tau) \rangle \right ]
\end{equation}
where the angle brackets indicate integration in time. The last term on the right hand side is an interference term, which appears implicitly as a modulation of detected power, but with no direct access to the delay term~$\tau$.
Fringe visibility therefore had to be inferred indirectly from total power variations as the geometry changed.

LRI uses the same approach but benefits from modern digital signal processing: the summed voltage is digitized, and the two-element correlation is produced by computing the voltage autocorrelation function directly:
\begin{equation}
R_{VV}(t') = \langle V(t) V^*(t-t') \rangle \equiv \frac{1}{2T}\int_{-T}^T V(t)V^*(t-t')dt 
\end{equation}
where $T$ is the integration time. Applying this to the LRI case, we have
\begin{align}
R_{V_{\mathrm{tot}}V_{\mathrm{tot}}}(t') &= \langle V_{\mathrm{tot}}(t) V^*_{\mathrm{tot}}(t-t') \rangle \\
  &= \langle ( V(t) + \alpha V(t-\tau) ) ( V^*(t-t') + \alpha^*V^*(t - \tau - t') )\rangle \nonumber
\end{align}
where we have introduced the {\it lag}, or autocorrelation delay variable, $t'$, to distinguish it from the geometric delay~$\tau$. 
This equation then yields several terms in the complex correlation: (direct $\times$ direct), (reflected $\times$ reflected), 
(direct $\times$ reflected), and (reflected $\times$ direct):
\begin{equation}
\label{RVV-eq}
R_{V_{\mathrm{tot}}V_{\mathrm{tot}}}(t') = ( 1 + |\alpha|^2) R_{VV}(t') + \alpha R_{VV}(t' - \tau) + \alpha^* R_{VV} (t' - \tau)
\end{equation}
where the geometric delay $\tau$ is now an explicit parameter in the correlation processing. It is evident here that, in comparison to the earlier sea-cliff case, the interference terms in $R_{V_{\mathrm{tot}}V_{\mathrm{tot}}}(t')$ produce peaks with Hermitian symmetry at both positive and negative lags: $t' = \pm \tau$.
From Figure~\ref{fig:LRIgeom}, the baseline component that is perpendicular to the incoming plane waves is $B(h, \theta) \simeq 2 h\sin\theta$, and the delay between arrivals at each interferometer element is $\tau(\theta) \simeq (2 h/c ) \cos \theta$. Neglecting background noise, a plane-wave source thus produces two peaks in the autocorrelation function at delay $\pm \tau$. The geometric delay is now explicitly accessible through $R_{V_{\mathrm{tot}}V_{\mathrm{tot}}}(t')$.

As with any two-element interferometer, the instantaneous ``sky fringe'' pattern can localize the source only in one of two orthogonal dimensions.
In LRI's case, the sky fringe is centered at the zenith, where the projected baseline approaches zero length, and the lack of any constraint on the azimuthal knowledge of the source direction means that each fringe is effectively a concentric ring around the zenith. 
As the source zenith angle increases, the projected baseline grows rapidly, and thus the sky fringe spacing is quite nonlinear, decreasing as $(\cos{\theta})^{-1}$.

The maximum delay $\tau_{\mathrm{max}} = 2h/c$ occurs for signals arriving from the local zenith, reflecting off the nadir point, and returning to the spacecraft from there. The spacecraft must be nominally in a relatively low lunar orbit (LLO), such that the altitude above the surface is small compared to the lunar radius; otherwise, surface curvature effects over the reflection region would destroy the specular coherence.
This condition is satisfied for LLOs up to about 150~km, with lower orbits being better, as we will show in a later section.
Our investigations have led us to focus on LLOs of $h \simeq 100$~km, for which $\tau_{\mathrm{max}} \simeq 667\,\mu$s. The voltages at the spacecraft antenna contain both direct and reflected signals with delays that depend inversely on the source zenith angle: a source at zenith ($\theta=0^\circ$) has the maximum delay $\tau_{\mathrm{max}}$, and the minimum baseline length. Sources at zenith angles $\theta>0^\circ$ produce a correlation amplitude at $\tau < \tau_{\mathrm{max}}$, and also have a longer projected baseline, leading to higher resolution at larger zenith angles. Also, since the autocorrelation function (ACF) is the Fourier dual of the Fourier power spectrum, which is more commonly used as the domain for radio interferometry, all operations can be transposed into the dual space without loss of information. 

In Appendix~\ref{appendix2}, we present a detailed analysis of the application of the van Cittert-Zernicke theorem in the context of LRI methodology and show that, once the effects of the Fresnel coefficient and the partial coherence inherent in the imperfect surface reflection are accounted for, this approach is fully consistent with traditional radio astronomical interferometry. Because the ACF in our case takes the role that a correlator would in a multi-element interferometer, we will center our discussion on the ACF as an imaging tool, and refer the reader to the appendix for more detailed information.

\subsection{Mapping with Lunar Surface Reflections.}\label{sec:mapping}

In the ACF delay domain, as outlined above in equation~\ref{RVV-eq}, there is a direct map from ACF lag to zenith angle on the sky, but since the spacecraft is always in motion, we must define a limited time duration over which the correlation products can be computed without loss of coherence. For LRI, as for other radiometers, the signal-to-noise ratio (SNR) scales as the square root of the product of integration time $T$ and frequency bandwidth $\Delta f$: $SNR \propto \sqrt{T \Delta f}$. 

\begin{figure}[tb]
\vspace{-2mm}
\centerline{~~\includegraphics[width=0.435\textwidth]{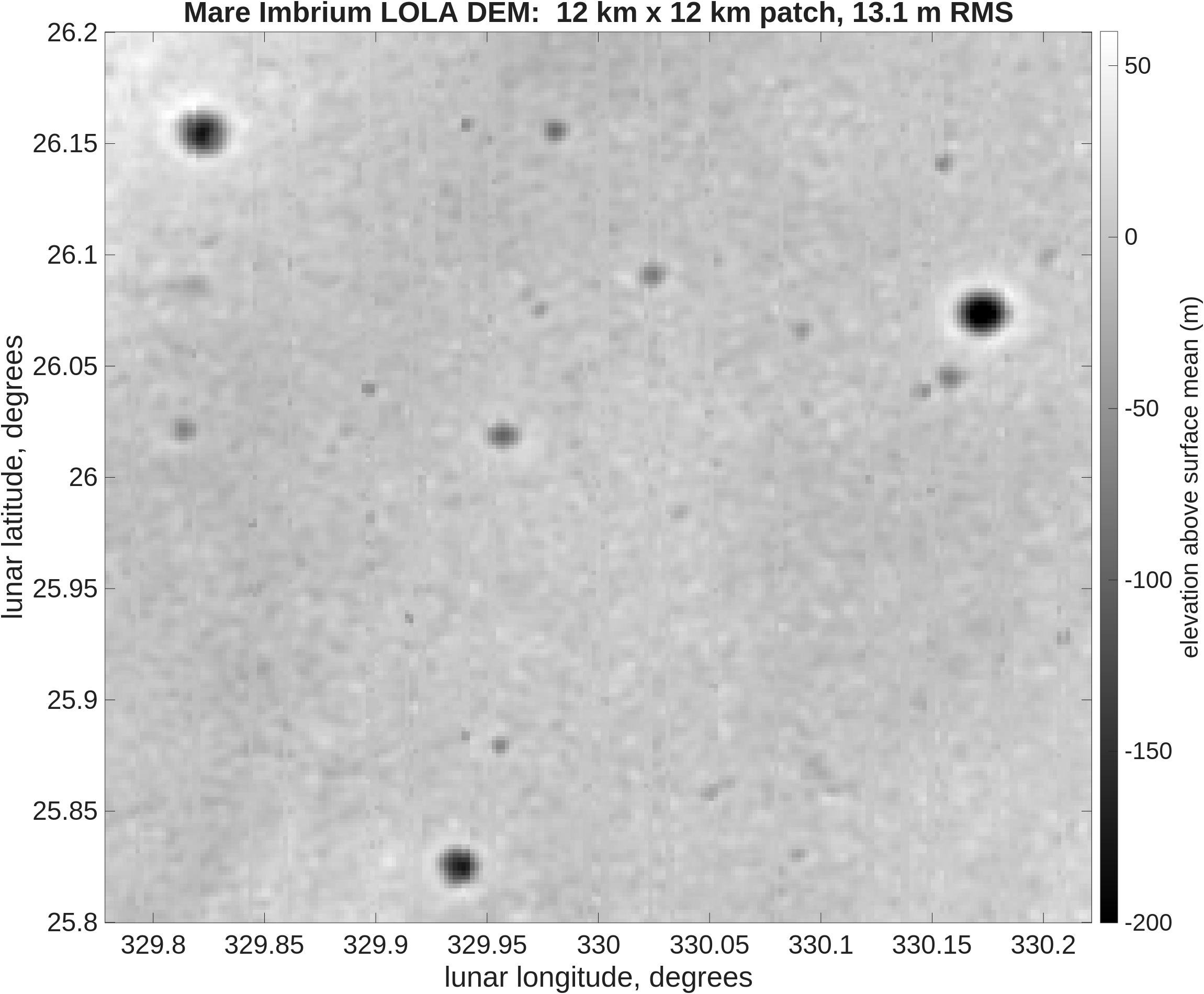}~~~~\raisebox{6mm}{\includegraphics[width=0.4\textwidth]{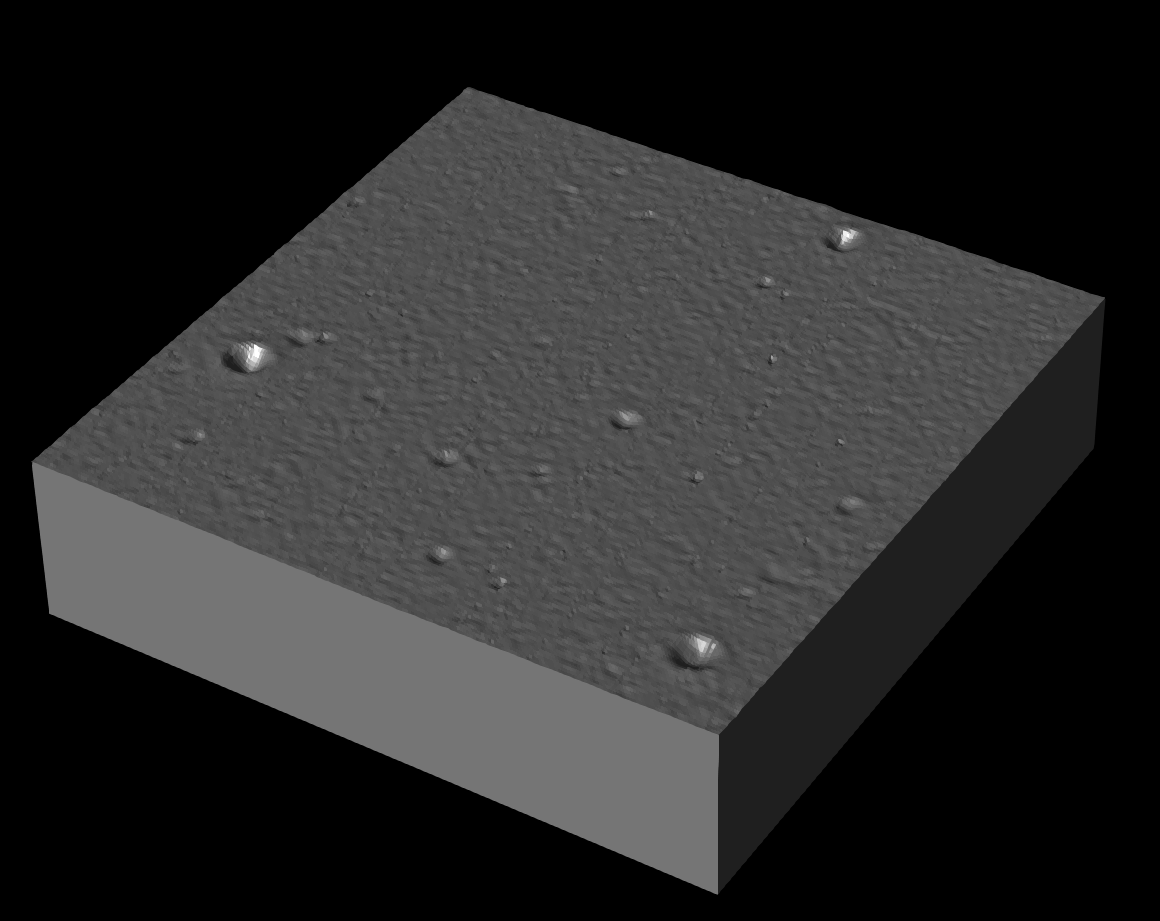}} }
\vspace{3mm}
\centerline{~~\includegraphics[width=0.85\textwidth, trim=0mm 0mm 0mm 0mm, clip]{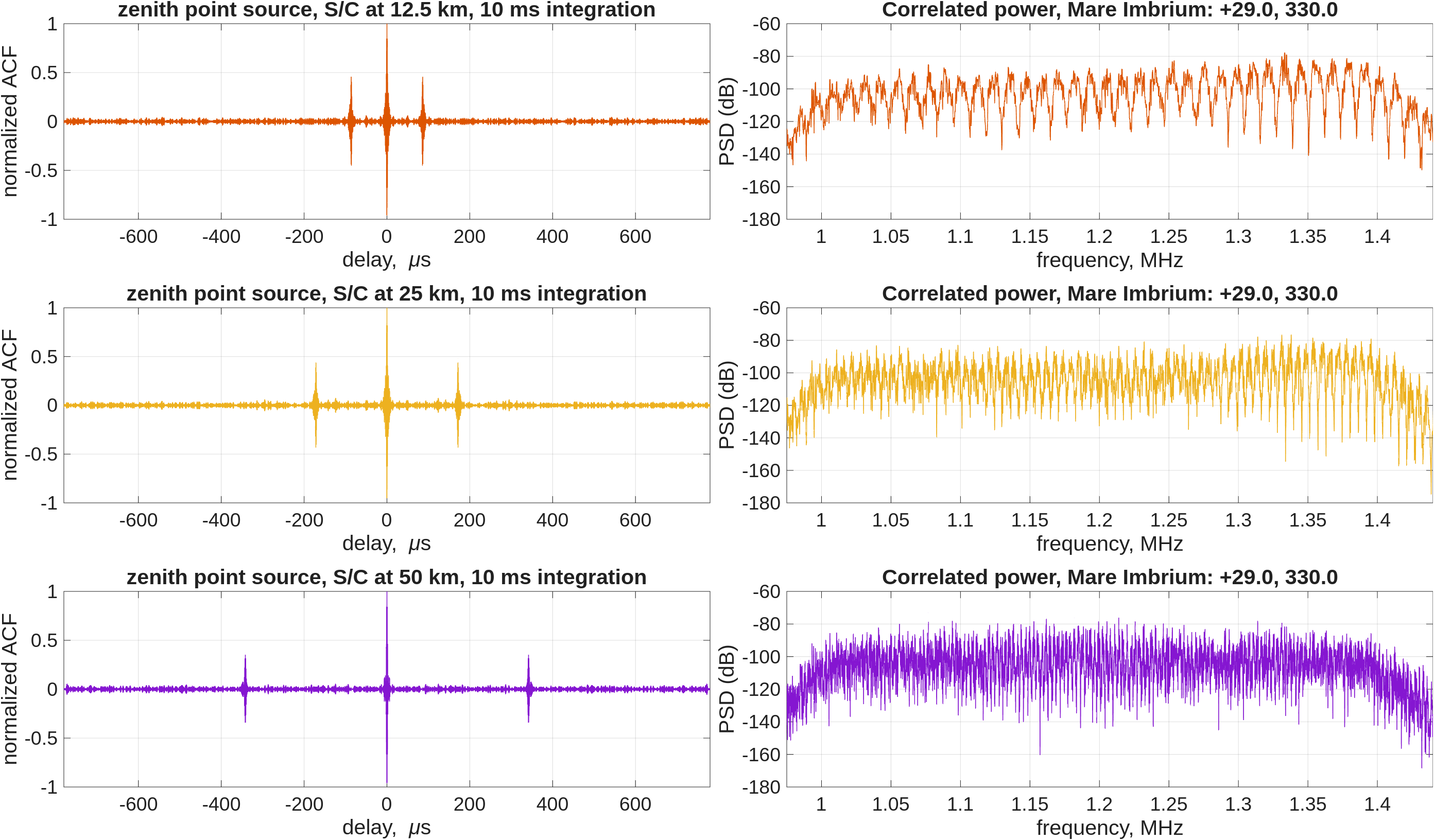}}
\vspace{-3mm}
\caption{Left top: DEM plot of lunar surface data for a test region in Mare Imbrium. Right top: 12 km x 12 km by 5 km depth solid model used for FDTD simulation. Bottom: Full-wave FDTD simulation of LRI strong-point-source zenith reflection from this Mare Imbrium patch at lunar latitude +29.0, longitude 330.0, for S/C altitudes from 12.5 to 50\,km. The left panels show the autocorrelation functions, and the right panels the corresponding power spectral density (PSD).
\label{ACFplots} }
\vspace{-1mm}
\end{figure}

We illustrate the correlation using a simulation of the thermal noise arriving from an infinitely distant point source at the zenith (plane-wave source), using full-wave electromagnetic (EM) simulations. For high fidelity, we use a commercial Finite-Difference Time Domain (FDTD) \citep{RemcomXF7}, and a LOLA digital elevation model for a portion of the lunar maria in the Mare Imbrium region. The spacecraft sensor, which has a nearly isotropic response and therefore sees both the incoming and reflected wave with equal sensitivity, is placed  above a 12~km wide surface model with lunar regolith permittivity characteristics. Due to gridding memory limitations, for this simulation we use a passband of 1.0\,MHz to~1.4\,MHz, and we are limited to spacecraft altitudes below 50~km.

Figure~\ref{ACFplots} (top row) displays the DEM data and the solid model implemented using it in the FDTD simulation. The bottom row shows the ACF and corresponding power spectrum with $T=10\,\mathrm{ms}$. The Galactic thermal sky background noise is suppressed to illustrate the ACF features and corresponding magnitude spectrum. For the lower altitudes, the PSD fringes are evident since the fringe spacing is coarser in frequency; at 50~km, the autocorrelation at the zenith lag is weakened by the increased size of the Fresnel zone footprint, and the corresponding increase in the RMS roughness of the surface, but the ACF peak is still strong and frequency fringes are evident in the PSD.

Galactic noise will dilute the SNR at short integrations such as this one.  At delays less than the zenith delay, the Galactic noise and its reflections from a wide range of angles over the lunar surface will lead to enhanced ACF power at lags below the zenith lag, and uncorrelated noise at higher lags. Point sources may still appear as discrete peaks in the ACF if they are strong compared to the Galactic noise at that lag.

In a circular LLO, the spacecraft orbits in about 2 hours, and the ground position moves at about $v_{gnd} \simeq 1.6$~km/s. The horizontal scale for the reflection region for broadband radio waves is given by the first Fresnel zone. For LRI geometry, the diameter of this zone is given by $D_{FZ} = 2\sqrt{\lambda h}$. For Figure~\ref{ACFplots}, this means that the FFZ has a diameter ranging from 3.8~km at the 12 km altitude, to almost 11~km at 100~km altitude. 

\begin{figure}[tb]
\centerline{~~\includegraphics[width=0.85\textwidth, trim=0mm 0mm 0mm 0mm, clip]{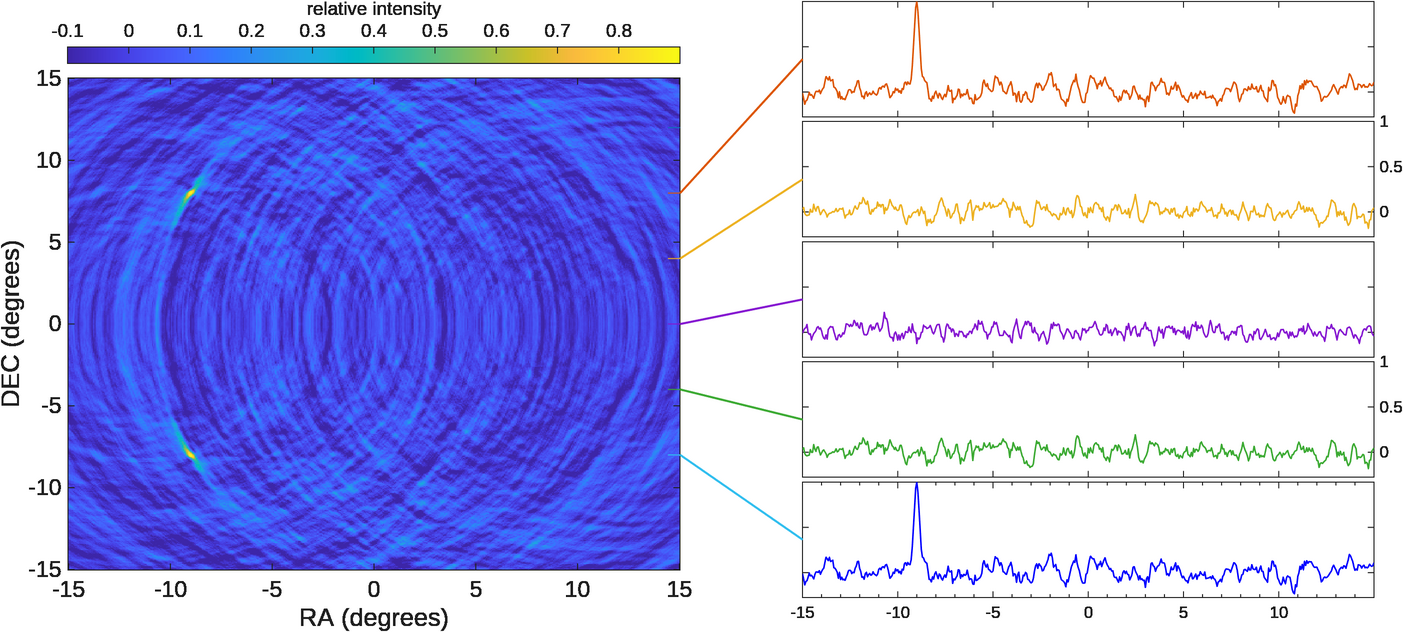}}
\vspace{-2mm}
\caption{(left) Stacked $30^\circ$ ``dirty map'' based on delay-tracking of 600 one-second integrations in a single orbital pass. (right)~1D slices at various DEC values, indicated by the connecting line for each of the slice plots. The zenith projection of the spacecraft path passes horizontally across the field at~$0^\circ$ declination.
\label{Snapshot} }
\vspace{-3mm}
\end{figure}

Figure~\ref{Snapshot} shows a simulated map, with frequency range 0.3--3\,MHz, using a $30^\circ$ portion of a single orbit of the lunar surface. The zenith above the spacecraft thus scans a $30^\circ$  along-track region of the celestial sky, and we map out to an equal angle in the cross-track direction, leading to a square map that, in this case, contains a single point source with an instantaneous intensity equal to the uniform sky background intensity, before reflection off the surface. Such an event could be consistent with a Jovian radio burst, or potentially a bright stellar radio burst as well. Here we have used the full orbital geometry for a spacecraft in a 100~km LLO, and assumed a conservative $\alpha\simeq 0.1$ corresponding to the combination of the expected Fresnel coefficient and an additional $-10$~dB power loss due to surface structure. In a 2-hour orbit, this sky patch is crossed in about 600 seconds, setting the time resolution of the 30$^\circ$ image tiles. We use 1\,s coherent integrations, which ``freeze'' the reflection geometry to adequate phase stability, as discussed below. 

The map is constructed in a manner that conforms to the constraints on LRI data acquisition. After each 1-second integration, we perform delay tracking to recenter the phases for the next integration, relative to the center of the map. At each spacecraft position, the uniform sky background is resolved out by the interferometer, and the source delay is well defined, but the azimuth of the source for each snapshot is undetermined until multiple azimuthal positions are measured.  The point spread function (PSF) for this single-pass map is strongly affected by the ring geometry of the sky fringes, which is evident in the map, for which no further PSF deconvolution has been performed. Because the spatial frequency coverage is limited in this case by the 1-D track across the field, the final map still leaves a mirror ambiguity in the source azimuth. For transient sources visible only during a single orbital pass, the ambiguity could be resolved with external source information. For non-transient sources, this ambiguity is resolved as subsequent orbits precess across the sky patch. Figure~\ref{Snapshot} thus illustrates LRI mapping effectiveness on short time scales; in practice, LRI would observe the same region of sky over many orbits, with a corresponding increase in the SNR and resolution of the map. We will return to this in a later section.


\subsection{Fresnel diffraction.}
For accurate determination of lunar reflections, Fresnel diffraction must be considered at the surface. 
For a narrow-band signal, the regions of phase coherence define a set of zones, called the Fresnel zones, each of which contributes to the far-field intensity, commonly evaluated using the Kirchhoff scalar wave approximation. In this formalism, the first central Fresnel zone (FFZ) contributes the largest fraction of intensity, and for broadband signals, the FFZ is the dominant contribution.
Phase perturbations due to a rough or uneven surface, or tilts due to slopes, are typically evaluated by integrating their effects over the FFZ.

We find that conservative quasi-static temporal coherence is achieved for integration times such that the spacecraft nadir point does not move more than $1/2$ of the \hbox{FFZ}. At $h=100$\,km, the spacecraft nadir position moves at about 1.6\,km\,s${}^{-1}$. Thus at 1~MHz, the FFZ diameter $\sim 11$\,km, and $T \lesssim 3.4$\,s; at 7~MHz the FFZ diameter is $\sim 4.14$~km, and the constraint is thus $T \lesssim 1.3$\,s.

LRI does require real-time orbit knowledge arising from the need for spacecraft ground position knowledge to within half a FFZ diameter, to ensure telemetry is optimized for coherence; at 7~MHz, this is $|\Delta \vec{r}_{nadir}| \lesssim 2$~km. For postprocessing, the angular resolution/accuracy requirements imply a baseline knowledge requirement of $\delta h \simeq h(\tan{\theta}) \Delta\theta$, which is most stringent near the zenith, giving $\delta h \simeq 100$~m.  

LRI would rely on post-processed precision orbit determination using radiometric tracking and GRAIL lunar gravity models, strengthened by surface-referenced geometric constraints that emerge from the autocorrelation delay mapping. Post-processed lunar orbit determination (OD) has been demonstrated: in one LRO summary, radio-only OD performed at about 20\,m, and with high-accuracy GRAIL gravity the radio-only reconstruction improved to $\leq 10$\,m \citep{Mazaricoetal2018}. This error was also predominantly horizontal; the radial error was typically about 0.5 m. 

LRI may not have the same level of ground-tracking coverage available as \hbox{LRO}, but can benefit from a much higher frequency (X-band $3^{\mathrm{rd}}$ generation radio compared to earlier generation S-band), recent updates to GRAIL gravity maps, and the lunar surface knowledge that LRO achieved. Residual phase/delay errors in the LRI effective interferometric baseline can be calibrated empirically during intervals when bright Jovian decametric bursts are in view~\citep{CassiniJupiter}, typically of order 5-10\% of the time. Because these bursts can dominate the sky brightness, they provide high-SNR phase references (e.g., Figure~\ref{Snapshot} above) that can be used to solve low-order residual baseline and instrumental delay terms and propagate those corrections into adjacent science intervals and into post-processed orbits. 


Geometrical errors in altitude also affect phase matching between snapshots. A fixed altitude bias $\delta h$ produces
$$
\delta\phi_j=
-\frac{4\pi\nu}{c}\,\delta h\cos\theta_j,
$$
which changes as the viewing geometry changes. Thus individually coherent snapshots can still combine incorrectly if the bias is omitted from the synthesis model. For example, in the most stringent case for 10~MHz mapping near zenith, 
$$
|\delta\phi|\simeq
0.419\,\frac{|\delta h|}{1\,{\rm m}}\ {\rm rad}.
$$
and requiring coherence $>0.9$ over gaussian random errors implies $\delta \phi \lesssim 0.46$~rad, giving $\delta h \lesssim 1.1$~m. This is a stringent requirement, and while LRO orbital tracking results provide plausible evidence that it can be met, it will require careful attention in post-processing.

Short-time-scale orbital stability is also important to maintain coherence. As we will discuss in a later section, detailed orbital modeling with high-order accurate lunar gravity is now possible. Our simulations measure the RMS deviations from straight-line orbital segments over the $\lesssim 1$~s  timescales for Fresnel-zone coherence, and for orbits at 100 km altitude, we find no more than centimeter-scale RMS deviations, far less than the meter-scale RMS deviations required to degrade coherence at 10~MHz. 

The anticipated LRI observing mode thus consists of a continuous sequence of snapshots of duration $T\lesssim 1$~s. These are processed according to the digital architecture described in a later section, and the primary data products, the channelized complex Fourier spectra, are collated and transmitted on telemetry for each integration time $T$. These spectra can then be combined to build up the integration once corrections for the surface reflectivity (based on LRO mapping) and delay tracking for the field-of-view phase center are performed. 

However, we note that surface coherence alone does not determine the permissible integration time, and our generic estimate of $T\simeq 1$~s must be qualified by the actual effects of fringe rate smearing vs. frequency. Motion through the sky fringes imposes an additional, direction-dependent
constraint. For a differential propagation delay changing at rate $\dot\tau$, a rectangular integration attenuates the fringe amplitude by

$$A=\frac{|\sin(\pi f_{fringe} T)|}{|(\pi f_{fringe} T)|}$$

with the fringe frequency $f_{fringe} = \nu\dot\tau$. In a circular 100~km orbit, spherical specular geometry gives
$|\dot\tau|\simeq0.36\,\mu\mathrm{s}\,\mathrm{s}^{-1}$ for the most extreme case we consider,  a source $\sim 40^\circ$ from zenith in the along-track direction. Limiting amplitude loss to 5\% in this case therefore requires short-time-scale integrations $T_s\lesssim0.49$, $0.10$, and $0.049$~s at 1, 5, and 10~MHz, respectively. The rate is smaller toward the cross-track direction and vanishes to first order at zenith for a circular orbit. Larger loss factors may also be an acceptable trade for higher frequency coverage.

Accordingly, wide-field observations require frequency- and geometry-dependent elementary accumulations of complex spectra at $T_{int}\lesssim T_s(\theta)$, retained before longer integration $T_s \rightarrow T$  by forward modeling of the evolution of the orbital geometry. Delay corrections applied only after temporal averaging cannot restore fringe contrast lost within an accumulation.
Because the fringe-rate has angular dependence, with a high rate along-track, and much smaller in the cross-track direction, averaging of short-time-scale coherent spectra will require tailoring for direction-dependent accumulation. Methods such as fringe-rate filtering~\citep{Parsons_2016} can be implemented to address such issues if telemetry is saturated, at the cost of added complexity in onboard processing.

\begin{figure}[tb!]
\vspace{-2mm}

\centerline{~~\includegraphics[width=\textwidth, trim=0mm 0mm 0mm 0mm, clip]{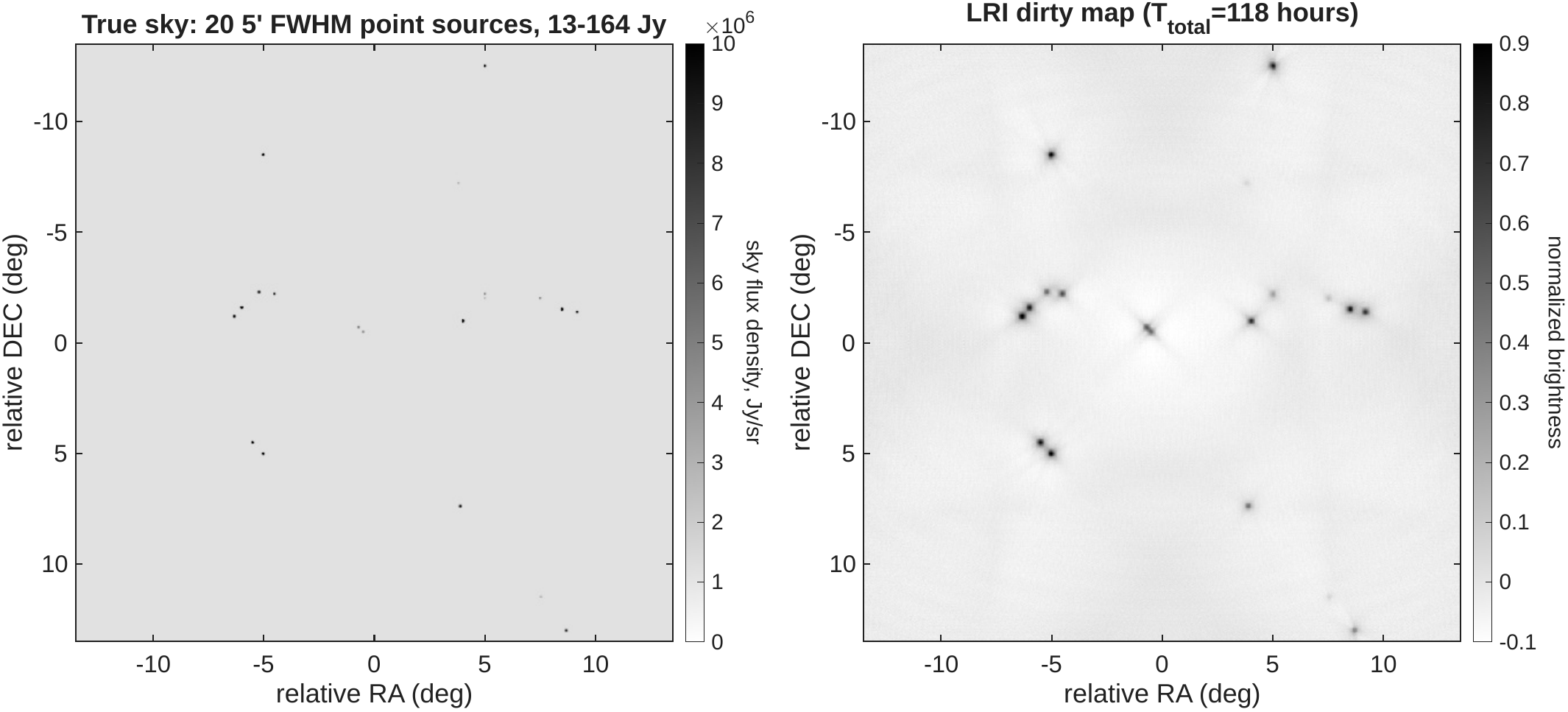}}
\centerline{~~\includegraphics[width=\textwidth, trim=0mm 0mm 0mm 0mm, clip]{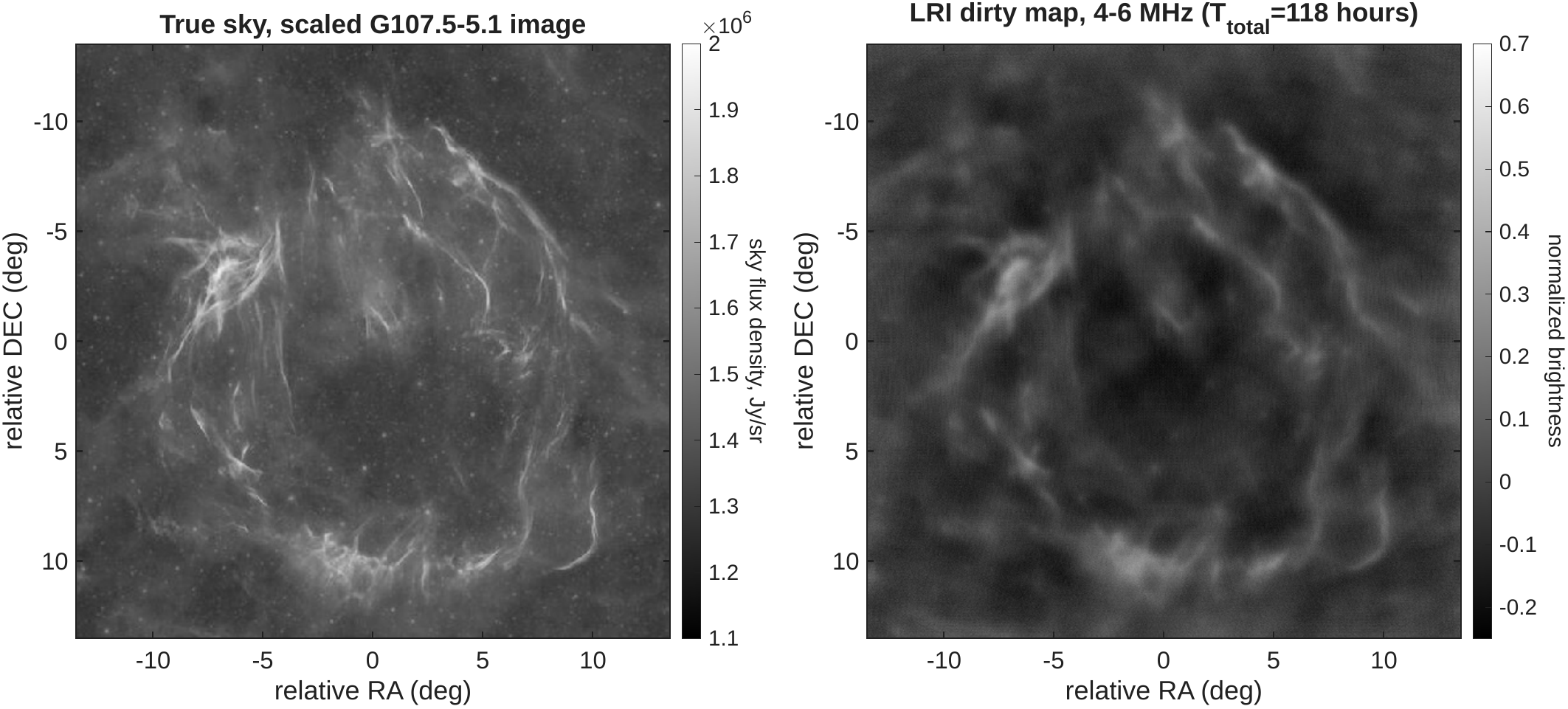}}
\vspace{-3mm}
\caption{(top left) Synthetic 20 point source true-sky image, with a uniform $1.2 \times 10^6$ Jy/sr background included in the map . Source intrinsic size is about 5 arcmin FWHM, and individual flux densities range from about 13 Jy to 164 Jy.  Top right: Normalized dirty map using a 16384-frequency channel complex integrated power spectrum, for about 118 hours (approx. 500 orbit passes for the field of view). Mapping was done via ACF backprojection.
(bottom left) Synthetic extended low-surface brightness source, with integrated flux density of about 50 KJy, on a uniform background of  $1.2 \times 10^6$ Jy/sr, including many weak compact sources. (bottom right) LRI dirty map using the same observation sequence as for the upper row point sources.
\label{LRImaps} }
\end{figure}

\subsection{Correlation and Mapping}\label{sec:map}

In post-processing, there are a variety of ways that mapping can proceed, but a simple algorithm that we have implemented uses the ACF amplitudes at each lag within a sky patch to map to a fixed location on the celestial sphere. Each ACF lag channel defines a ring of proportional amplitude on the sky; see appendix~\ref{appendix2} for more details. Since each snapshot may have a different zenith angle due to spacecraft motion, each ACF has a delay-tracking algorithm applied to ``re-center'' the rings on the appropriate celestial reference point. 

This procedure then applies to any snapshot in which the local zenith angle of the spacecraft is within a set angular range (expected to be $<45^\circ$ due to potential bandwidth smearing, but still to be optimized) of the celestial map center. Once the delays have been corrected, the ring of the appropriate fringe width and zenith center is added to the map with the partial amplitude defined by the snapshot ACF. Over the summation of many thousands of such snapshots, with fringe ``rings'' overlain from many different directions and radii, a dirty map emerges with good fidelity and dynamic range. Figure~\ref{Snapshot} illustrates this development clearly for a single-pass map. 

Once many passes over the same lunar surface region have been made, with gradual precession of the orbit so that the passes cover both orthogonal baseline directions, the map fidelity and noise levels improve significantly. Figure~\ref{LRImaps} shows the results of two simulations of LRI mapping for a group of point sources and a large diffuse extended source. These studies are done with attention to all of the fundamental physical details, with observations taking place over the equivalent of order 120 hours of integration. Since LRI provides only a single baseline, the visibility plane must be gradually filled as the spacecraft's orbit precesses, a process that takes many weeks or even months depending on the sky coordinates relative to the Mare reflectors. 

For this simulation all reflector baselines within a square sky tile $\pm 21^{\circ}$ around the zenith (assumed to coincide with the image field phase center) are used, for some arbitrary celestial reference point that lies above the lunar maria at the beginning of the mapping period. We map two cases, one with multiple point sources against a uniform Galactic background, and a second case with a large, low-surface brightness structure, also over a uniform Galactic noise background. The point sources are assigned a range of brightnesses corresponding to estimates of the LRI sensitivity for an integration of this duration. The diffuse source is modeled after optical images of a known supernova remnant, G107.5-5.1 (the Nereides Nebula), but the scale is changed from the $3^{\circ}$ actual diameter of the remnant, to approximately $20^{\circ}$ diameter, covering most of the imaged field of view; we opt for a synthetic source in this case to validate very wide-field mapping. A true-scale mapping example of the large-angular-scale radio source Cen A is provided in a later section.

In each case we image over the passband from 4-6 MHz, and use the Galactic noise at 5 MHz, roughly $1.2 \times 10^6$ Jy/sr, as the background level. 
The supernova remnant is assigned an integrated flux density of 50 kJy over the displayed input image, corresponding roughly to an average brightness of about $1.2 \times 10^5$ Jy/sr, around 10\% of the uniform Galactic background. A spacecraft lunar orbital period at $h=100$~km is just under two hours, and it precesses $\sim 0.8^{\circ}$ per day in inertial space for a non-polar orbit. Assuming we use the known frozen orbit with an inclination of $50^{\circ}$, the rotation of the Moon will scan across the maria, so it takes several months to acquire complete coverage of the celestial tile, as discussed in more detail below.

In these studies we include only baselines where the ground track passed over the corresponding nadir locations beneath this tile; thus, the largest zenith angle that could contribute to the map was along the diagonals at $\theta=\sqrt{2} \times  21^{\circ} \simeq 30^{\circ}$, corresponding to a baseline of $B = 2 h_{orb} \sin\theta = 100$~km for $h_{orb} = 100$~km.  The reflected signal was reduced in amplitude by the Fresnel factor for the effective permittivity and loss tangent observed by Kaguya LRS~\citep{Kaguya2009,Kaguya2015,kumamoto2021}, $\epsilon \simeq 6,~\tan\delta = 0.02$, and we assume a conservative $-10$~dB power loss due to the coherence.

We display only the so-called dirty maps here, since development of an accurate library of point-spread functions, necessary for deconvolution using CLEAN (e.g.\ \cite{hogbom1974clean}) or other algorithms, is beyond the scope of our study. Despite this limitation, it is evident that for both point sources and large-scale diffuse structure, LRI is able to produce robust maps under these conditions, with the well-motivated assumption that the surface coherence and Fresnel coefficients are well-represented by our current models.

For the compact sources, we set the intrinsic size to about 5 arcmin FWHM, based on expectations for scattering in the interplanetary and interstellar medium. Individual flux densities for the sources range from about 13 Jy to 164 Jy, and all of them appear above the background in the dirty map, although not all are clearly evident without CLEAN deconvolution, and artifacts of the point spread function are apparent, specifically the diffraction spikes due to the square image used to define the baselines. 

For the synthetic supernova remnant, while the uniform Galactic noise is completely resolved out, the non-uniform background haze or nebulosity is still reconstructed, along with the filamentary structure of the remnant. This can be attributed to the fact that LRI, which operates as a kind of ``end-fire'' two-element interferometer, does not lack short baselines; for the sky brightness imaged at or near the zenith, the signal reflects off the nadir, and the baseline extends down to zero length. 

Since LRI thus depends critically on the nature and coherence of the surface reflections, we will turn to this in the following section.


\subsection{Coherence properties of the lunar surface as a reflector.}\label{sec:surface}

Even to the unaided eye, a fundamental property of the lunar surface is apparent: the lunar maria, regions of past volcanism, are darker patches of basaltic material that contrast strongly with the lighter colored highlands (Figure~\ref{fig:mariamap2}).
The maria also have an important property with regard to our investigation: they are far smoother in surface topography than the highlands, both in terms of their distribution of surface slopes and the root-mean-square (RMS) deviation of heights from the average surface elevation.

Altimetric and radar data acquired from the Lunar Reconnaissance Orbiter (LRO~\cite{Smith2010LOLA,Scholten2012LROC}) and the Kaguya Lunar Radar Sounder ~\citep{Araki2009Kaguya} have been used to construct digital elevation models (DEMs) of~98\% of the lunar surface at better than 60\,m horizontal resolution and~1\,m vertical resolution ~\citep{Barker2016SLDEM}. These DEMs enable the level of coherence of reflections over virtually the entire lunar surface, and specifically over all maria regions, to be determined. A study of the global slope distribution~\citep{Rosenburgetal2011} of the lunar surface highlights the coherence issue: absolute median slopes $S_{med}$ over an effective baseline of $L_{\mathrm{min}}=17$~m are estimated to be $7.5^{\circ}$ for the lunar highlands and $2.0^{\circ}$ for the maria. The absolute median slope is directly related to the height RMS: for Gaussian slope distributions, 
\begin{equation}
S_{med} \simeq 0.6745~ \sigma_h~ L^{-1}~ (L / L_{min})^{1-H}
\end{equation}
where $L$ is the outer scale for the averaging, $L_{min}$ is the inner scale, and $H$ is the Hurst parameter; expressing this in terms of the height RMS, we have
\begin{equation}
\label{sigma_h-eq}
\sigma_h \simeq 1.4826 \, S_{\text{med}} \, L^H \, L_{\text{min}}^{1-H}
\end{equation}
where the coefficient 0.6745 and its inverse arise from the median of a standard normal distribution.

The standard model for loss of coherence by reflection from a surface with an RMS height deviation of $\sigma_h$ is the Ruze equation~\citep{Ruze1966, RochblattSeidel1992}, based on a Gaussian phase-screen approximation.
The coherent field reflection coefficient in this approximation is given by 
\[
R_{\mathrm{rough}} = R_{\mathrm{smooth}}  \exp \left ( -\frac{1}{2} \left [ \frac{4 \pi  \sigma_h \cos\theta}{ \lambda}  \right ]^2 \right )
\]
where $R_{\mathrm{smooth}}$ is the Fresnel coefficient for a specular reflection, $\theta$ is the angle from normal incidence, and $\lambda$ is the wavelength.
Thus the factor of~3.5 larger absolute median slope of the highlands compared to the maria, translated to a similar factor in RMS height, precipitously decreases the usefulness of the average highlands as a partially reflective surface, except at very low frequencies.

\begin{figure}
\centerline{~~\includegraphics[width=0.75\textwidth, trim=0mm 0mm 0mm 0mm, clip]{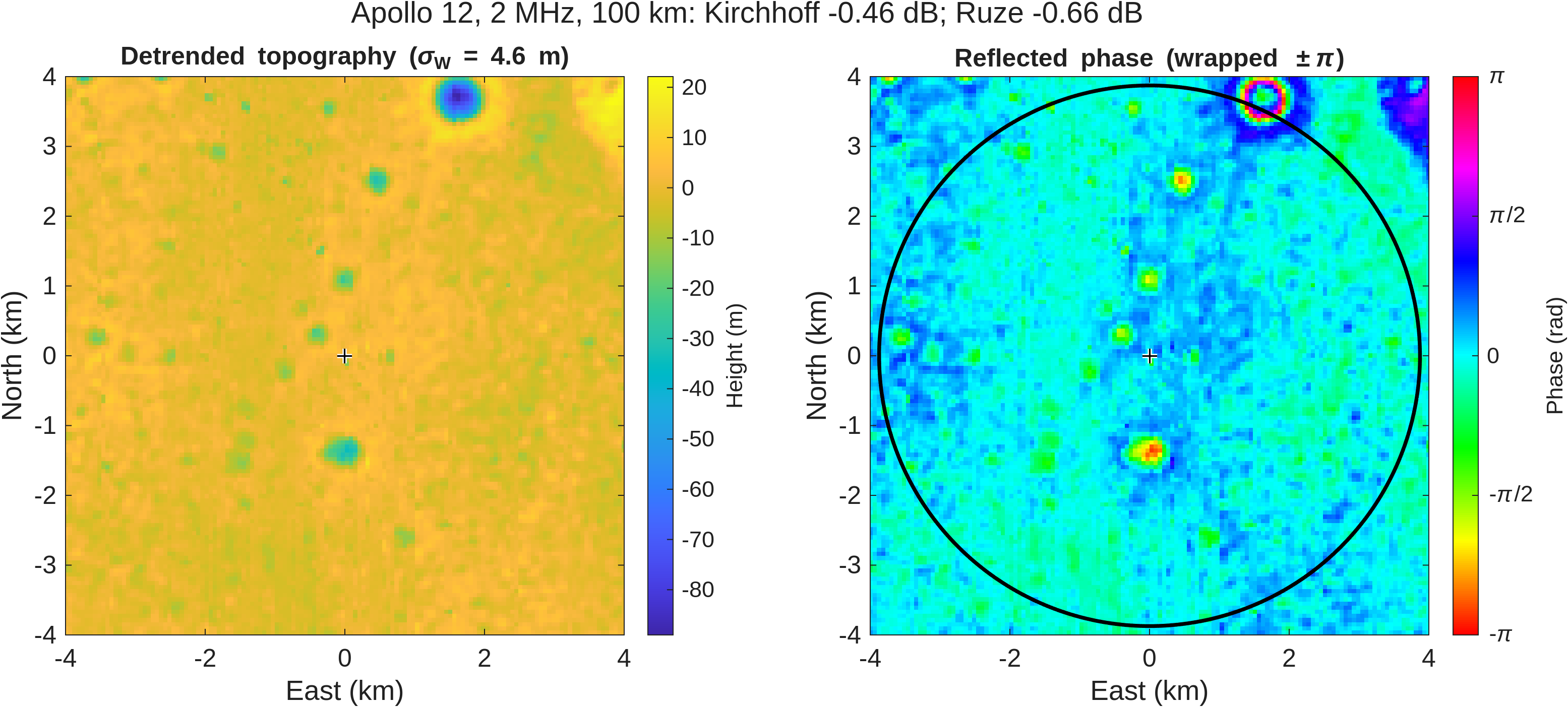}}
\centerline{~~\includegraphics[width=0.75\textwidth, trim=0mm 0mm 0mm 0mm, clip]{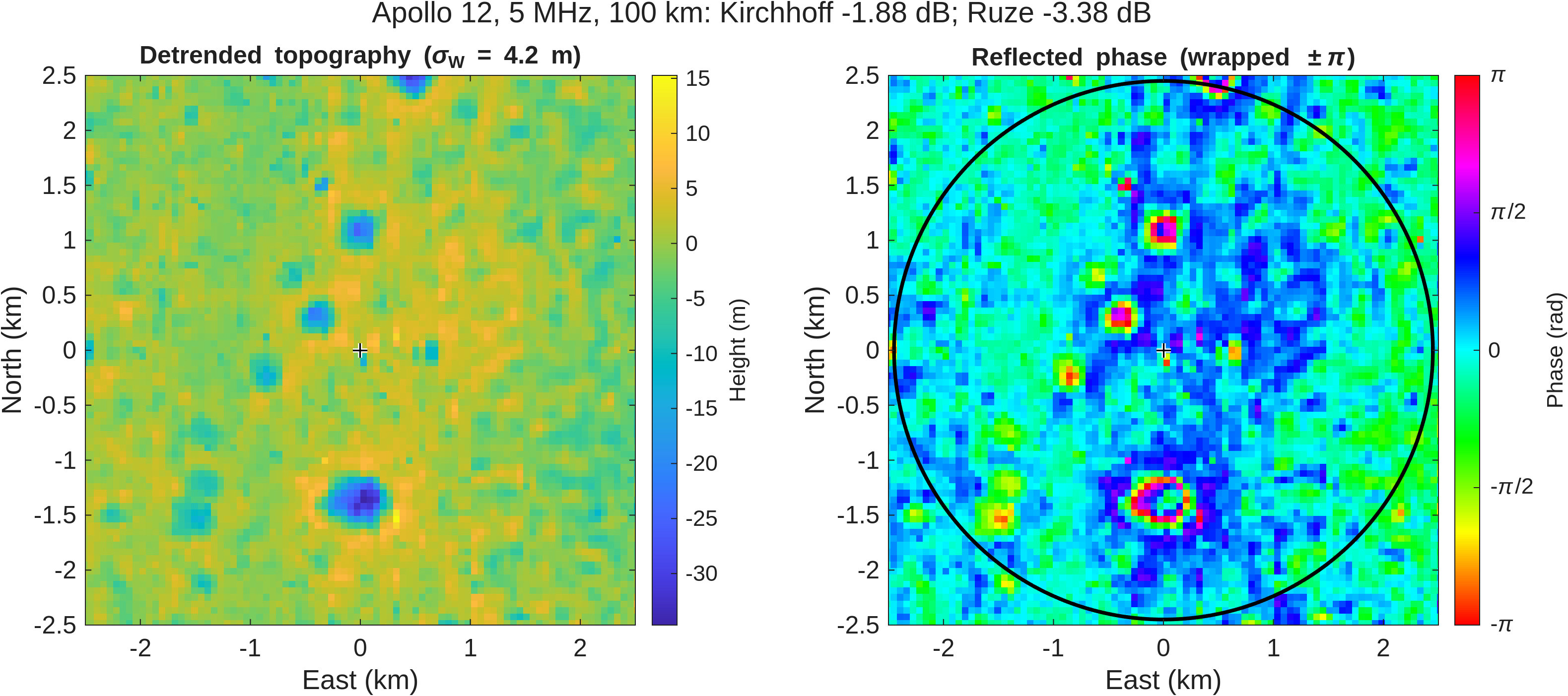}}
\vspace{-2mm}
\caption{Right column: LOLA digital elevation model (DEM) patch at the Apollo 12 site in the extensive maria region Oceanus Procellarum, for  2 MHz (top) and 5 MHz (bottom. Left column: Kirchhoff phase and first Fresnel zone ring at~2\,MHz (top) and~5\,MHz (bottom). Comparative power loss in each figure is computed from both the Kirchoff integral over the first Fresnel zone, and the Ruze equation based on the RMS height.
\label{Kirchoff1} 
}
\end{figure}

The primary limitation of the Ruze model in the context of the lunar surface is the assumption of uncorrelated random roughness. 
Natural surfaces are frequently self-affine (fractal), characterized by a Hurst parameter $H \in [0, 1]$. For the lunar surface as revealed by LOLA data, the surface height distributions are well-represented by a self-affine model with $H \approx 0.76$ for the maria and $H \approx 0.95$ for the highlands. By these models, the height correlation function $C(\tau)$ follows a power-law in the Hurst parameter.

Because the 60~m LRO sampling interval and the 17~m baselines utilized by \cite{Rosenburgetal2011} in his lunar roughness model are both non-negligible relative to the wavelengths of interest below 10~MHz, we note that  the RMS given in equation~\ref{sigma_h-eq}, extrapolated from the 60~m LRO sampling interval down to $\sim 1$~m using the fitted Hurst parameter values and median slopes above, is $\sigma_h \leq 2$~m RMS for the maria, and $\sigma_h \leq 10$~m RMS for the highlands, so the LRO sampling interval and derived self-affine $L_{min}$ does not impact coherence at sub-60~m scales within the maria for any wavelength in our range, nor for the highlands at lower frequencies. We can thus use the LRO DEM at its current sampling interval with confidence in what follows. 

High values of the Hurst parameter $H$ (persistent surfaces) imply that the power spectrum of the roughness is heavily weighted toward low spatial frequencies. In such cases, the ``roughness'' perceived by the wavefront is not a diffuse scatterer, but rather a series of local quasi-specular facets or ``glints.'' In this regime, very characteristic of the lunar surface, the Ruze equation tends to strongly underestimate the coherence of the reflection.

For LRI, the relevant surface area for any reflection is the FFZ on the surface, as introduced above. The FFZ in our context is the region over which the path difference for reflection is less than a half-wavelength. At the radius where the path difference is precisely a half-wavelength, there is destructive interference, and a null in the angular spectrum.  For normal incidence at center, the diameter of the FFZ  is given by $D_{F} = 2 \sqrt{\lambda h}$ where $h$ is the height of the orbit above the lunar surface. Although the higher order annular zones outside of this FFZ do make important contributions for narrow band measurements, for broader bandwidths, the FFZ provides the dominant contribution.

Because of the limitations of Ruze equation modeling for \hbox{LRI}, a more rigorous approach that directly integrates the DEM surfaces using the Kirchhoff Integral (based on scalar wave theory) over the detrended DEM is used,
\begin{equation}
    E_s = \frac{-i e^{ikR}}{2\pi R} \iint_{S} W(x,y) \exp[i 2k z(x,y) \cos \theta] \, dx dy.
    \label{eqn:ki}
\end{equation}

where $R$ is the distance from the receiver to the surface, $k=2\pi/\lambda$ is the wave number, $W(x,y)$ is the Fresnel zone weighting function, and $z(x,y)$ is the digital surface model. Because the radius of curvature ($R_c$) of the self-affine lunar terrain is significantly larger than $\lambda$, the stationary phase points (specular glints) within the Fresnel zone dominate the integral, resulting in a reflected intensity that may far exceed the Ruze prediction.

\begin{figure}
\centerline{~~\includegraphics[width=7in, trim=0mm 0mm 0mm 0mm, clip]{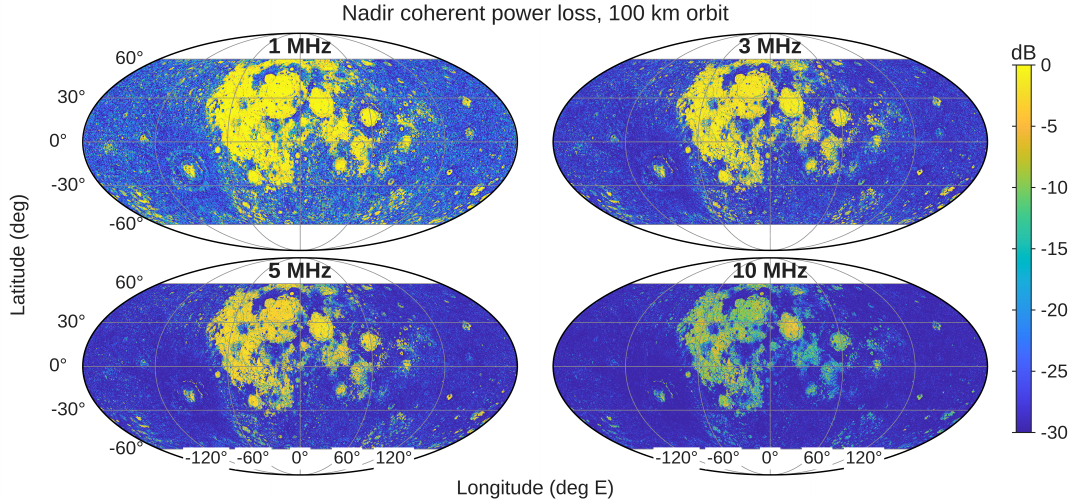}}
\vspace{3mm}
\centerline{~~\includegraphics[width=3.5in, trim=0mm 0mm 0mm 0mm, clip]{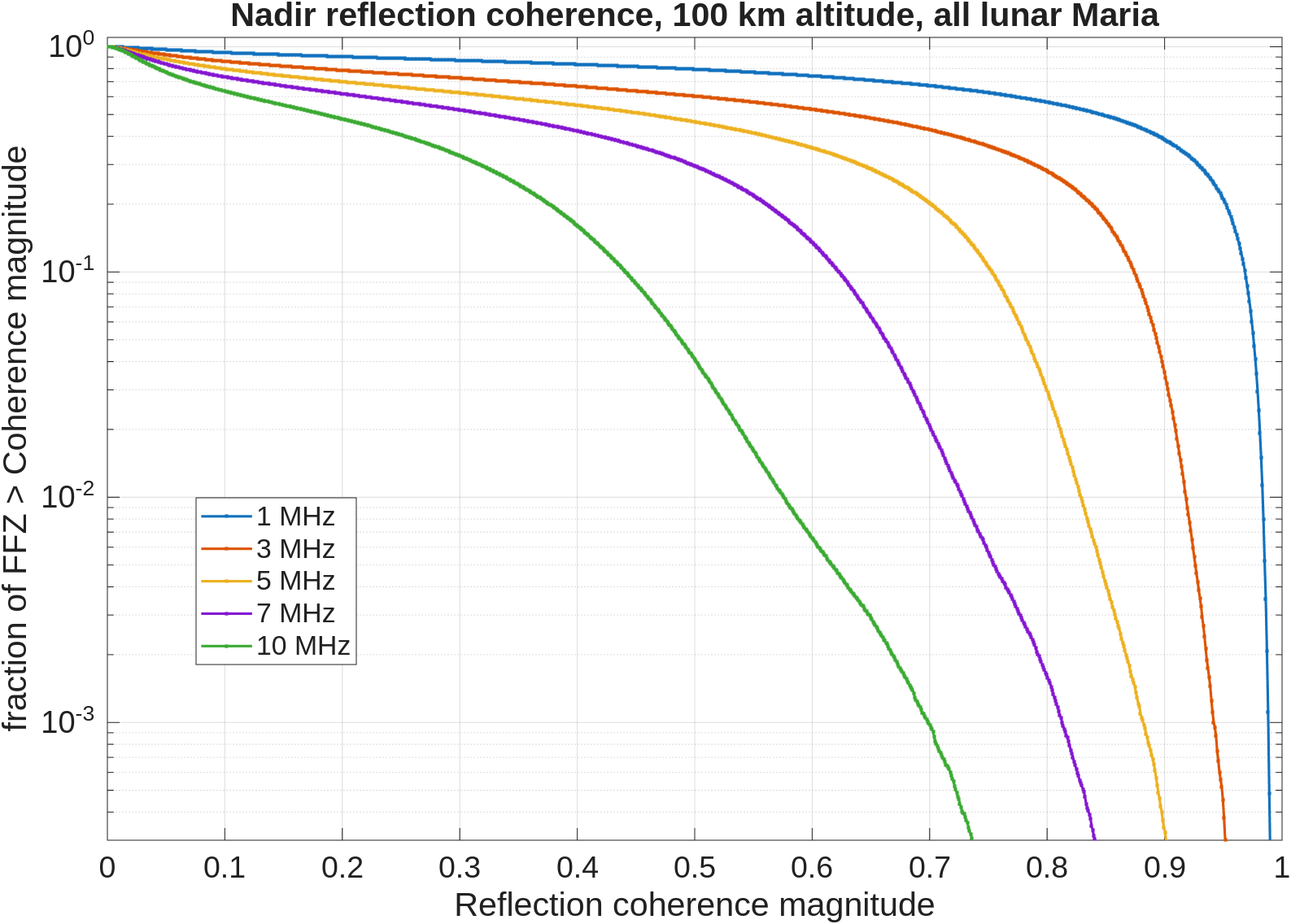}\includegraphics[width=3.5in, trim=0mm 0mm 0mm 0mm, clip]{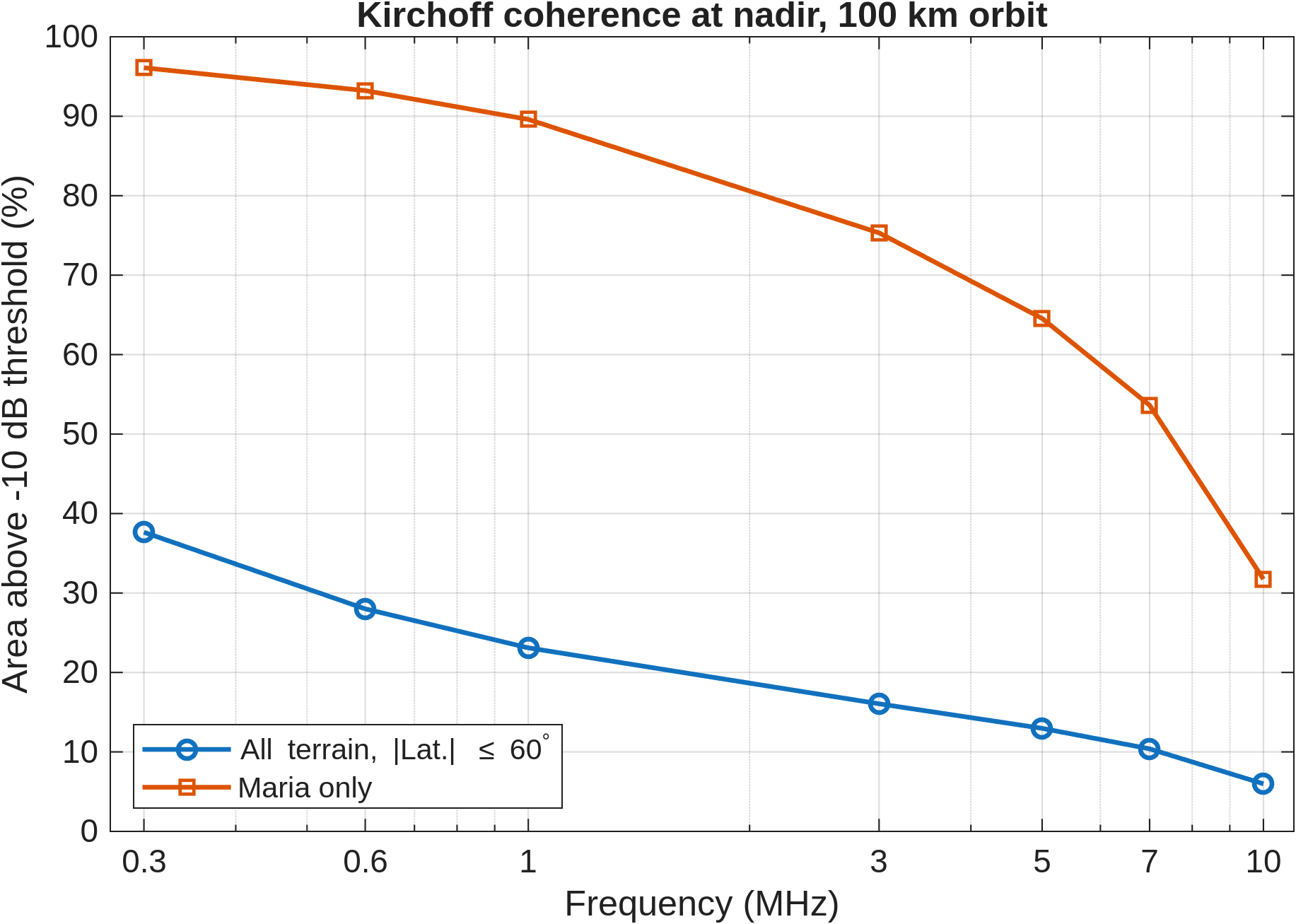}}
\vspace{-2mm}
\caption{Top four panels: Reflectivity, in excess power loss (dB) above Fresnel, for several frequencies over the entire Moon between latitudes 60S and 60N, covering virtually all the lunar Maria, in a Mollweide equal-area projection. Bottom left: Cumulative distributions of Kirchhoff integral reflected coherence factors from 1--10\,MHz over all lunar Maria. Our nominal mapping threshold requirement of -10 dB in power reflection corresponds to a coherence magnitude of $\sim 0.32$. Bottom right: Area above the -10 dB coherence threshold for all terrain within latitude $\pm 60^{\circ}$, and for the maria only within that same range.
\label{Kirchoff2} 
}
\end{figure}

Figure~\ref{Kirchoff1} shows two examples of Kirchhoff integral results for Fresnel zone patches in the Oceanus Procellarum mare region, centered on the Apollo 12 landing site (marked with a cross), at 2 MHz (top row), and 5 MHz (bottom row). In each case the areas of the patches are scaled to enclose the first Fresnel zone for the wavelengths of interest, assuming a spacecraft height of 100~km, and the RMS of the height is indicated at the top by $\sigma_W$. False color images of the detrended (overall slope removed) LOLA data are shown in the left column, and on the right side the phase, in units of $\pi$ radians is plotted over each region, for the wavelength considered. At the top of each plot the results of the loss of coherence for the Kirchhoff integral and the Ruze equation are shown.

Kirchhoff theory was used effectively for the Lunar Radar Sounder on the Kaguya/Selene spacecraft to model lunar reflections at~5\,MHz, for the removal of clutter and sharpening of the broad radar main beam~\citep{Kobayashi2010,Ono2009}.
In actual observations, the typical losses above Fresnel were 2--3\,dB for the maria, which were also found to have an effective dielectric permittivity of up to $\epsilon_r \simeq 6$, rather than the typical near-surface regolith permittivity of $\epsilon_r \simeq 3$, indicating the 5\,MHz reflections were responding to bulk higher-density materials below the regolith, which has typical depths of 3--5~m, a small fraction of a wavelength at these frequencies. 


\begin{table*}[t]
\centering
\caption{Propagation of lunar-reflection uncertainties into LRI imaging
and source recovery.}
\label{tab:reflection-uncertainties}
\small
\setlength{\tabcolsep}{4pt}
\renewcommand{\arraystretch}{1.12}
\edef\LRIlabelwidth{\the\dimexpr0.12\textwidth\relax}
\edef\LRIeffectwidth{\the\dimexpr
  (\textwidth-\LRIlabelwidth-6\tabcolsep)/3\relax}
\newcommand{\LRIuncertaintycell}[2]{%
  \begin{minipage}[t]{#1}\raggedright
  #2\par
  \end{minipage}%
}
\begin{tabular}{@{}llll@{}}
\toprule
\LRIuncertaintycell{\LRIlabelwidth}{
\textbf{Uncertainty}
}
&
\LRIuncertaintycell{\LRIeffectwidth}{
\textbf{Coherence effects}
}
&
\LRIuncertaintycell{\LRIeffectwidth}{
\textbf{Mapping effects}
}
&
\LRIuncertaintycell{\LRIeffectwidth}{
\textbf{Source-recovery performance}
} \\
\midrule

\LRIuncertaintycell{\LRIlabelwidth}{
\textbf{Unresolved topography}
}
&
\LRIuncertaintycell{\LRIeffectwidth}{
For Gaussian residual heights, the mean coherent field is reduced by
\[
  \eta_u=\exp\!\left[-\frac{q^2\sigma_u^2}{2}\right],
  \quad \gamma_u=\eta_u^2,
\]
where $\gamma_u$ is the coherent-power factor.
Suppression increases with frequency and residual height variance.
}
&
\LRIuncertaintycell{\LRIeffectwidth}{
Height errors introduce phase errors
$\delta\phi=q\,\delta h$.
Spatially correlated errors can distort the PSF, change sidelobes,
and bias positions.
RMS height alone does not specify map blur; the spatial correlation
or roughness spectrum is also required.
}
&
\LRIuncertaintycell{\LRIeffectwidth}{
For uniform attenuation with otherwise unchanged noise,
\[
  \begin{gathered}
    \mathrm{SNR}/\mathrm{SNR}_0=\eta_u,\\
    S_{\min}/S_{\min,0}=\eta_u^{-1},\\
    t_{\rm req}/t_0=\eta_u^{-2}.
  \end{gathered}
\]
Unmodelled spatial variations can leave bright-source residuals
that obscure faint sources.
} \\
\midrule

\LRIuncertaintycell{\LRIlabelwidth}{
\textbf{Dielectric properties}
}
&
\LRIuncertaintycell{\LRIeffectwidth}{
At normal incidence,
\[
  \begin{gathered}
    r=\frac{1-\sqrt{\epsilon_r}}{1+\sqrt{\epsilon_r}},\\[2pt]
    \frac{\delta r}{r}\simeq
    \frac{\delta\epsilon_r}
         {\sqrt{\epsilon_r}(\epsilon_r-1)}.
  \end{gathered}
\]
A uniform dielectric change alters amplitude and possibly phase
without necessarily causing decoherence.
}
&
\LRIuncertaintycell{\LRIeffectwidth}{
A common gain error gives
\[
  \frac{\delta S}{S}\simeq
  \operatorname{Re}\!\left(\frac{\delta r}{r}\right).
\]
Frequency-dependent phase contributes delay
$\tau_r=-(2\pi)^{-1}\partial_\nu\arg r$.
Spatial and polarization variations can produce flux, spectral,
positional, and polarization errors.
}
&
\LRIuncertaintycell{\LRIeffectwidth}{
For an amplitude-only change,
\[
  S_{\min}\propto\lvert r\rvert^{-1},
  \qquad t_{\rm req}\propto\lvert r\rvert^{-2}.
\]
Phase errors introduce additional mismatch.
A common unknown gain is degenerate with source flux;
near a reflection null, fractional-error approximations fail.
} \\
\midrule

\LRIuncertaintycell{\LRIlabelwidth}{
\textbf{Subsurface structure}
}
&
\LRIuncertaintycell{\LRIeffectwidth}{
Multiple reflections give
\[
  \Gamma_{\rm eff}
  =r_s+\sum_m a_m e^{-2\pi\mathrm{i}\nu\tau_m}.
\]
For a buried interface near normal incidence,
$\tau_b\simeq2n'd/c$, with field attenuation $e^{-2\alpha d}$.
Stable echoes can enhance or suppress the net coherent field.
}
&
\LRIuncertaintycell{\LRIeffectwidth}{
Surface--subsurface interference produces spectral ripples with spacing
\[
  \Delta\nu_{\rm ripple}\simeq1/\tau_b
\]
and displaced lag responses.
Unmodelled echoes can generate displaced rings or false structure.
Near interference minima, a single effective delay is inadequate.
}
&
\LRIuncertaintycell{\LRIeffectwidth}{
Depth and index errors give
\[
  \delta\tau_b\simeq
  \frac{2}{c}(n'\delta d+d\,\delta n').
\]
For positive template overlap $\rho$,
$\mathrm{SNR}_{\rm used}/\mathrm{SNR}_{\rm optimal}=\rho$;
the detection threshold increases by $1/\rho$.
Destructive interference can also reduce the optimal SNR itself.
} \\
\bottomrule
\end{tabular}

\vspace{0.5em}
\begin{minipage}{\textwidth}
\footnotesize
\textit{Notation.}
$q=4\pi\cos i/\lambda$, where $i$ is local incidence angle and
$\lambda=c/\nu$; $\sigma_u$ is unresolved residual height RMS;
$\epsilon_r$ is complex relative permittivity; $r$ is a Fresnel
\emph{field} coefficient; and PSF denotes point-spread function.
$S_{\min}$ is the limiting source flux at fixed significance and
$t_{\rm req}$ is the integration required for a specified SNR.
$r_s$ is the surface field coefficient; $a_m$ and $\tau_m$ are the
complex amplitude and excess delay of a buried contribution.
$d$ is layer depth, $n'$ its real refractive index, and $\alpha$
its field attenuation coefficient.
$0<\rho\leq1$ denotes the normalized noise-weighted overlap between
assumed and true source templates.
The delay convention uses fields proportional to
$e^{+2\pi\mathrm{i}\nu t}$.

\end{minipage}
\end{table*}

The Kirchhoff integral method tends to underestimate the coherence since the true reflective surface is an average over a subwavelength portion of the subsurface materials and structure. For detailed LRI calibration, we anticipate that the simpler scalar Kirchoff method we have employed in our initial investigation will be replaced by a full vector Kirchoff treatment, including full preservation of transverse electric, transverse magnetic, and cross-polarization effects, and the complex permittivity of the regolith. Since the coherent reflection arises over the Fresnel zone regions, covering square kilometers of surface, there are multi-scale averages involved. Such effects will certainly require careful accounting to fully calibrate the measured visibilities in post-processing. In Table~\ref{tab:reflection-uncertainties} we summarize some of the possible effects to be addressed in calibration, and outline their impact on the observables.

The maria, covering about 16\% of the lunar surface, are natural regions to consider for LRI reflective coherence, and the Kaguya results reinforce this conclusion. The highlands, covering the bulk of the remaining surface, present a much less coherent reflector.  With the exception of a number of large crater floors on either the near or the far side, typical highlands regions will suffer a much larger loss of coherence; even if the Ruze estimates are pessimistic, usable reflections over the highlands will be constrained to lower frequencies, and thus different science targets for LRI.

Figure~\ref{Kirchoff2}(top) shows a four-panel, wider-scale view of Kirchhoff-estimated reflectivity for the entire lunar surface between latitudes of 60S and 60N, which cover virtually all of the maria, based on LRO data~\citep{Barker2016SLDEM}. The coherence losses here are expressed as the excess reflected power loss (dB) above the Fresnel reflection loss, for LRI frequencies from~1\,MHz--10\,MHz.
Even with an excess power loss of~$-10\,\mathrm{dB}$ or more, we find that robust maps are produced.  In all of our mapping simulations here, we fix the reflection coherence to ~$-10\,\mathrm{dB}$ for conservative purposes, although clearly at low frequencies, reflection coherence will often be quite high. In practice, areas that garner high orbital coverage will allow mapping to weaker levels of reflectivity, and the archived data will retain this information regardless, allowing for later improvements in mapping.

Figure~\ref{Kirchoff2} (bottom left) shows cumulative distributions of reflection coherence amplitude from 1--10\,MHz from Kirchhoff integrals over the Fresnel zones of many thousands of individual patches covering all lunar Maria, both near and far side, within the 60S to 60N bounds. At the lower frequencies, the major fraction of the maria will produce very minimal reflective losses and high dynamic range. In Figure~\ref{Kirchoff2} (bottom right) we show the fraction of all terrain vs. frequency above the -10 dB threshold, and also the fraction within the maria, both within the expected $\pm 60^{\circ}$ latitude range accessible to the anticipated $i=50^{\circ}$ frozen orbit.



\subsection{Sensitivity}\label{sec:science.sensitivity}

Before discussing the specific approach to sky mapping, it is useful to set the sensitivity scale for LRI based on traditional radio interferometry analysis.  Unlike a conventional two-element radio interferometer with identical dishes, the LRI configuration features unequal elements due to dielectric and scattering losses at the reflection interface. The physical antenna possesses an effective collecting area~$A_{\mathrm{e1}}$, which for an electrically short dipole in free space is bounded by its Hertzian limit:
\begin{equation}
A_{\mathrm{e1}} = \frac{3\lambda^2}{8\pi}.
\label{eqn:hertzdipole}
\end{equation}
For  reflections away from the zenith, the projection of the dipole pattern degrades the available collecting area by a geometric factor, modifying the response to $A_{\mathrm{e1}}(\theta_z) = A_{\mathrm{e1}} \cos^2\theta_z$.
In practice, since many baselines with different zenith angles will contribute to the sensitivity, some average value $\langle \cos\theta_z \rangle$ obtains.

The LRI virtual image antenna's effective collecting area $A_{e2}$ is attenuated by the surface properties.
We define an effective power reflection coefficient $|R_{\text{eff}}|^2 = \gamma |R_{\text{Fresnel}}|^2$, where $R_{\text{Fresnel}}$ is the standard complex Fresnel reflection coefficient evaluated at the incidence angle $\theta_i = \theta_z$ for a medium with complex permittivity $\epsilon_r$, and $\gamma$ is an empirical coherence factor ($\gamma \le 1$) capturing surface roughness and scattering decoherence. In the most conservative case, $\gamma$ would follow the Ruze equation above:
$$\gamma(f,\theta_i)=
\exp\left[-\left(\frac{4\pi\sigma_h f\cos\theta_i}{c}\right)^2\right]$$
and we find reasonable agreement at frequencies below $\sim 5$~MHz with $\sigma_h \simeq 6.0$~m for $\theta_i = 30^{\circ}$. For these parameters, $R_{eff} \simeq 0.17$ at 1~MHz, 0.10 at 3~MHz, 0.05 at 5~MHz, but drops rapidly above this, in a manner that is inconsistent with the more accurate Kirchoff results, as we have also noted above. We have used a softened exponential here to extend to 10~MHz, where $R_{eff} \simeq 0.01$

The image area scales as $A_{e2} = |R_{\text{eff}}|^2 A_{e1}$. The cross-correlated effective area of the interferometer is the geometric mean of the two apertures:
\begin{equation}
A_{\text{eff}} = \sqrt{A_{e1} A_{e2}} = |R_{\text{eff}}| A_{e1} \langle \cos\theta_z \rangle^2
\end{equation}

\subsubsection{System Temperature and Sky Noise}
In the $0.1 - 10\text{ MHz}$ band, the total noise budget is dominated by the combination of Galactic synchrotron background radiation ($T_{\text{sky}} \sim 10^5 - 10^7\text{ K}$) and separate effect of solar illumination and the solar wind. The solar wind plasma creates a region  of enhanced noise around the antenna in an effect known as the  {\it quasi-thermal noise} (QTN)~\citep{QTN1989, QTN2017}. Separately, direct solar ultraviolet illumination ejects photoelectrons from the antenna surface, leading to additional noise at low frequencies. 

\begin{figure}[tb!]
\vspace{-2mm}
\centerline{~~\includegraphics[width=0.98\textwidth, trim=0mm 0mm 0mm 0mm, clip]{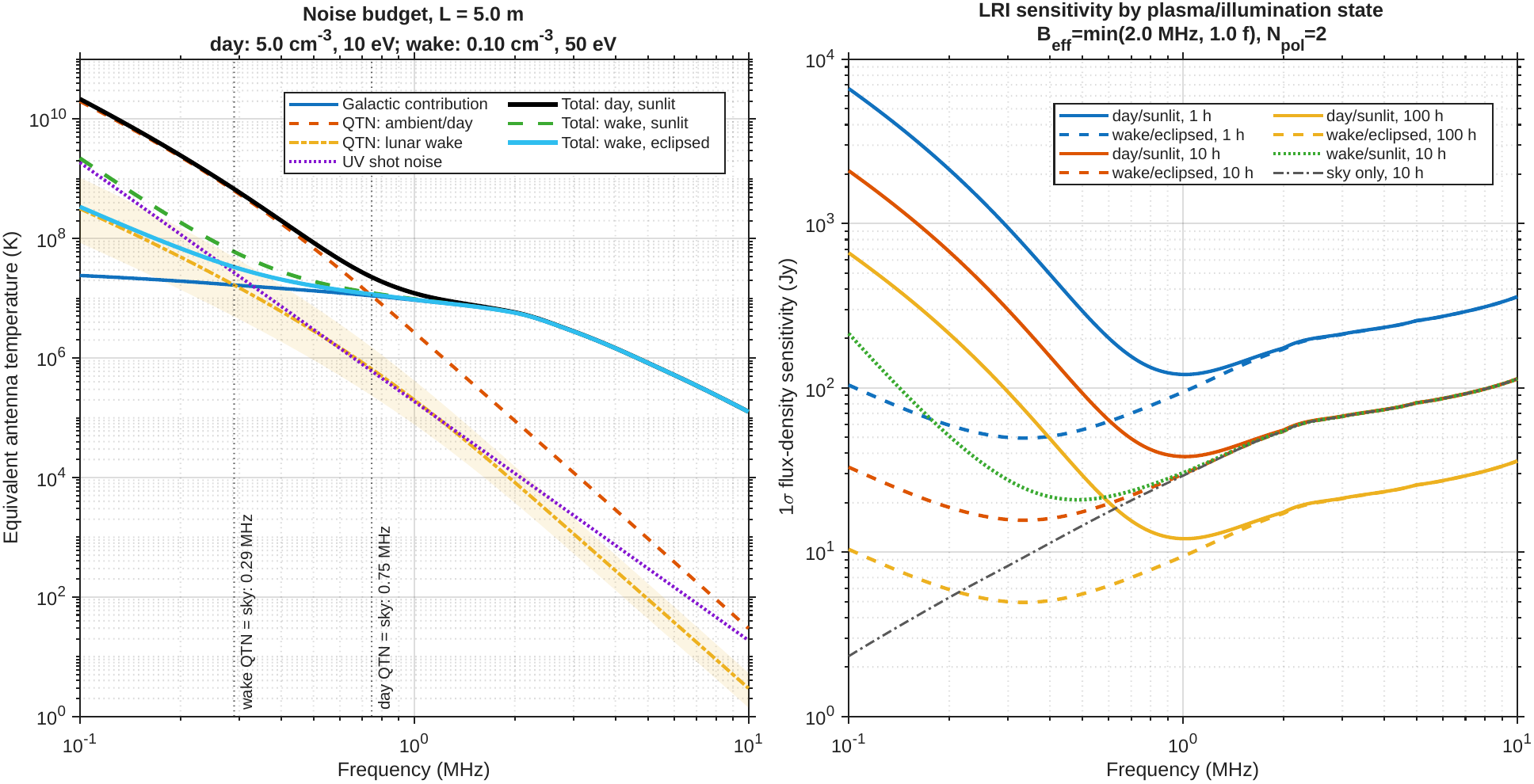}}
\vspace{-3mm}
\caption{(Left) Antenna noise temperature vs. frequency for various components and their sum. (Right) Interferometer sensitivity in Jansky vs. frequency for observation durations of 1, 10, and 100 hours, and several cases of solar illumination. 
\label{fig:LRIsense} }
\end{figure}

For high-quality receivers, these external thermal noise sources render internal thermal receiver noise negligible with a careful choice of antenna and high impedance amplifier to sense the antenna open-circuit voltage. The single receiver terminal records the voltage superposition of both the direct Galactic+(solar-induced noise)  background $T_{sky}$ and the back-scattered reflected sky background. The total system noise temperature $T_{\text{sys}} $ seen by the receiver as a function of frequency $f$ is written as:
\begin{equation}
T_{\rm sys}^{(s)}(f) \simeq \frac{T_{\rm gal}(f)}{2} \left[1+\left|R_{\rm eff}(f)\right|^2\right]
 +T_{\rm QTN}^{(s)}(f)
 +\chi_\odot^{(s)}T_{\rm ph}(f)
 +T_{\rm rx}(f),  ~~~\chi^{(s)}_\odot=
\begin{cases}
1,&\text{sunlit antenna},\\
0,&\text{lunar eclipse}.
\end{cases}
\end{equation}
where (for now) we have neglected the receiver noise term which is a small fraction of the Galactic noise. The leading factor of 1/2 on $T_{sky}$ arises from the approximation that the Moon blocks approximately half the sky, and then partially reflects the uniform sky noise back into the antenna, leading to an overall lower antenna noise temperature than if the Moon were not present. The QTN and photoelectron terms $T^{(s)}_{QTN}, T_{ph}$ depend on solar illumination, and are computed using standard engineering estimates given by the references noted above.

\subsubsection{Modified Sensitivity Estimator}
The minimum detectable flux density $\Delta S$ of a radio interferometer is dictated by the radiometer equation. To account for the unequal elements of the LRI geometry and to include the sensitivity improvement gained by averaging two orthogonal, independent polarization states, the classical equation is structured as:
\begin{equation}
\Delta S = \frac{2 k_B T_{\text{sys}}}{A_{\text{eff}} \sqrt{2  \Delta\nu  \tau}}
\end{equation}
where $k_B$ is the Boltzmann constant, $\Delta\nu$ is the pre-detection bandwidth, $\tau$ is the post-detection integration time, and the factor of $\sqrt{2}$ in the denominator stems from dual-polarization combining. Substituting the explicit expressions for $T_{\text{sys}}$ and $A_{\text{eff}}$ yields:
\begin{equation}
\Delta S = \frac{ k_B [ 2(T^{(s)}_{QTN}+ \chi^{(s)}_\odot T_{ph}) + T_{\text{sky}} \left(1 + |R_{\text{eff}}|^2\right)]}{|R_{\text{eff}}| A_{e1} \langle \cos\theta_z \rangle^2 \sqrt{2 \cdot \Delta\nu \cdot \tau}}~.
\end{equation}

This is the most general form for the sensitivity, but in fact QTN and photoelectron noise rapidly fall with frequency and become negligible above $\sim 1$~MHz. For that case we can isolate the instrument and environmental parameters from the measurement metrics, and the LRI sensitivity for most of the passband becomes:
\begin{equation}
\Delta S = \left[ \frac{8\pi k_B T_{\text{sky}}}{3\lambda^2 \langle \cos\theta_z \rangle^2 \sqrt{2  \Delta\nu  \tau}} \right] \cdot \left[ \frac{1 + |R_{\text{eff}}|^2}{|R_{\text{eff}}|} \right]  ~~~(\nu \gtrsim 1~\rm{MHz})
\end{equation}

The second bracketed term operates as a distinct asymmetric penalty factor. If the lunar surface were a perfect lossless reflector ($|R_{\text{eff}}| = 1$), this term would minimize to $2$. For realistic, lossy regolith environments where $|R_{\text{eff}}| < 1$, this penalty scales inversely with the reflection magnitude, illustrating how surface decoherence directly impairs the minimum detectable flux threshold.

Figure~\ref{fig:LRIsense} shows several components of the sensitivity estimate, here based on a dipole antenna with 1~cm diameter and a tip-to-tip dipole length of 5~m. On the left, the antenna noise temperature is shown, along with its various components as functions of frequency.  Galactic noise rises steeply from 10~MHz down to lower frequencies, but has a turnover near 2--3\,MHz due to free-free absorption in the Galaxy. Antenna temperature is dominated by this noise at the higher part of the LRI band, but below 1~MHz, the solar plasma QTN and photoelectron effects dominate.

The right side of Figure~\ref{fig:LRIsense} shows the sensitivity as a function of frequency for a source at a zenith angle of $30^{\circ}$,  using a bandwidth of 2~MHz at and above 2~MHz, decreasing linearly with frequency below 2 MHz, but remaining at 100\% fractional bandwidth. The three integration times considered correspond roughly to a single orbital pass ($\sim 1$\,hr), of order 1 full day of combined observations ($\sim 10$\,hr), and several weeks of integrated observations ($\sim 100$ hr). 
The difference in sensitivity in the lowest frequency band is more than an order of magnitude in the sunlit and deep lunar eclipse phases.

At fixed frequency, sky temperature, and integration time, if we define \(r=|R_{\rm eff}|\) and \(\mu=\langle\cos\theta_z\rangle\), then for small, independent uncertainties, the fractional uncertainty in the predicted sensitivity is
\[
\left(\frac{\sigma_{\Delta S}}{\Delta S}\right)^2
\simeq
\frac14\left(\frac{\sigma_{\Delta\nu}}{\Delta\nu}\right)^2
+4\left(\frac{\sigma_\mu}{\mu}\right)^2
+\left(\frac{1-r^2}{1+r^2}\right)^2
\left(\frac{\sigma_r}{r}\right)^2.
\]
Thus, in the weak-reflection limit \(r\ll1\), a 10\% uncertainty in effective bandwidth, angular factor, or reflection amplitude contributes approximately 5\%, 20\%, or 10\%, respectively, to the sensitivity uncertainty.

\subsection{Sky Mapping Methodology}\label{sec:tech.skymapping}

Having established that the lunar maria will provide adequate coherence for LRI mapping, we now discuss the mapping strategy.  Critical to this is the choice of orbit.  At altitudes $\lesssim 100$\,km, which are required to maintain adequate reflection coherence, lunar mass concentrations strongly affect orbit stability. For our purposes, the figures of merit for an orbit are the dwell time spent over the maria, combined with steady inertial precession of the orbit, so that the mapping gradually covers the celestial sky.  The planned orbit for LRI has an altitude $h = 100\,\mathrm{km}$. This altitude achieves a balance between the various factors mentioned above.

To help validate an LRI mission, a high-fidelity lunar orbit propagation and Monte Carlo stationkeeping simulator has been developed that numerically integrates the spacecraft equations of motion under the GRAIL-derived GRGM1200A spherical harmonic gravity field~\citep{Goossensetal2020}, with event-driven impulsive burns for altitude maintenance. The model reproduces expected secular behaviors (e.g., ascending node drift rates and perilune evolution consistent with published frozen-orbit studies) and yields stationkeeping $\Delta v$ budgets in the same order of magnitude as published results for Lunar Prospector, providing an internal physics sanity check and validation anchor.

\begin{figure}[tb!]
    \centering
\centerline{\includegraphics[width=\linewidth]{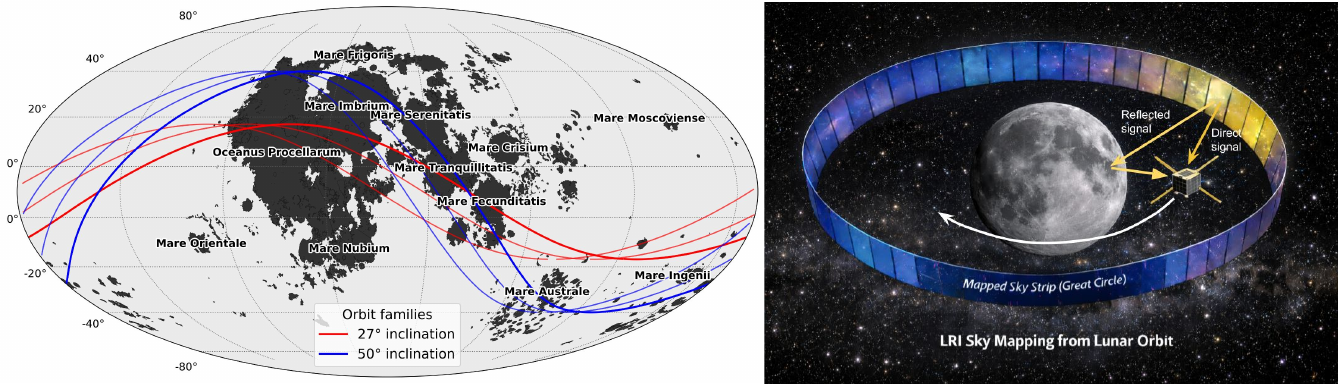}}
    \caption{(left) Equal area projection of the lunar surface, centered on the Maria. The orbits at $i=27^\circ$ and~$50^\circ$ are shown in bolder solid lines, with two additional precessed orbit ground tracks shown after one and two lunar sidereal months. (right) Illustration of LRI mapping methodology.}
    \label{fig:mariamap2}
    \vspace{0mm}
\end{figure}

\begin{figure}
\centerline{~~\includegraphics[width=0.85\textwidth, trim=0mm 0mm 0mm 0mm, clip]{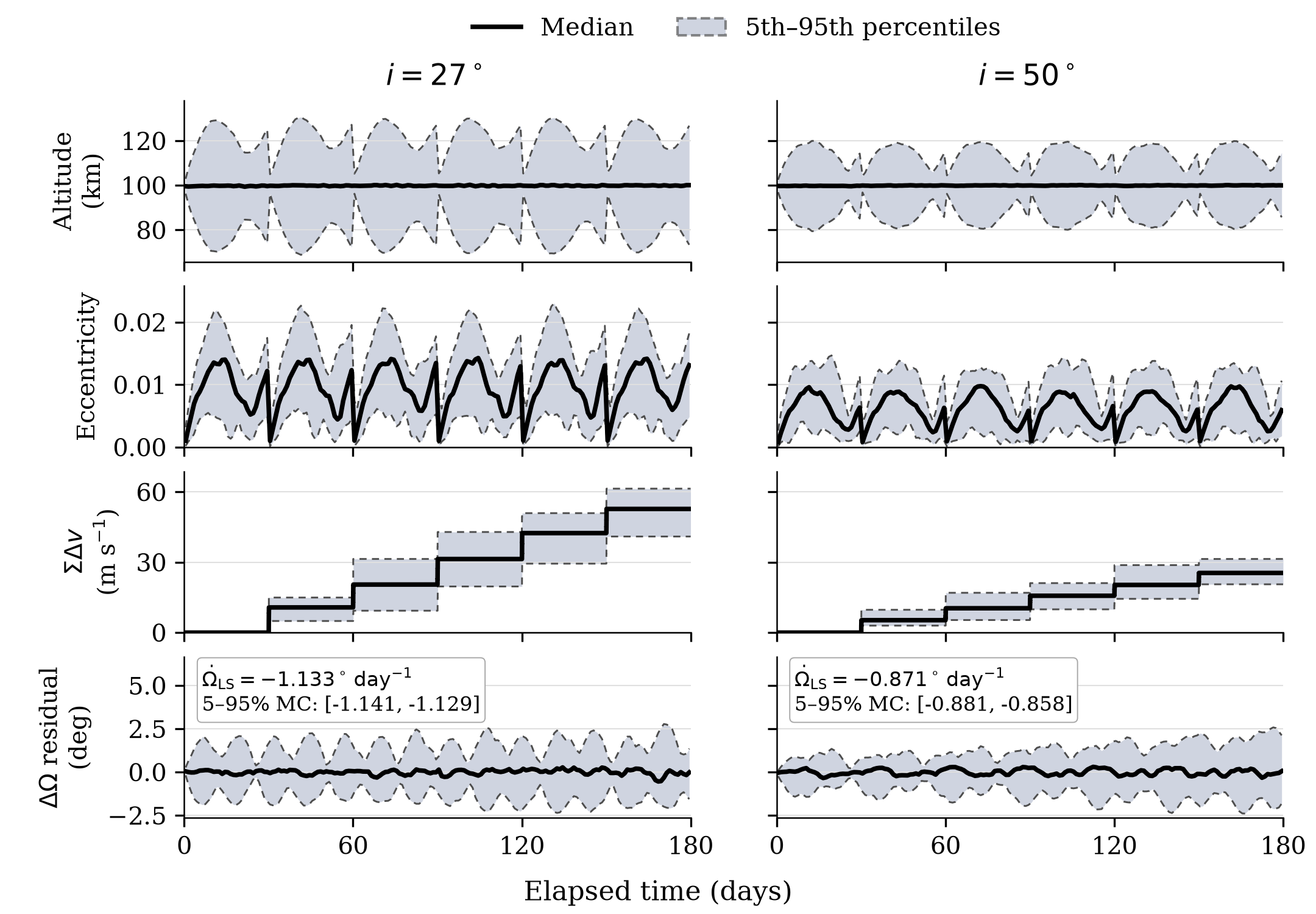}}
\vspace{-2mm}
\caption{Examples of 180~day monte-carlo orbital studies for the $i=27^\circ$ (left column) and $i=50^\circ$ (right column) frozen orbits, 100~km altitude, with monthly orbital maintenance thruster $\Delta v$ to maintain orbital altitude. 100 trials were performed for each inclination, varying the longitude of ascending node over the full range. Each pane shows median values and 5-95 percentile envelopes.
(top)~median altitude; (upper middle) eccentricity; (lower middle)~cumulative $\Delta v$, and (bottom) nodal precession residual with the fitted value subtracted, and the best-fit precession rate shown in the legend.
\label{LRIorbits1} 
}
\vspace{0mm}
\end{figure}

The maria extend as far as $\sim 60^{\circ}$ in both N and S lunar latitude. Detailed orbital studies~\citep{Lara2009a, Lara2009b,ElipeLara2012, Sirwah2020ASO} indicate the existence of nearly circular ``frozen orbits'' that are quasi-stable to the lunar perturbations, at $i = 27^{\circ}$, $50^{\circ}$, $76^{\circ}$, and $85^{\circ}$. The latter two orbits spend significant time at latitudes higher than the maria, while the lowest inclination orbit misses the southern maria entirely. Figure~\ref{fig:mariamap2} plots the lunar surface in an equal-area projection, showing the two lower inclination orbits. The boldface line orbit is shown with the ascending node position that provides the largest maria coverage, and two additional tracks are shown with their precessed ascending nodes after 1 and 2 lunar sidereal months.

The original investigation of the stability of these near-circular low lunar orbits used lunar gravity models as of 2009~\citep{Lara2009a}.
For this present study, the models have been updated  with simulations based on more accurate GRAIL data, using Monte Carlo methods to verify their stability and understand the range of excursions from the ideal orbits. Figure~\ref{LRIorbits1} shows results from a 100-trial, 180-days-per-trial simulation of the $i=27^{\circ}$, 100~km orbit. For this study, the orbital code was first validated against tabulated data from the LRO spacecraft during a recent tracking period,  March 15–22, 2022. Examples of parameters measured during this validation comparison are shown in Table~\ref{tab:lro_validation}. The final results shown in Figure~\ref{LRIorbits1} were all computed at GRAIL degree 200, where agreement is excellent.

\begin{deluxetable*}{crrr}
\tablecaption{Orbital simulation validation against reconstructed LRO trajectories.
\label{tab:lro_validation}}
\tablehead{
\colhead{GRAIL degree} &
\colhead{\shortstack{6 hr position\\RMS error (m)}} &
\colhead{\shortstack{7 day maximum\\node error (arcsec)}} &
\colhead{\shortstack{7 day position\\RMS error (m)}}
}
\startdata
 50 & 48.75 & 13.477 & 560.80 \\
100 &  8.55 &  1.815 & 134.02 \\
200 &  6.66 &  0.729 & 142.69
\enddata
\tablecomments{
\footnotesize{Position errors are three-dimensional separations from the
reconstructed LRO ephemeris. Nodal errors compare complete-orbit
averages obtained from position-derived orbital planes.
Each simulation propagates freely from a single reference state,
without subsequent state resets or fitted force parameters.}
}
\end{deluxetable*}


The orbit studies described here included a station-keeping controller. 
It was used if the orbit was predicted to evolve below 75\,km on perilune, and a $\Delta v$ correction was then applied at apoapsis. The orbits were allowed a $\Delta v$ correction every 30~days in the 180~day period. It is evident that the perturbations in the orbit are periodic; the period is a bit shorter than the lunar synodic period due to the nodal precession, but they are evidently due to gravitational perturbations from Earth; solar gravitational effects are several orders of magnitude smaller. 

The orbital nodal precession rate was very tightly grouped for all orbits at $\dot{\Omega} = -0.87^{\circ}$ per day for the $i=50^\circ$ orbits, and $\dot{\Omega} = -1.13^{\circ}$ per day for the $i=27^\circ$ orbits, implying $157^{\circ}$ to $203^{\circ}$  of inertial frame rotation for a 6 month mission. Precession varies with the cosine of orbit inclination; thus we could potentially gain in sky coverage by using a lower inclination orbit, at the expense of celestial sky coverage around the lunar poles, and lower fractional coverage of the maria per orbit. 

However, in almost all respects other than precession rate, the higher inclination orbit at $i = 50^{\circ}$ appears to be more favorable: it covers more of the maria and corresponding sky above it, the orbits are more stable, and require significantly less maintenance. 
In a single orbit, LRI mapping will cover a great circle on the celestial sky (Figure~\ref{fig:mariamap2}).
Maria occupy about~16\% of the lunar surface, but the selected frozen LRI orbits both help to optimize maria coverage.
The typical fraction of usable tiles is projected to be at least 30\%.
During the monthly lunar rotation, certain portions of the month will yield high fractional coverage and other portions relatively low coverage, but every orbit will contribute to a portion of the celestial sky. 

\begin{figure}
\centerline{\includegraphics[width=0.95\textwidth, trim=0mm 0mm 0mm 0mm, clip]{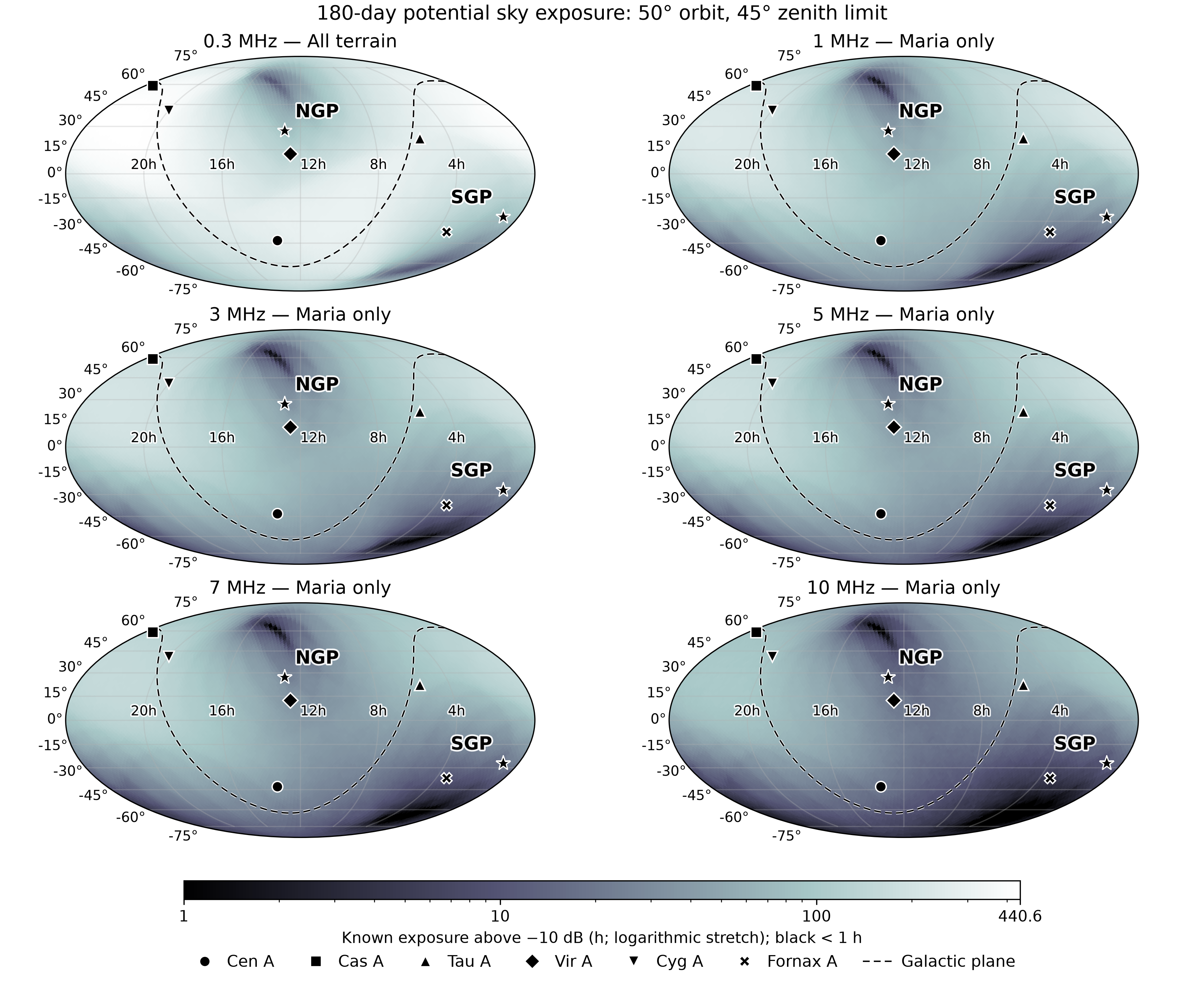}}
\vspace{-3mm}
\caption{LRI sky coverage dwell time in hours, for six different frequency bands in equatorial coordinates, using a Mollweide equal-area projection. The six brightest ``A-team'' radio sources are indicated with markers, along with the galactic poles (NGP, SGP) and plane. } 
\label{skyCover}
\end{figure}


Figure~\ref{skyCover} shows the projected sky coverage in equatorial coordinates on a Mollweide equal-area projection, with Galactic features and bright radio sources annotated, for the  $i = 50^{\circ}$ frozen orbit in 180 days. These estimates use the simulated data directly, sampling the sky via point reflections from the spacecraft location, modified for the off-nadir Fresnel effects, for incidence angles out to $45^{\circ}$, which is still within the expected Brewster angle region. The RMS projected height of the surface decreases off the nadir, but the Fresnel zone area and RMS height of that area increases -- effects that compete with regard to coherence. Our simulations indicate that on average the height projection effects win over the increased RMS of the Fresnel zone area at all frequencies above 1~MHz, and the result is a modest improvement of 0-3 dB in coherence from 1-10 MHz at $45^{\circ}$.

We have not yet explored mapping large fields-of-view using tiling or facets; this is beyond the scope of our work here. As a scale estimate, a tile centered at $45^{\circ}$ zenith with a $5^{\circ}$ radius has delays differing from its center by at most about $40\mu$s. After appropriate delay referencing, an ideal 500 Hz spectral channel average (see below) would introduce less than {0.1\% additional fringe-amplitude loss} across that tile. Each region of sky is generally visited at least six times in a 6 month mission (6.6 lunations), and since the orbital precession rate is $0.87^{\circ}\,\mathrm{day}^{-1}$ compared to $\sim 13^{\circ}\,\mathrm{day}^{-1}$ for lunar rotation, the latter dominates the coverage on the shorter time scales.

\begin{figure}
\centerline{\includegraphics[width=0.75\textwidth, trim=0mm 0mm 0mm 0mm, clip]{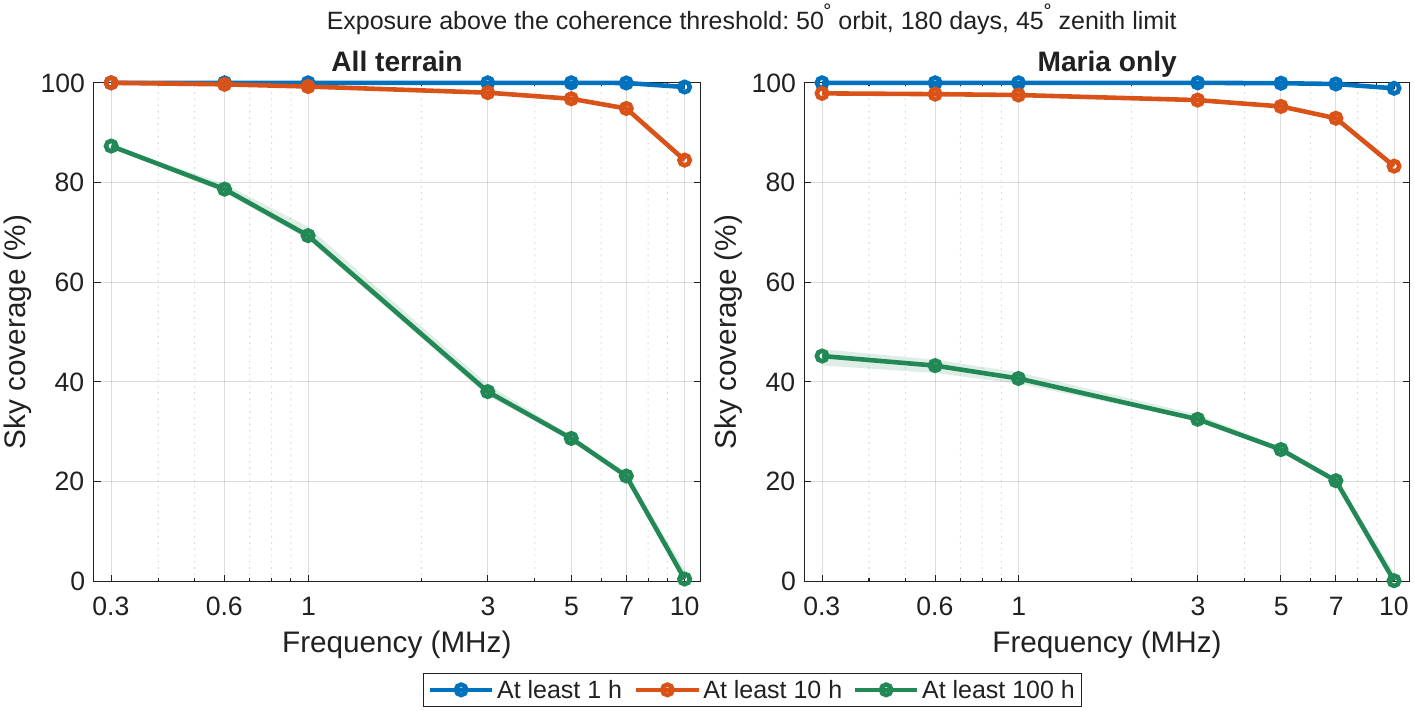}}
\vspace{-1mm}
\caption{Sky coverage vs. frequency for all terrain (left) or maria only (right) above the -10 dB coherence threshold validated in our mapping studies. } 
\label{skyCover2}
\end{figure}

Figure~\ref{skyCover2} shows summary plots of the sky coverage vs. frequency based on all terrain (left) and maria only (right), for areas with estimated coherence above the -10~dB threshold established previously in our mapping studies. Virtually the entire sky is covered by at least a short-duration measurement of 1 hour or more, and more than 80\% of the sky for all frequencies has at least 10 hours of exposure. Between 10 and 100 hours, the coverage declines but remains substantial over frequencies up to 5~MHz, dropping quickly above that.

\begin{figure}
\centerline{\includegraphics[width=0.75\textwidth, trim=0mm 0mm 0mm 0mm, clip]{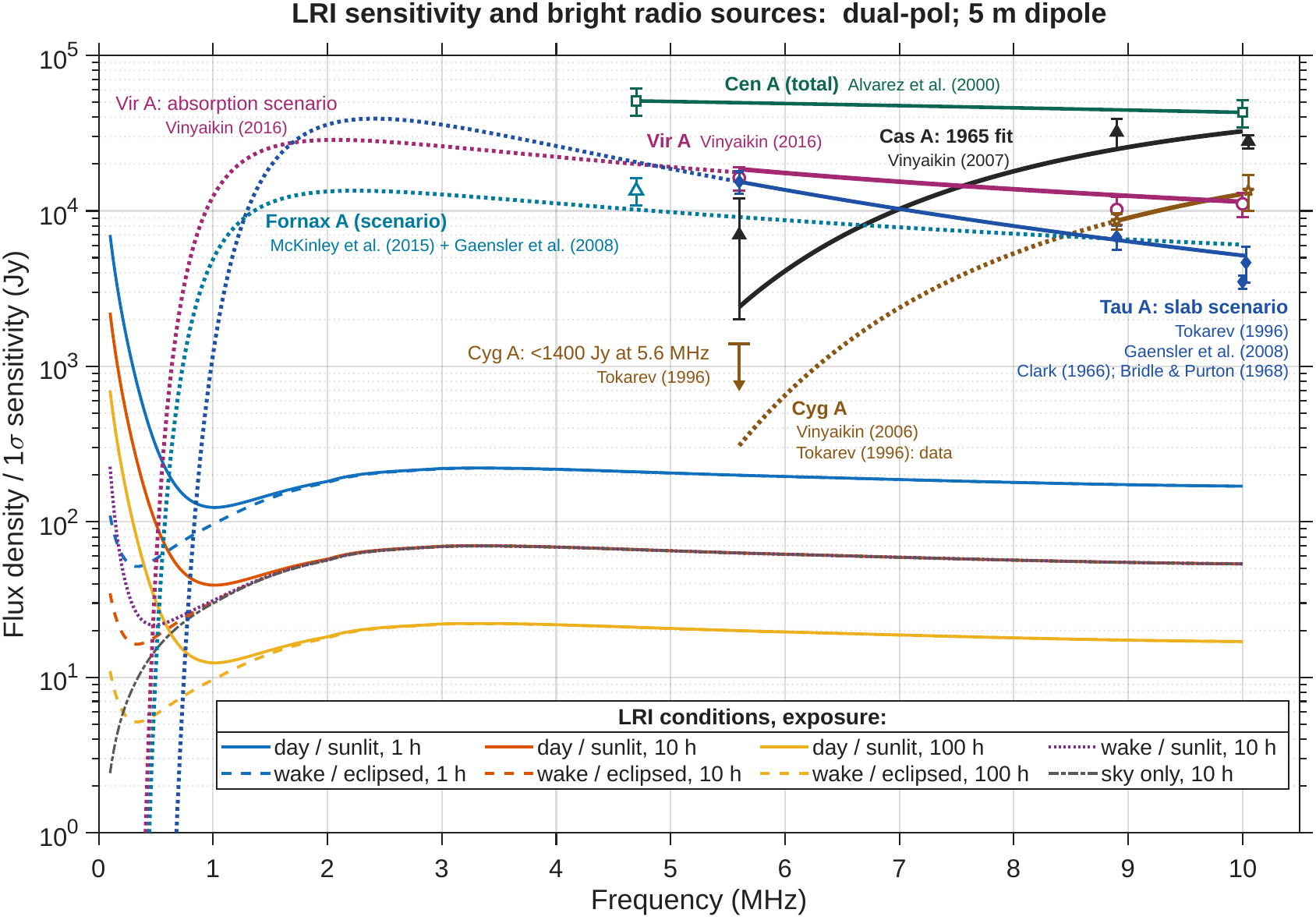}}
\vspace{-1mm}
\caption{ LRI sensitivity from Figure~\ref{fig:LRIsense} above, plotted on a linear frequency scale and logarithmic flux density scale, along with measurements and predictions for the six brightest ``A-team'' radio sources.} 
\label{LRI_Ateam}
\end{figure}

Figure~\ref{LRI_Ateam} synthesizes the sensitivity of LRI with prior measurements and predictions for spectral behavior for the six brightest low-frequency radio sources indicated above in Figure~\ref{skyCover}: Cen A, Vir A, Cas A, Cyg A, Tau A and Fornax A. Based on prior data, LRI will provide good spectral overlap with all of the previous measurements, and will be able to test galactic and intrinsic absorption models for all of these sources. 

The source overlays show integrated flux densities from historical
low-frequency measurements and compilations
\citep{Alvarez2000,Bridle1967,bridlePurton1968,Baldwin1971,
VinyaikinEtAl1987,Tokarev1996,McKinley2015}.
The Cyg~A, Cas~A, and Vir~A spectral curves follow
\citet{Vinyaikin2006,Vinyaikin2007,Vinyaikin2016}, respectively;
Cas~A retains its 1965 reference epoch, while the Vir~A extension below
5.6~MHz uses the published illustrative absorption scenario.
The Cen~A curve linearly interpolates its two measurements.
For Tau~A and Fornax~A, we construct illustrative absorbed extrapolations
using the compact-component slope and 5.6~MHz normalization of
\citet{Tokarev1996} and the 154~MHz normalization and higher-frequency
slope of \citet{McKinley2015}, respectively.
These extrapolations adopt a warm Galactic free--free screen motivated by
\citet{Gaensler2008}, with a finite path length of 1.90~kpc for Tau~A
\citep{Lin2023}.
Dotted curves indicate extrapolations whose low-frequency behavior remains
model dependent. The 5.6~MHz Cyg~A datum is an upper limit.
The measurements retain their historical epochs and flux scales;
source extent and scattering must be considered when converting these
total flux densities to correlated flux.

\subsection{Illustrative Spacecraft Payload}\label{sec:tech.payload}



\begin{figure}[ht!]
\vspace{0mm}
 \centerline{\includegraphics[trim={0 0 0 0},clip, width=0.85\textwidth]{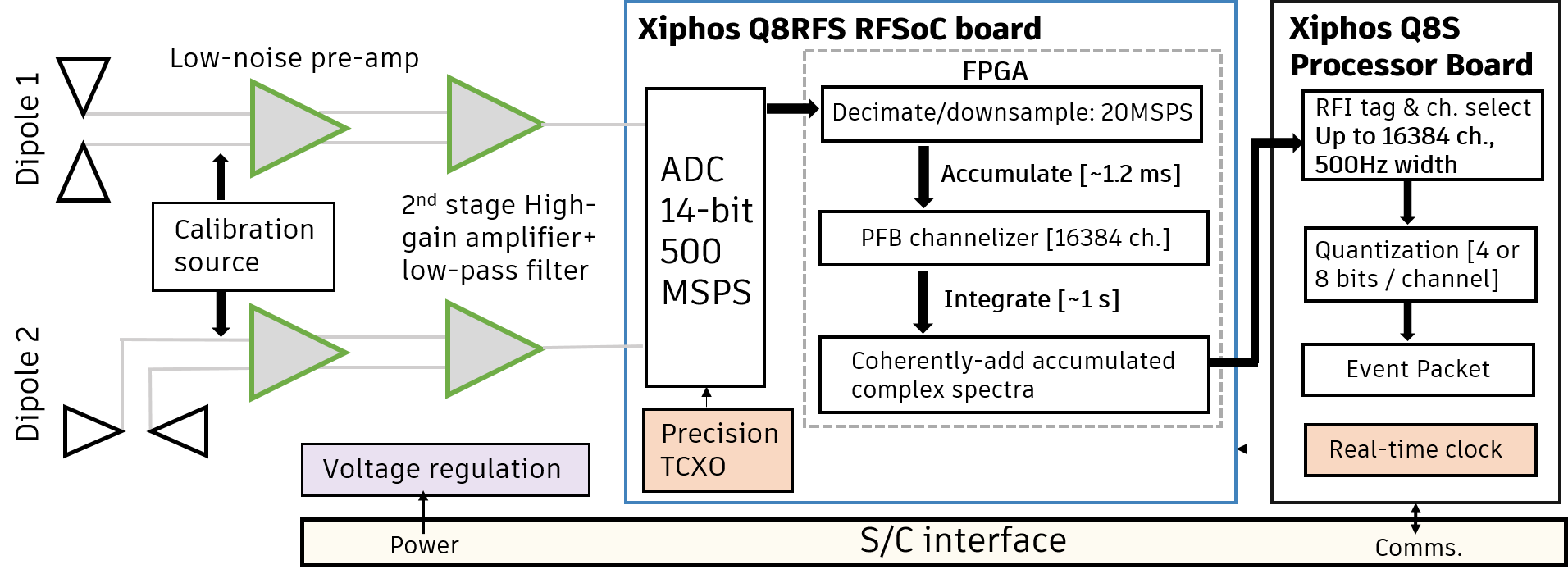}}
\caption{An example block diagram of an LRI instrument with a modern digital implementation.
\label{fig:payloaddiagram}}
\vspace{-2mm}
\end{figure}

Figure~\ref{fig:payloaddiagram} shows a block diagram of a spacecraft payload that could implement the LRI technique.
We do not attempt to provide a full description, as a number of similar spacecraft payloads have been implemented previously~\citep{James2015RRI, SWAVES, SunRISEPayload}. Rather, our intent is to illustrate that many aspects of an LRI spacecraft payload are well-understood.

The sky brightness electric field would be incident on a pair of crossed-dipole antennas, measuring the two linear polarizations (X and~Y).
These antennas could be electrically short, as the sky is sufficiently bright at these frequencies that a high-efficiency antenna is not required. Typical lengths of dipole antennas for similar missions have been several meters.  Further, relatively short antennas are easier to stow on small spacecraft, and there is a demonstrated history of deployable antennas of this scale. Low-noise receivers, while less critical in this case due to Galactic noise than for higher frequencies where receiver noise can dominate, must still be chosen and implemented with care. 

\begin{figure}[tb]
\centerline{\includegraphics[width=0.85\textwidth]{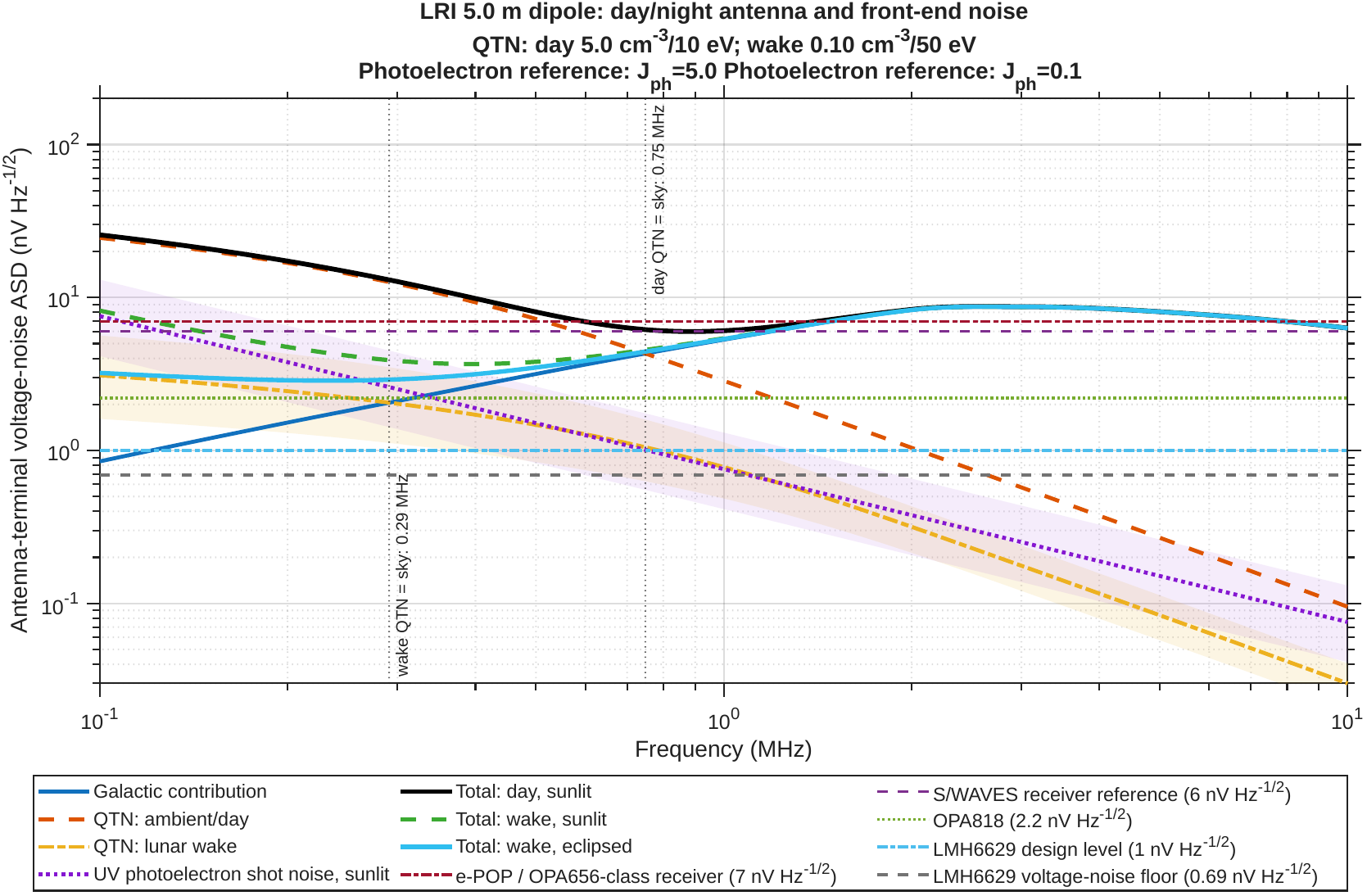}}
\vspace*{-1ex}
\caption{ Expected receiver noise for three choices of front-end amplifier, compared to Galactic, QTN, and photoelectron noise for both lunar dayside and nightside operation, referenced to the LRI antenna amplitude spectral density (ASD).
\label{fig:receiverNoise} 
}
\end{figure}

Figure~\ref{fig:receiverNoise} shows the expected noise levels and preamplifier performance reference to antenna output voltage for the various noise contributions as described above, and performance of heritage systems and a proposed preamplifier based on currently available devices and high-impedance amplifier designs. Although some historical receiver configurations as noted in the figure did in fact create non-negligible noise contributions, more recent amplifier technology will enable receivers with sensitivity that is more optimal for this frequency range, as indicated in the figure.

The antenna signals are conditioned by an analog receiver, which provides a high impedance input to the antenna terminals, filter, and amplify the signals for further digital processing. A typical implementation of a digital processing board combines an analog-to-digital converter (ADC) with a field-programmable gate array (FPGA); recent commercial devices now implement combined ADC+FPGA into a single radio-frequency system-on-chip (RFSoC) architecture.  RFSoC devices and supporting electronics are now available as space-flight-qualified systems, and afford very high bandwidth and processing power for such applications.

Because of the challenges of local and other narrow-band radio-frequency interference (RFI), it is also prevailing wisdom to convert the input voltage time series to high spectral resolution spectra using a polyphase filter bank \citep[\hbox{PFB},][]{2008PASP..120.1207P}. The use of a PFB provides both high spectral resolution and suppressed spectral sidelobes to mitigate any interference appearing at nearby frequencies. The nature of the adding interferometry imposes certain requirements on the channelization, e.g., the channel bandwidth cannot exceed the inverse of the maximum delay between the direct and reflected signals.  For our target altitude of~100\,km, this is 667\,$\mu$s, implying channel bandwidths no greater than 0.75~kHz, including the required Nyquist sampling of the fringe spacing in the power spectral domain (Figure~\ref{ACFplots}).
In order to allow margin for excursions to higher altitudes within an acceptable range, a robust PFB design requires approximately 0.5\,kHz channel bandwidths.
In practice, a PFB requires a power-of-two number of channels. A 16\,384-point spectrum, providing access to frequencies as high as 8.192\,MHz, could be implemented with the current generation of FPGAs. Subsequent processing and filtering of interference, along with selection of subbands according to surface coherence, will reduce the number of channels sent via telemetry.

An instrument flight computer supervises the science data delivery to the host spacecraft and manages operating modes and housekeeping data.
For this configuration, a point-design of an LRI payload has been performed, and we find a mass of approximately 12\,kg, with power requirements of $\sim 50$~W at maximum, not including spacecraft operations such as telemetry and orbit maintenance.

\subsection{Data Transmission}\label{sec:data-rate}

LRI does not require on-board cross-correlation of the PFB channels, since the additive signals from the direct and reflected fields in each antenna are already combined in each of the antenna data streams. 
Sampled voltages in 2\,ms integrations are channelized by the \hbox{PFB}, and then complex spectra are summed coherently up to O(1) second integrations consistent with the ground track coherence time.
At this point, offending RF interference and unused bandwidth are excised, and the spectra are sent directly to telemetry. 

Telemetry bandwidth from cis-lunar space to Earth is generally constrained due to the high demand for time on NASA's Deep Space Network telemetry antennas. We thus evaluate the viability of an aggregate telemetry bandwidth limit of $\leq 2$Gbit/day, which can be accomplished with a standard X-band radio with 2\,Mbps of throughput in a $\leq$ hour pass per day using a non-DSN commercial telemetry antenna. In order to meet these telemetry bandwidth requirements, the cadence at which different subbands are sent is metered according to the coherence qualities of the lunar surface. Per the results of Figure~\ref{Kirchoff2}, the sub-band from~0.1 to~1.1\,MHz (2048 channels) will be telemetered at a cadence of 0.5\,Hz (corresponding to a coherence time of 2\,s) for all regions of the maria.


In contrast, the subband from~6--7\,MHz, with a coherence time of order 0.5~s will achieve adequate coherence over a few percent of the maria, and the effective cadence will thus be $\leq 0.05$\,Hz. Intermediate bands will have intermediate telemetry weights, pre-computed for each spacecraft nadir location on the lunar surface. Over the highlands, only the 0.1--0.5\,MHz portion of the spectrum will yield coherent reflections, adding a small fraction to the overall science data which is concentrated over the maria, averaging about 20\% per orbit. This telemetry selection can be done dynamically by observing autocorrelation levels and transmitting appropriate buffered data. 

In practice, mission requirements may afford higher bandwidth telemetry or may constrain it further, but we provide these representative values to illustrate that LRI data is not incommensurate with standard spacecraft telemetry requirements.





\section{Potential Systematic Effects}\label{sec:systematic}

\subsection{The Lunar Ionosphere}\label{sec:systematic.iono}

The Moon supports an ionized exosphere, often termed the lunar ``ionosphere,'' due to near-surface ionization and solar-wind entrainment \citep[][and references within]{Halekas2011,2026MNRAS.546ag077M}.
Estimates for its density have ranged from below $10\,\mathrm{cm}^{-3}$ to as high as $10^4\,\mathrm{cm}^{-3}$, depending upon the method used and the extent to which the region sampled was exposed to sunlight.
Similarly, estimates for its spatial extent have ranged from being significant only within a few meters of the surface to extending 10\,km or higher.

The presence of a lunar ionosphere could have two potential effects for the LRI technique. 
First, the ionosphere would have a plasma frequency~$f_p \simeq 9~{\rm kHz}\sqrt{n_e}$, where $n_e$ is the plasma (electron) density per cubic cm. For the highest possible day-side inferred densities, the lunar ionospheric plasma frequency could approach 900\,kHz, but, for lines of sight on the night side, the plasma frequency is likely well below 100\,kHz. 
Thus, LRI observations occurring over the night side would be unaffected. 

On the day side, frequencies at or below the plasma frequency will undergo a complex interaction with the plasma, with the potential for significant absorption of the wave energy, as well as a plasma-depth-dependent reflection. If the plasma presents a relatively thin layer with a small fraction of the radio wavelength of high electron density, reflection from the surface may still occur with only modest effects, if the evanescent field depth exceeds the plasma height. Conversely, plasmas with many-km heights may render such observations unreliable for the lowest LRI frequencies. In that case, the lowest frequencies might only be observable on the lunar night side.

The second effect is that the lunar ionosphere will introduce a dispersive delay~$\tau_{\mathrm{iono}} \propto \nu^{-2} \int n_e d\ell$ that affects the reflected ray but not the direct ray (Figure~\ref{fig:LRIgeom}).
Multiple approaches could be taken to determine and remove this dispersive delay.
A spacecraft could carry a dedicated swept-frequency transmitter, akin to an ionosonde.
The dispersive ionospheric delay would be determined from the measured round-trip light travel time in combination with knowledge of the spacecraft's position.
Likely the highest precision, the use of such a transmitter would have the most operational complexities, as it likely would have to be interfere with the science measurements and would have additional demands for power and mass on the spacecraft.

An alternate approach would be to use broadband natural radio sources, such as Jovian or solar radio bursts. For even the modest level of sensitivity of LRI in the 1-7 MHz band, we can expect to detect several Jovian bursts per hour when Jupiter is in view.
The spacecraft's position could then be used to estimate the geometric delay for the broadband natural radio source.
The dispersive delay could be inferred from the measured ACF at multiple frequencies. Because of the high fidelity of the lunar DEM, and the excellent accuracy possible for ground-based spacecraft tracking, such delay measurements can be refined in post-processing, and should provide a robust alternative to active delay tracking if such instrumentation is not available.

\subsection{Lunar Surface and Subsurface Topography and Refractive Index}\label{sec:tech.topo}

The lunar surface contains long-wavelength undulations and fluctuations, some on the approximate size of the Fresnel scale for the LRI technique.
These may appear as tilts, thus effectively ``steering'' the beam as the spacecraft moves through its orbit. More complex surface curvature beyond the simple global component can magnify or otherwise distort the response.  If not compensated, these changes in pointing position would smear all resulting images. Fortunately, as we have noted above, the remarkable accuracy and completeness of lunar surface digital elevation models will allow for very detailed modeling of these effects and postprocessing of the data to correct for them.

While the surface elevation is known to excellent accuracy across the lunar globe, the dielectric response of the surface as a function of radio frequency is less well known. There is detailed information from returned samples at the Apollo sites, and more recently from the Chang'E landers at several other sites. There are also parametric estimates of global dielectric quantities available, but the uncertainties on these maps are not negligible. In the section above, we evaluated the effects of loss of coherence on the complex reflected amplitude received by the spacecraft; uncertainties in the effective refractive index of the reflection region will also translate to uncertainties in the reflected amplitude. 

LRI will have at least one method to constrain the refractive index of the surface over its frequency range, by making use of the effect of the Brewster angle on the polarization ratio for reflections as a function of frequency and angle. Analogous methods were used in Apollo bistatic radar experiments to estimate the surface refractive index at 13~cm and 116~cm wavelengths. The Brewster angle arises from a consideration of the Fresnel coefficients for two linear polarization components.  The Fresnel reflection coefficients at the lunar surface for electric fields polarized perpendicular to the plane of incidence (the $s$ or TE field) are given by
\[
r_s = \frac{n_1 \cos\theta_i - \sqrt{n_2^2 - n_1^2 \sin^2\theta_i}}{n_1 \cos\theta_i + \sqrt{n_2^2 - n_1^2 \sin^2\theta_i}}
\]
while those in the plane of incidence (the $p$ or TM field) are:
\[
r_p = \frac{n_2^2 \cos\theta_i - n_1 \sqrt{n_2^2 - n_1^2 \sin^2\theta_i}}{n_2^2 \cos\theta_i + n_1 \sqrt{n_2^2 - n_1^2 \sin^2\theta_i}}
\]
where in each case $\theta_i$ is the angle of incidence with respect to the normal, and $n_1,~n_2$ are the refractive indices of the incident medium (vacuum) and the regolith, respectively. It is straightforward to show that for an angle $\theta_B = \tan^{-1}(n_2/n_1)$, the reflection coefficient $r_p$ has a zero, while $r_s$ continues monotonically increasing with larger incidence angle. The ratio of $r_p/r_s$ as a function of incidence angle for a source can thus be used to determine the Brewster angle and therefore $n_2$. 

LRI does not have a controlled source for bistatic measurements, but the polarization ratio of bright planetary, celestial, or Galactic sources can be tracked as a function of the zenith angle of the source reflection over many orbital passes. Since the high angle-of-incidence measurements are not required for interferometric imaging, those data can be repurposed specifically to map the Brewster angle over the lunar surface. Such data will allow both for a post-analysis calibration of the interferometric data, as well as yielding a uniquely direct global lunar measurement of the low-frequency refractive index of the near-surface regolith.

We note also that at the lowest frequencies considered, the presence of a tenuous plasma near the surface would modify $n_1$ from its vacuum value of~1 to a value $n_1 = \sqrt{1 - (f_p/f)^2} \leq 1$, where $f_p$ is the plasma frequency, and $f$ is the radio frequency of observation. If the value of the Brewster angle were found to have a strong frequency dependence, it could signal the presence of these near-surface plasma effects, which could then in principle be calibrated directly from the data.

\subsection{Angular Broadening}\label{sec:angbroad}

\begin{figure}[tb]
\centerline{~~\includegraphics[width=0.5\textwidth, trim=0mm 0mm 0mm 0mm, clip]{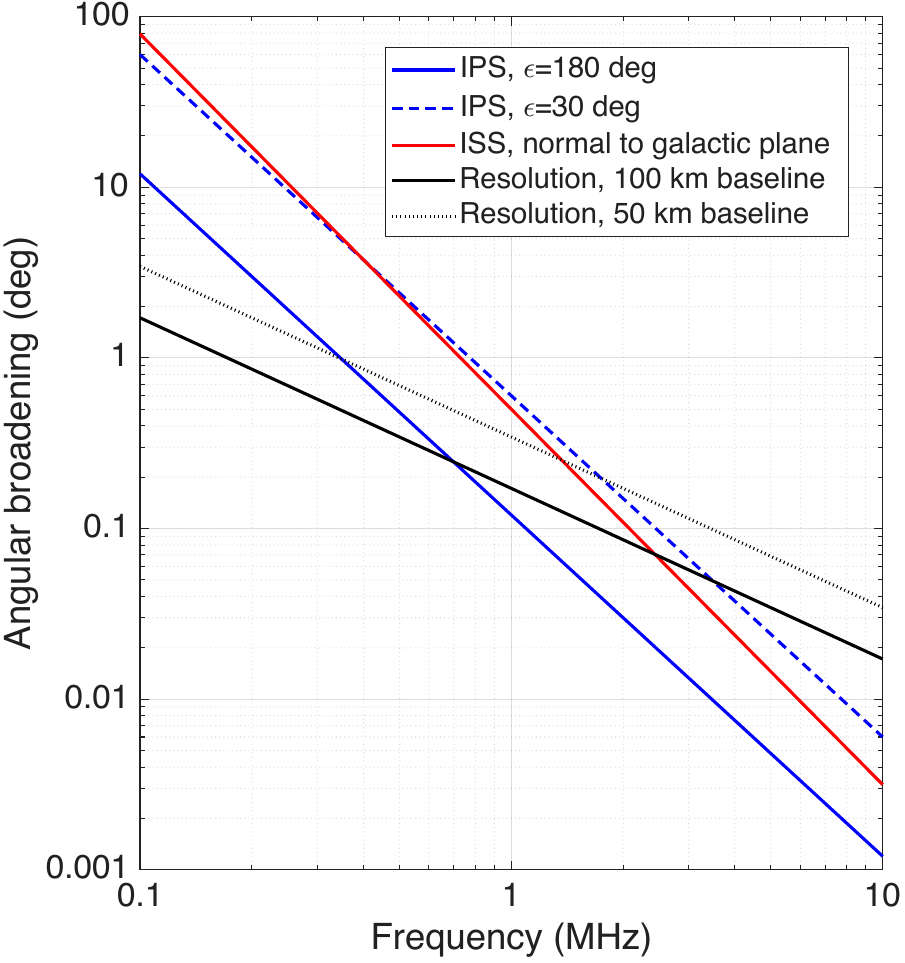}}
\vspace{-1mm}
\caption{Angular broadening by interplanetary and interstellar scattering (\hbox{IPS}, ISS) over the 0.1 to~10\,MHz radio frequency range.  (Adapted from\citealt*{RickettColes2000}.)  Also shown is the diffraction-limited angular resolution for~50 and~100\,km baselines.
\label{fig:LRIscatter} }
\end{figure}

In addition to the requirement that the spacecraft's orbit be low enough that surface curvature effects are unimportant~(\S\ref{sec:tech}), IPM and ISM scattering limit the usefulness of long baselines at lower frequencies and may affect the number of discrete sources that could be detected with the LRI technique (\S\ref{sec:ISMbroaden}). 
\cite{RickettColes2000} provide estimates for these two effects (Figure~\ref{fig:LRIscatter}): 
The typical angular broadening through the \hbox{IPM} and the ISM is 
\begin{eqnarray}
 \theta_{\rm IPS} &\sim&  0.2\arcdeg\,f_{\rm MHz}^{-2},\nonumber\\
 \theta_{\rm ISM} &\sim&  0.5\arcdeg\,f_{\rm MHz}^{-2.2}.
 \label{eqn:broaden}    
\end{eqnarray}
These expressions assume a solar elongation of~90\arcdeg\ for IPM angular broadening and a high Galactic latitudes ($|b| \gtrsim 60^{\circ}$) for the ISM angular broadening.
In directions opposite the Sun, the magnitude of the IPM angular broadening can be lower by a factor of~0.6, while, toward the Sun, it can increase by up to a factor of~5.  
Further, due to solar activity, there can be transient enhancements in the amount of IPM angular broadening by factors of several.
For lower Galactic latitudes, the ISM scattering can lead to angular broadening 1--2 orders of magnitude higher.
As a consequence, observations in directions closer than about~30\arcdeg\ to the Sun are feasible only at the highest frequencies.

For a spacecraft at an altitude of~100\,km observing at $\sim 5$~MHz, a source zenith angle  $\theta_z \lesssim 30^{\circ}$ obtains baselines of up to approximately 100\,km and an angular resolution of order 1\,arcminute.
Such observations would not be limited by IPM or ISM angular broadening, over most of the sky.
In comparison to the best current maps at this frequency ($\simeq 5^{\circ}$ angular resolution and limited sky coverage), LRI observations would represent a factor of 300 improvement;
in comparison to the RAE-2 maps at this frequency ($\simeq 70^{\circ}$ angular resolution over most of the sky), they represent a factor of~4000 improvement.

\subsection{Auroral Kilometric Radiation (AKR)\label{sec:systematic.ak}}

Auroral kilometric radiation (AKR) is likely to be a first-order contaminant for the LRI technique at frequencies below about $1$~MHz, because its characteristic band overlaps the planned $0.1$--$0.7$~MHz range. Classic measurements place strong terrestrial kilometric emission in roughly the $100$--$600$~kHz band, often with a spectral peak near 250\,kHz, while more recent reviews describe AKR more broadly as intense, highly variable auroral radio emission extending over tens to hundreds of kilohertz. In practice, AKR should not be treated as a weak, stationary background. Its intensity depends strongly on geomagnetic activity, and the low-frequency foreground seen by LRI may therefore be set not only by the diffuse sky but also by terrestrial space-weather state and the instantaneous Earth-Moon-spacecraft geometry \citep{gurnett1974,kaiser1977terrestrial,baumjohann2022akr}.

The contamination is also strongly anisotropic. AKR is observed most often and most intensely from viewing geometries that sample the terrestrial nightside and evening auroral source region, and is much less common in dayside viewing sectors. This makes AKR a geometry-dependent systematic rather than a uniform sensitivity penalty: there should be relatively clean intervals for low-frequency mapping, but the usable duty cycle will vary with orbit phase, Earth visibility, and geomagnetic conditions. For LRI, this implies that observing strategy will matter as much as raw radiometric sensitivity at the low-frequency end \citep{kaiser1977terrestrial,fogg2022wind}.

This conclusion is consistent with the heritage of lunar low-frequency radio astronomy. RAE-2 was placed in lunar orbit specifically to reduce terrestrial low-frequency contamination, and lunar occultation of the Earth was central to recovering astrophysical measurements. Modern farside-environment studies reach the same basic conclusion: the lunar body is a uniquely effective shield against intense terrestrial emission, especially below $\sim 0.5$--$1$~MHz. For LRI, Earth-occulted or deep-farside geometries are therefore not merely advantageous but may be operationally important for the deepest integrations in the lowest-frequency channels \citep{novacoBrown1978,bassett2020,Burns_2021}. This does create tension with the desire to map preferentially above the near-side maria, but AKR will not affect frequencies above 1 MHz, and will be a removable stochastic contaminant in near-side data at low frequencies. 

At the same time, Kaguya/SELENE provided a useful cautionary precedent. In passive mode, Kaguya observed terrestrial AKR with clear direct-plus-lunar-reflected interference structure in the $\sim 100$--$500$~kHz range, demonstrating that the Moon can preserve a phase-coherent reflected component at frequencies directly relevant to the low end of LRI. This result is encouraging for the basic reflection concept, but it also shows that strong AKR illumination can imprint structured delay/phase signatures that are lunar-geometry dependent and therefore can masquerade as celestial structure if not modeled or excised. In other words, AKR is not only a source of added power; it can generate coherent measurement artifacts in exactly the kind of direct-plus-reflected geometry that LRI intends to exploit \citep{Goto2011,Wahlund_2025}.

Accordingly, AKR should be treated in LRI operations as an observing-state variable rather than as a stationary additive noise term. A reasonable mitigation strategy is to (i) prioritize Earth-occulted or farside segments for the deepest low-frequency integrations, (ii) maintain burst-detection and excision in the time--frequency plane, (iii) tag data by Earth-viewing geometry and geomagnetic context, and (iv) assess whether exceptionally bright AKR intervals can be repurposed as calibration opportunities rather than discarded outright. In this sense AKR is both a risk and, potentially, a calibration resource; but unless it is explicitly handled in the observing concept, it is one of the clearest mechanisms by which the $0.1$--$0.7$~MHz channel could underperform relative to a simple sky-noise-limited expectation \citep{kaiser1977terrestrial,Goto2011,Burns_2021,fogg2022wind}.

\section{Science Use Cases}\label{sec:science}

Having demonstrated the technical feasibility of the LRI approach, we turn to a selection of several potential applications, to illustrate the scientific reach of \hbox{LRI}.
\cite{1990LNP...362.....K}, \cite{2000GMS...119.....S}, and \cite{2009NewAR..53....1J}, and references within, contain extensive discussion of the possibilities of space-based observations at frequencies below roughly 10\,MHz.
We do not discuss heliophysics or solar system planetary applications here; they have been treated in detail elsewhere \citep{Zarka2012}.

\subsection{Centaurus~A}\label{sec:cena}

Centaurus~A is the strongest and most extended extragalactic target for LRI imaging.
LRI imaging of Cen~A has the potential to bridge the gap between the classic sub-10\,MHz era, for which only very coarse maps and integrated source fluxes exist, and modern low-frequency interferometry, for which MWA and PAPER have mapped the giant lobes at~100–-200\,MHz but still far above the ionospheric cutoff.
Ellis \& Hamilton’s 4.7\,MHz survey detected Cen~A as the brightest discrete source in their southern sky sample, with a listed flux density of 51\,kJy, using an 11\arcdeg\ $\times$ 3\arcdeg\  beam~\citep{ellis1966survey}.
They already recognized that the low-frequency source spectra were curved relative to extrapolations from higher frequencies, with a hint of curvature in excess of that expected from free-free absorption, and that observations at~1-–2\,MHz would distinguish intrinsic synchrotron curvature from absorption, since optically thick free-free absorption and optically thick synchrotron emission have different spectral slopes.

The historical context below 10 MHz is sparse but highly relevant. Reber/Ellis-era ground observations reached roughly 0.9–2.1 MHz with very coarse beams~\citep{reberEllis1956}; Ellis, Bessell, Waterworth, and related Tasmanian surveys covered 1.5–10 MHz~\citep{ellis1962galactic, ellis1966survey}; and RAE-2 later produced lunar-orbit all-sky maps at 1.31, 2.20, 3.93, 4.70, 6.55, and 9.18 MHz~\citep{novacoBrown1978}. RAE-2 demonstrated that below 10 MHz the sky is not simply a lower-frequency continuation of high-frequency Galactic synchrotron emission: free-free absorption increasingly shortens the path length, so by 1 MHz the observed emission is dominated by the local interstellar environment rather than the full Galactic disk. This matters for Cen A because its line of sight is at moderately low Galactic latitude ($b=+19^\circ$), so the observed 1–10 MHz spectrum is a convolution of intrinsic lobe physics, Galactic foreground absorption, possible local absorption, and scattering.

\begin{figure}[tb!]
\vspace{-2mm}
\centerline{~~\includegraphics[width=\textwidth, trim=0mm 0mm 0mm 0mm, clip]{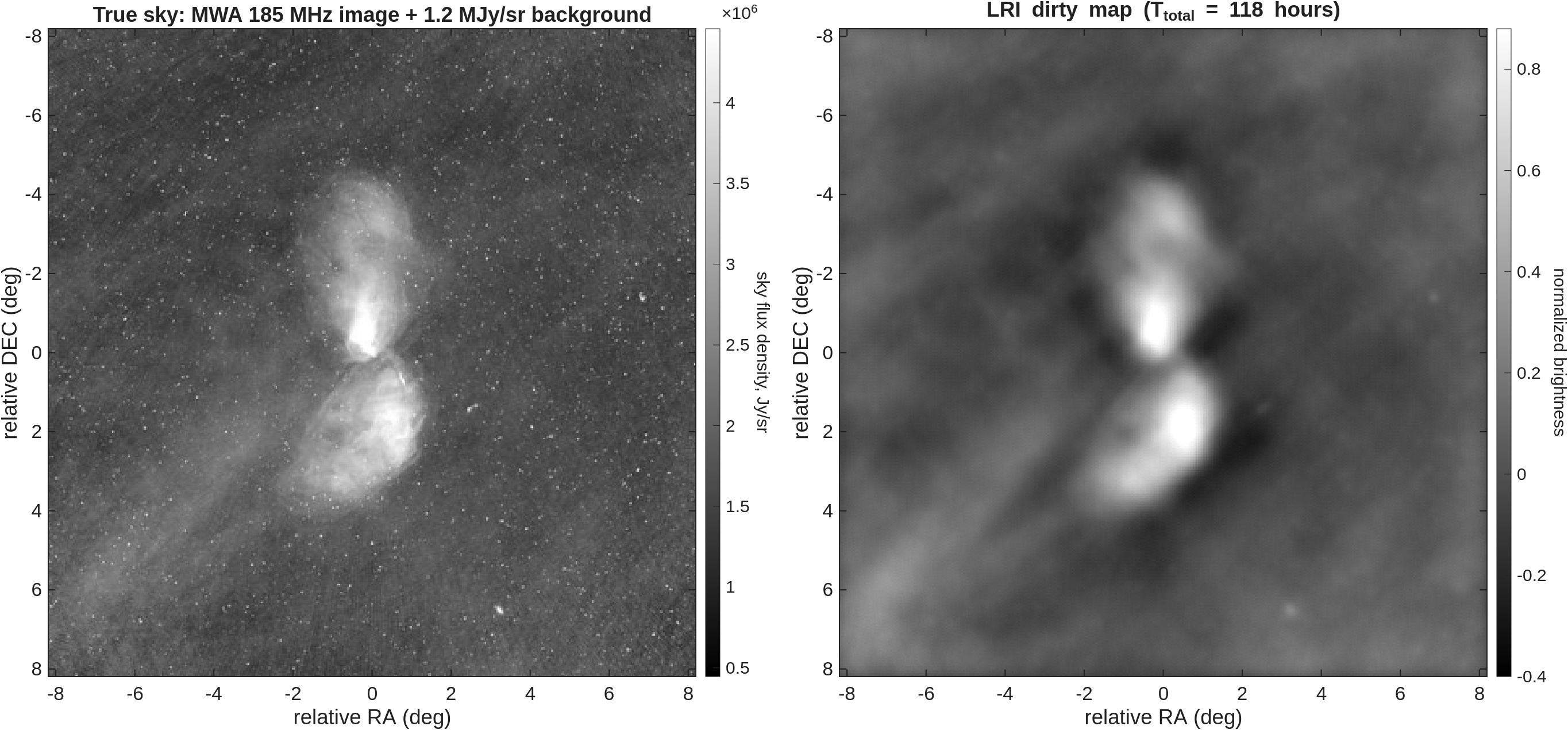}}
\vspace{-3mm}
\caption{(Left) 185\,MHz Murchison Widefield Array map of Cen A (J2000 map center at RA $=201.36506^{\circ}$, DEC $=-43.019113^{\circ}$), modified to include the expected uniform Galactic noise at~5\,MHz. (Right)~Simulated LRI dirty map of the input brightness distribution from the left image, reconstructed over several months of orbital coverage. 
\label{LRICenA} }
\end{figure}

Modern low-frequency work has transformed the higher-frequency anchor point. PAPER imaged Cen A at 148 MHz with ~15–25 arcmin resolution and found an integrated flux density of about 4100 Jy, explicitly emphasizing the suitability of wide-field instruments for the extended outer lobes~\citep{stefan2013imaging}. MWA observations at 118 MHz and especially the later 185 MHz imaging pushed this further: McKinley et al. used MWA to image the full source at 1.5 arcmin resolution, detecting previously unseen filamentary structure in the diffuse outer lobes and emphasizing Cen A’s role as the nearest laboratory for AGN feeding/feedback across many physical scales~\citep{mckinley2013giant, mckinley2021multi}. 

LRI would map Centaurus A's synchrotron-emitting electrons at much lower electron energies than this previous work. 
The electron energy range observed for a given radio frequency and magnetic field strength is given by
\begin{equation}
E_e \simeq 0.25~{\rm GeV} \left ( \frac{\nu}{1~{\rm MHz}} \right )^{1/2} \left ( \frac{B_{\perp}}{1~\mu{\rm{G}}}  \right )^{-1/2}
\label{Ee_equation}
\end{equation}
where $E_e$ is electron energy, $\nu$ is radio frequency, and $B_{\perp}$ is the perpendicular magnetic field component relative to electron velocity.
For Cen~A, the likely field strengths are $B_{\perp} \lesssim 1\mu$G in the outer radio lobes, so LRI would sample electron energies in the tens to hundreds of MeV range, well below the several to tens-of-GeV range sampled by MWA or GHz observations.

In the multi-GeV energy regime ($\nu \gtrsim 100\text{ MHz}$), relativistic electrons have short radiative lifetimes ($\tau_{\mathrm{rad}} \sim 10^7\text{ yr}$). Consequently, high-frequency emission traces recent particle acceleration, active jet channels, and localized re-acceleration zones (e.g., internal shocks or shear layers).
For $\nu < 10\text{ MHz}$, the lower-energy electrons observed possess long radiative lifetimes ($\tau_{\mathrm{rad}} \sim 10^8$--$10^9\text{ yr}$). They thus act as an integrated \emph{fossil archive} of the total outflow history and energy injection from Centaurus A over $10^8\text{--}10^9\text{ yr}$, revealing aged giant lobe structures where high-energy electrons have long since cooled. By comparing LRI spectral-turnover maps with MWA morphology, the mission will determine whether the lowest-frequency emission traces diffuse fossil plasma, localized particle re-acceleration, or foreground/intervening absorption, thereby using Cen A as a nearby laboratory for AGN feedback and cosmic-ray transport in radio lobes.

Figure~\ref{LRICenA} shows a high-fidelity simulation of LRI imaging of Cen A, using the 185 MHz MWA image as a reference. The MWA input brightness distribution is modified to include the additive uniform 1.2 MJy/sr Galactic noise expected at 5 MHz, and the LRI map is made at 5~MHz with a $\pm 1$~MHz bandwidth over several months of orbital passes, using baselines taken from reflections whenever the spacecraft was within a $\pm 21^{\circ}$ square region around the source, thus providing projected baselines (along the diagonals of the square) of up to 100~km at the 100 km altitude assumed. Coherence magnitude for reflections was fixed conservatively at -10~dB, even when higher coherence was possible.  The instrinsic source size for reconstruction was constrained by the IPS and ISS scattering parameters noted previously.  The map shows angular resolution already in the $0.2-0.4^{\circ}$ range, and several of the brighter compact sources in the original brightness map are evident, even prior to CLEANing of the map. The broader structure in the background and some finer detail in the radio lobes are evident as well. Because of the complexity of the sky-position-dependent point-spread function (PSF), we have not yet implemented CLEAN or any other PSF-based deconvolution for the mapping. These initial maps already show useful science value, and deconvolution algorithms will likely significantly improve image quality, resolution, and dynamic range.

\subsection{Large-scale Galactic Structures}\label{sec:galactic}

Large-scale Galactic structures such as the North Galactic Spur / Loop~I and Cetus Arc sit at the intersection of several major ISM questions.
They may be old nearby supernova remnants or larger, more distant superbubble shells. If they are local supernova remnants, they provide laboratories for how supernovae inject cosmic rays, amplify and compress magnetic fields, and structure the local cavity.  For a plausible local diffuse-ISM $B_{\perp} \sim 3$--$5\,\mu$\hbox{G}, synchrotron emission in the 1--10\,MHz band corresponds to electron energies $E \sim 0.1$--0.45\,GeV (eqn.~[\ref{Ee_equation}]), an order of magnitude or more below the energies probed with maps in the 100--1000\,MHz range.
These lower energy electrons are the population most affected by ionization, Coulomb losses, diffusion, and local confinement, and are thus essential for understanding the role of cosmic ray feedback in galaxy evolution.
A spatially-resolved spectral break or turnover across the North Galactic Spur / Loop~I and Cetus Arc would constrain whether they are filled by a similar electron population as throughout the Local Bubble or whether the loops are distinct acceleration or compression sites.

The importance of low frequency mapping of large-scale Galactic structure was highlighted by the results of RAE-2, which has to date made the only near-complete maps of the Galaxy in the 0.1--10\,MHz range. These maps showed features combining complex synchrotron emission with frequency-dependent free-free absorption, modulated by Galactic magnetic fields, but then ultimately smeared by the $\sim 1$\,sr angular resolution that RAE-2 afforded. 
They provided suggestive data on the low-frequency behavior of the North Galactic Spur / Loop~I and Cetus Arc as a diagnostic of the (likely) nearby cosmic-ray electron population, magnetic-field geometry, and warm ionized absorbing medium on scales of roughly tens to a few hundred parsecs, but lacked the resolution to answer any of the questions posed.

The key conclusion from RAE-2’s maps was that below 10\,MHz the observed sky is no longer a simple projection through the whole Galaxy \citep{novacoBrown1978}. The emission is synchrotron radiation from cosmic-ray electrons in the Galactic magnetic field, but the line of sight becomes progressively opaque because of free-free absorption by low-density ionized gas (the warm ionized phase of the interstellar medium). \citet{novacoBrown1978} explicitly noted that different observing frequencies sample the ISM over a range of opacities, and that even low thermal-electron-density regions become efficient absorbers at the frequencies probed by RAE-2.
At~0.3\,MHz, they estimated path lengths only about 10–-50\,pc in the Galactic plane, while near the Galactic poles at the high frequency end of the band they suggest that the path length extends beyond the Milky Way. This absorption allows low frequency sky maps to give three-dimensional information about the interstellar medium. 



The distance frequency- and direction-dependent distance \(s_{\tau=1}(\nu,l,b)\) at which the cumulative free-free optical depth
reaches unity is related to the free-free absorption coefficient $\alpha_{ff}$ by
\begin{equation}
    \tau_{\rm ff}(\nu,s,l,b)
    =
    \int_0^s \alpha_{\rm ff}(\nu,s',l,b)\,ds'
    \simeq 1 ,
\end{equation}
Inferring such a distance requires either
comparison to a model of the unabsorbed Galactic synchrotron emissivity and
warm ionized medium, or comparison to structures whose distances are known
from other data.  In the case of the latter approach, in which the location of a structure that should be optically thick is known in advance (e.g., lines of sight towards known massive star forming regions), the absorber separates foreground from background synchrotron emissivity  at that known distance.  In this sense the LRI maps act as an opacity-weighted
tomographic filter: as frequency decreases, distant synchrotron structures are
progressively suppressed, nearby loops and shells can become relatively more
prominent, and finally even those structures disappear when the optical-depth
horizon moves inside their distances.  The primary Galactic LRI observables are therefore
the spatially resolved spectral turnovers, absorption silhouettes, and
frequency-dependent changes in morphology; the conversion of these observables
to distances is performed through foreground/emissivity modeling and through
cross-identification with structures of independently estimated distance.
Multi-frequency LRI maps would therefore estimate
the effective radio horizon as a function of frequency and direction when combined with models of cosmic ray electrons and the Galactic warm ionized medium, and provide new input to those models by measuring the synchrotron emission column on lines of sight with absorption features at known distances.

The North Galactic Spur and Cetus Arc are especially valuable because RAE-2 saw them behave differently with frequency \citep{novacoBrown1978}. At higher RAE-2 frequencies, the bright region was a blend of inner-Galaxy emission plus the North Galactic Spur. As the frequency decreased, the maximum shifted toward $l\sim40^\circ$, because the distant inner Galaxy was suppressed by absorption and the Spur became more prominent. At 2.20 MHz, RAE-2 found the maximum south of the plane near $l\sim40^\circ$, and interpreted this as blending between the North Galactic Spur and the Cetus Arc, with the Cetus Arc possibly becoming more dominant at lower frequency. 
\citet{novacoBrown1978} suggested that if the Cetus Arc dominates at a lower frequency than the North Galactic Spur, a simple interpretation would be that the Spur lies farther away, though they correctly warned that this assumes similar intrinsic spectra and similar absorbing columns in the two directions. 

That old ambiguity of distance estimates to the Galactic loops remains a topic of active research in a broader form. Modern work on the North Polar Spur / Loop I still debates whether it is primarily a nearby supernova/superbubble shell or related to much larger Galactic-center/eROSITA/Fermi-bubble-scale structures; work using Gaia distances to clouds has argued for a relatively nearby North Polar Spur distance, about 500 light-years, i.e. roughly 150 pc~\citep{NPolarSpur2020Gaia}, but a 2022 review concludes that both contradictory pictures are still defended~\citep{CRPHYS_2022__23_S2_1_0}.   LRI would not definitively resolve this question alone, but it would provide an updated low-frequency constraint unavailable from ground-based surveys.

RAE-2’s poor angular resolution limits the interpretation of its maps. Its V-antenna main lobe varied from about 1 sr near 10 MHz to nearly hemispheric near 1 MHz, so the features were fundamentally blended.  RAE-2 could infer that Cetus and the Spur influenced the 2.2 MHz maximum, but it could not determine whether the maximum was a ridge, a shell limb, a superposition of multiple arcs, or an absorption-modulated projection.
LRI observations could be used to determine whether the low-frequency brightness maximum near $l\sim40^\circ$ is produced primarily by intrinsic synchrotron emissivity in local loop SNR structures, by foreground free-free absorption windows, or by superposition of the North Galactic Spur and Cetus Arc at different effective distances.

If the North Galactic Spur and Cetus Arc are genuine shell structures, higher-resolution maps should show limb-brightened arcs whose curvature, width, and substructure can be compared to higher-frequency radio, $H_{\alpha}$, dust, soft X-ray, and polarization data. RAE-2 could only see broad maxima and minima; LRI could test whether the 2.2 MHz maximum decomposes into two intersecting arc systems.

To support this, LRI could measure the spatially resolved spectral turnover from roughly 1–7 MHz, and potentially to 10 MHz. For each pixel or resolved region, the spectrum can be fit to a synchrotron emissivity plus foreground/mixed absorption model. A simplified form would be

\begin{equation}
T_b(\nu)=T_{\rm fg}(\nu)+T_{\rm loop,0}
\left(\frac{\nu}{\nu_0}\right)^{-\beta}
e^{-\tau_0(\nu/\nu_0)^{-2.1}} 
\end{equation}

or, for mixed emitting/absorbing plasma,

\begin{equation}
T_b(\nu)\propto j_\nu\frac{1-e^{-\tau_\nu}}{\alpha_\nu}.
\end{equation}

Key observables are the turnover frequency and whether the turnover is coherent along an arc, follows $H_{\alpha}$/EM structures, or varies patchily with foreground absorption. The synchrotron emission itself may become optically thick in regions of intense emission.  Optically thick synchrotron radiation from cosmic ray electrons with a power-law spectral energy distribution has a spectral index of 2.5 regardless of the spectral index of the electron energies, and thus synchrotron self-absorption can be distinguished from free-free absorption by the spectral index of the optically thick portion of the spectrum.    

In some lines of sight, morphology offers a clear way to distinguish a foreground free-free absorption structure. \cite{polderman2019HII}, following a method proposed by \cite{Kassim1990}, used absorption silhouettes of HII regions against the diffuse Galactic background to map foreground and background synchrotron emissivities, combining several datasets from 22--231\,MHz. At frequencies relevant to \hbox{LRI}, \cite{novacoBrown1978} speculated that absorption by the Gum Nebula and I~Ori association contributed to the low-brightness region around $l\sim240^\circ$ seen with RAE-2, which they note may also be partly attributable to low magnetic field or cosmic-ray electron density as well as the ionized absorbing structures. LRI will have the angular resolution to distinguish the contribution of the Gum Nebula and I~Ori.  

LRI will also probe the magnetic fields in the very local ISM. In the 1.31\, MHz RAE-2 map, neither Cetus nor the North Galactic Spur were visible, and \cite{novacoBrown1978} concluded that the dominant remaining structure likely traced the very local interstellar magnetic field.  The LRI maps would test this hypothesis.


Galactic structure also matters as foreground physics to cosmology studies. Any future lunar-farside or space-based cosmology experiment below tens of MHz could benefit from accurate foreground models.
The 1–-10\,MHz foreground is not a smooth extrapolation of Haslam-like maps; it is a frequency-dependent, absorption-filtered, local-ISM map. LRI could produce the empirical basis for that model.

\subsection{Probing the Interstellar Medium with Angular Broadening of Compact Sources}\label{sec:ISMbroaden}


While angular broadening represents a potential systematic for the LRI technique (\S\ref{sec:angbroad}), observing angularly-broadened compact radio sources also provides a probe of the \hbox{ISM} \citep{1990ARA&A..28..561R,1990LNP...362..155S,cordes2000interstellar}:
Intrinsically compact sources acquire an apparent angular size that depends strongly on frequency and Galactic latitude.
Not only can ISM angular broadening be used to map the distribution of plasma turbulence in the Galaxy, characterizing the magnitude and distribution of interstellar scattering at high Galactic latitudes may benefit studies of fast radio bursts (FRBs) that seek to distinguish the contributions of various propagation effects along their lines of sight.
The LRI band is especially attractive for this experiment because the scattering law is steep,
so modest source-size constraints at meter wavelengths can become arcminute- to
degree-scale signatures at hectometric wavelengths.

We write the apparent angular width of a compact source as
\begin{equation}
    \theta_{\rm app}^2(\nu,l,b,t)
    = \theta_{\rm LRI}^2(\nu,t)
    + \theta_{\rm int}^2(\nu)
    + \theta_{\rm ISM}^2(\nu,l,b)
    + \theta_{\rm IPM}^2(\nu,\epsilon,t),
    \label{eq:theta_budget}
\end{equation}
where $\theta_{\rm LRI}$ is the instrumental response, $\theta_{\rm int}$ is the intrinsic
low-frequency source size, and $\theta_{\rm ISM}$ is the angular broadening due to the \hbox{ISM} and $\theta_{\rm IPM}$ is the angular broadening due to the IPM (eqn.~[\ref{eqn:broaden}]).
We have neglected the intrinsic size of the source in this expression consistent with the assumption of a compact source.
We assume observations at large solar elongations, such that $\theta_{\mathrm{IPM}} \ll \theta_{\mathrm{ISM}}$ (Figure~\ref{fig:LRIscatter}).

Representing the ISM as a turbulent plasma screen, $\theta_{\mathrm{ISM}}$ is expected to follow a steep power law,
\begin{equation}
    \theta_{\rm ISM}(\nu,l,b)
    =
    \theta_0(l,b)
    \left(\frac{\nu}{\nu_0}\right)^{-\beta},
    \qquad
    \beta \simeq 2.0~{\rm to~2.2} ,
    \label{eq:theta_powerlaw}
\end{equation}
with $\beta=2$ corresponding to the simple Gaussian-screen or inner-scale-limited case and $\beta=11/5$ corresponding to a Kolmogorov scaling.

Models to predict $\theta_0(l,b)$ have been developed, based on pulsar dispersion measures and angular broadening of a variety of Galactic and extragalactic sources \citep{2026ApJ..1002....3O}.
At high Galactic latitudes, a representative value for the angular broadening of an extragalactic source is
\begin{equation}
 \theta_{\rm ISM}(\nu,l,b) = 0\farcm15\,\csc{b}\,\left(\frac{\nu}{10\,\mathrm{MHz}}\right)^{-2.2},
    \label{eq:theta_highb}
\end{equation}
where the $\csc{b}$ term assumes that the line of sight is at high enough Galactic latitude that it does not intersect nearby spiral arms.
The characteristic value toward any high Galactic latitude pulsars would be lower by a factor that would account for the fact that a pulsar is immersed in the ISM while an extragalactic source is seen through its entire extent. 
Clearly, the expected ISM angular broadening can approach degree scales near~1\,MHz.
Many investigations of angular broadening of astronomical radio sources have required Very Long Baseline Interferometry (VLBI) to resolve the angular broadening.
These are not VLBI-scale effects in the LRI band; they are directly relevant to the angular response of a lunar-orbit low-frequency interferometric mapper.

\begin{figure}[tb!]
\vspace{-2mm}
\centerline{~~\includegraphics[width=\textwidth, trim=0mm 0mm 0mm 0mm, clip]{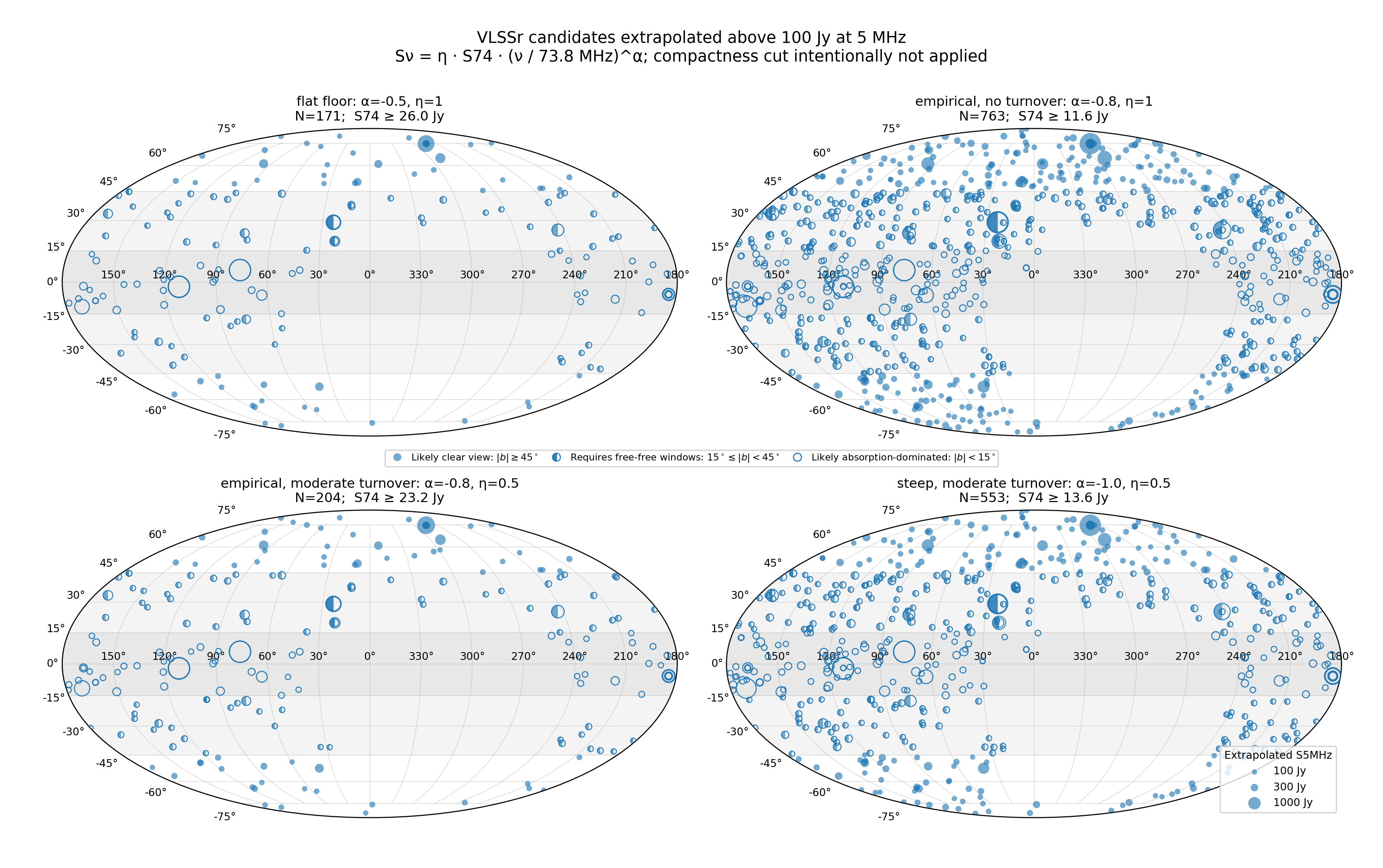}}
\vspace{-3mm}
\caption{Mollweide projections of candidate sources from the VLSSr radio source survey at~73.8\,MHz.  The four panels illustrate different choices of spectral index and low-frequency absorption turnover. 
\label{fig:vlssr_lri_candidates} }
\end{figure}

 Let \(t_j\) denote the midpoint of one short LRI integration
or ``snapshot,'' and let \(\hat{\boldsymbol{z}}_j\) be the local zenith direction of the
spacecraft during that snapshot.  The channelized autocorrelation can be transformed
into a frequency domain cross-spectrum with a
direct--reflected interference term for which the calibrated complex form may be written as
\begin{equation}
    D_j(\nu)
    =
    \int_{\Omega_j}
    I_\nu(\hat{\boldsymbol{s}})\,
    {\cal W}_j(\nu,\hat{\boldsymbol{s}})\,
    \exp\!\left[-2\pi i\nu\,\tau_j(\hat{\boldsymbol{s}})\right]\,
    d\Omega
    +
    n_j(\nu),
    \label{eq:lri_sky_integral}
\end{equation}
where \(I_\nu(\hat{\boldsymbol{s}})\) is the sky brightness distribution,
\(\hat{\boldsymbol{s}}\) is a unit vector on the sky, \(\Omega_j\) is the portion of the
sky visible to the direct and reflected paths during snapshot \(j\), and \(n_j(\nu)\)
is the thermal and systematic noise term.  The geometric delay is
\begin{equation}
    \tau_j(\hat{\boldsymbol{s}})
    =
    \frac{2h_j}{c}\,
    \hat{\boldsymbol{z}}_j\cdot\hat{\boldsymbol{s}}
    =
    \frac{2h_j}{c}\cos\theta_j(\hat{\boldsymbol{s}}),
    \label{eq:lri_delay_general}
\end{equation}
where \(h_j\) is the spacecraft altitude above the local reflecting surface and
\(\theta_j\) is the source zenith angle.  The factor
\({\cal W}_j(\nu,\hat{\boldsymbol{s}})\) contains the instrumental and lunar-reflection
weights,
\begin{equation}
    {\cal W}_j(\nu,\hat{\boldsymbol{s}})
    =
    G_{{\rm d},j}(\nu,\hat{\boldsymbol{s}})
    G_{{\rm r},j}^{*}(\nu,\hat{\boldsymbol{s}})
    \Gamma_j(\nu,\hat{\boldsymbol{s}})
    \operatorname{sinc}\!\left[\Delta\nu\,\tau_j(\hat{\boldsymbol{s}})\right].
    \label{eq:lri_weight}
\end{equation}
Here \(G_{{\rm d},j}\) and \(G_{{\rm r},j}\) are the voltage responses of the antenna
to the direct and reflected rays, respectively, \(\Gamma_j\) is the complex
Fresnel/coherence factor for the lunar reflection, \(\Delta\nu\) is the frequency-channel
width, and \(\operatorname{sinc}x\equiv\sin(\pi x)/(\pi x)\).  The sinc factor is the
finite-channel-width attenuation of a sinusoidal fringe across one spectral channel.
The raw autocorrelation or power-spectrum measurement contains the real
interference term corresponding to \(D_j\); writing the analytic complex form in
Eq.~(\ref{eq:lri_sky_integral}) is a compact way to represent the matched-filter
quantity used in the forward model.


For the scatter-broadening ($sbr$) measurement, an intrinsically compact source is not
treated as a delta function on the sky.  Instead, its flux density is distributed over
a small angular distribution function, or kernel,
\begin{equation}
    I_\nu^{\rm sbr}(\hat{\boldsymbol{s}})
    =
    \sum_q
    S_q(\nu)\,
    K_q\!\left(
        \hat{\boldsymbol{s}};
        \hat{\boldsymbol{s}}_q,
        \theta_{{\rm maj},q}(\nu),
        \theta_{{\rm min},q}(\nu),
        \psi_q(\nu)
    \right),
    \label{eq:broadened_source_sky}
\end{equation}
where \(K_q\) is normalized so that
\begin{equation}
    \int K_q(\hat{\boldsymbol{s}})\,d\Omega = 1 .
    \label{eq:kernel_normalization}
\end{equation}
Thus \(S_q(\nu)\) remains the total flux density of the source.  The parameters
\(\theta_{{\rm maj},q}\), \(\theta_{{\rm min},q}\), and \(\psi_q\) describe the major-axis
FWHM, minor-axis FWHM, and position angle of the apparent source image.  A
simple first model for \(K_q\) is an elliptical Gaussian, although other kernels could
be used for anisotropic or non-Gaussian scattering.

Substituting Eq.~(\ref{eq:broadened_source_sky}) into Eq.~(\ref{eq:lri_sky_integral})
shows explicitly what is meant by replacing a point source with an angular kernel:
\begin{equation}
    D_j^{\rm sbr}(\nu)
    =
    \sum_q
    S_q(\nu)
    \int
    K_q(\hat{\boldsymbol{s}})
    {\cal W}_j(\nu,\hat{\boldsymbol{s}})
    \exp\!\left[-2\pi i\nu\,\tau_j(\hat{\boldsymbol{s}})\right]
    d\Omega
    +
    n_j(\nu).
    \label{eq:lri_broadened_forward_model}
\end{equation}
Physically, the broadened source is treated as the coherent sum of many
infinitesimal source elements spread over a small region of sky.  Each element has
a slightly different delay and fringe phase.  As the source becomes broader, these
phases average together less coherently on long projected baselines, reducing the
fringe contrast and spreading the response in delay.  The angular broadening is
therefore measured by fitting the frequency- and baseline-dependent loss of compact
fringe response, not merely by inspecting the width of a reconstructed image.

A practical concern is whether there is a sufficiently large population of compact sources bright enough to be detected by the LRI technique.
We explored this using the Very Large Array (VLA) Low-frequency Sky Survey Redux (VLSSr) catalog at~73.8\,MHz as a parent sample \citep{2012RaSc...47.0K04L,Lane2014VLSSr}, assuming that all
selected catalog components satisfy a $<3\arcmin$ compactness criterion.  The source
selection was based on the simple extrapolation
\begin{equation}
    S_\nu(\nu)
    =
    \eta_\nu\,S_{74}
    \left(\frac{\nu}{73.8\,{\rm MHz}}\right)^{\alpha}
    \label{eq:flux_extrapolation}
\end{equation}
where $\alpha$ is the low-frequency spectral index and $\eta_\nu\leq1$ is a phenomenological
survival factor representing intrinsic turnover due to synchrotron self-absorption in the extragalactic source, free-free absorption in either the source galaxy or our own, and other losses below the
ground-accessible bands.
Requiring $S_\nu(\rm{5MHz}) \geq 100\,\mathrm{Jy}$, equivalent to approximately $5\sigma$ for a 10\,hr integration (Figure~\ref{fig:LRIsense}),
Figure~\ref{fig:vlssr_lri_candidates} illustrates the following four cases:
\begin{equation}
\begin{array}{ccl}
    \alpha=-0.5,\ \eta_5=1.0 &:& S_{74,{\rm min}} \simeq 26.0\ {\rm Jy},\\
    \alpha=-0.8,\ \eta_5=1.0 &:& S_{74,{\rm min}} \simeq 11.6\ {\rm Jy},\\
    \alpha=-0.8,\ \eta_5=0.5 &:& S_{74,{\rm min}} \simeq 23.2\ {\rm Jy},\\
    \alpha=-1.0,\ \eta_5=0.5 &:& S_{74,{\rm min}} \simeq 13.6\ {\rm Jy}.
\end{array}
\label{eq:threshold_cases}
\end{equation}
These choices of $\alpha$ deliberately bracket the uncertainty, based on prior 5~MHz observations 
of a number of bright point sources, subject to uncertainties 
in the low-frequency spectral index and the free-free absorption effects.  
The $\alpha=-0.5$ case is a flat-spectrum (pessimistic) floor, $\alpha\simeq-0.8$ is closer to the typical integrated low-frequency spectral index of synchrotron-dominated radio-source populations, and $\alpha=-1.0$ represents a steep-spectrum-selected sample.
The factor $\eta_5$ is essential:
without it, a steep-spectrum extrapolation can overstate the number of usable 5 MHz
sources, while with $\eta_5\simeq0.5$ the prediction remains optimistic but not
unphysical.  The factor $\eta_5$ is held constant over the whole sky.  In practice, $\eta_5$ will decrease toward the Galactic plane as free-free absorption increases.  
Figure~\ref{fig:vlssr_lri_candidates} shows the resulting source
distributions.  The upper left panel ($\alpha=-0.5$) is sparse, even without including the effect of spectral turnover ($\eta_5=1$), indicating that
a 5 MHz scatter-broadening experiment would not densely cover the sky under the flattest plausible extrapolation.  The empirical no-turnover case,
$\alpha=-0.8$, $\eta_5=1$, produces a much healthier population of potential targets.
The moderate-turnover case, $\alpha=-0.8$, $\eta_5=0.5$, collapses the potential source population back toward the flat-spectrum floor case.
The steep-spectrum moderate-turnover case,
$\alpha=-1.0$, $\eta_5=0.5$, recovers a more substantial population and is probably the most defensible optimistic case for a source sample selected explicitly for low-frequency steep spectra.  

The upper part of the LRI band provides the best compact-source yield and the
least-scattered reference measurements.  For high-Galactic-latitude lines of sight,
the expected ISM broadening at 6--7 MHz is comparable to or somewhat below the
nominal LRI synthesized beam for a 100 km projected baseline, and will therefore
be measured primarily through forward-model fitting of source kernels rather than
by simple image-plane deconvolution.  At lower Galactic latitudes, where the
angular broadening is expected to be one to two orders of magnitude larger, the
broadening remains directly resolvable even at the upper edge of the band.  The
principal frequency-leverage for high-latitude ISM scattering is therefore expected
to come from the 2--5 MHz range, with the 6--7 MHz data anchoring the intrinsic
source flux and weak-scattering limit.

This source-count result shapes this science case.
LRI can use existing
ground-based surveys above the ionospheric cutoff to define a high-confidence parent
sample, then produce the first measured compact-source and scatter-broadening catalog
in the hectometric band.  The program would begin with high-Galactic-latitude sources
whose intrinsic structure is minimized, then extend to lower latitudes where the
broadening itself becomes the measurement.  Repeated observations at different solar
elongations would separate the quasi-fixed Galactic scattering term from time-variable
interplanetary scattering, while the multi-frequency dependence would distinguish
instrumental broadening, intrinsic source structure, and ISM propagation. Note that scattering and free-free absorption provide complementary probes of the electrons in the ISM: scattering measurements are sensitive to the spectrum of electron density fluctuations, whereas free-free absorption is sensitive to the total electron column density.  LRI's 
catalog of $\theta_{\rm maj}(\nu)$, $\theta_{\rm min}(\nu)$, $\psi(\nu)$, and flux density would
provide an empirical low-frequency scattering foreground model for LRI imaging and a
new set of constraints on the turbulent electron-density distribution of the local and
large-scale Galactic \hbox{ISM}.

\subsection{Stellar Space Weather (Exospace Weather): Stellar Radio Bursts}\label{sec:stellarradiobursts}


Finally, we discuss an application that is motivated to address a gap in our understanding in the space weather generated by other stars or \emph{exospace weather} \citep{KISS}.
More than two decades of continuous observations by multiple spacecraft in multiple locations throughout the Solar System have significantly advanced our understanding of the Sun's space weather and its effects \citep{2008SoPh..247..171H,2021LRSP...18....4T,2022ApJ...926L...1B}.
The Sun generates space weather---a steady solar wind punctuated by impulsive, transient events generated by the rapid conversion of magnetic energy into kinetic energy, resulting in bulk mass motions, shocks, and energetic particles.
The steady solar wind contributes to mass and angular momentum loss over the Sun's lifetime, and both the steady solar wind and transient events can transform planetary atmospheres.
Solar flares produce ultraviolet and soft X-ray (XUV) emissions that heat the upper layers of an atmosphere and affect its chemistry, while the particle flux from both the steady solar wind and transient events can change the chemistry of a planetary atmosphere and even erode it.
Mars' atmosphere presents a dramatic illustration of (solar) space weather on a planetary atmosphere, as studied using observations from the Mars Atmosphere and Volatile Evolution (MAVEN) mission \citep{jgl+15,2023JGRA..12830884J}.

Characterizing {exospace weather} presents obvious challenges in comparison to solar space weather:
It is not possible to obtain multi-point or \textit{in situ} exospace weather observations. 
Nonetheless, there are clear indications that other stars generate exospace weather.
Nearby stars produce ``hydrogen walls'' or astrospheres, consistent with steady stellar winds interacting with the local \hbox{ISM}, analogous to how the Sun's solar wind produces the heliosphere \citep{2004LRSP....1....2W,2006SSRv..126....3W,2014ASTRP...1...43L}; these observations are key data for estimating how the strength of the Sun's solar wind has changed over its lifetime.
Transient brightenings of stars have been identified, consistent with them generating stellar flares \citep{1991ARAandA..29..275H,2010ARAandA..48..241B}.
Detected stellar flares are more luminous than solar flares, due to the stars being at larger distances than the Sun, but their energy distribution connects smoothly to that of solar flares.
The long-standing interpretation is that the same physical processes that operate on the Sun to generate solar flares and eruptive phenomena also must happen within (magnetized) stellar atmospheres.

The shocks and energetic particles generated in the Sun's corona produce solar radio bursts, intense plasma emissions at the local plasma frequency, which can be used to track both the steady solar wind and the transient events.
So-called Type~II bursts are relatively rare radio bursts that drift slowly from higher frequencies to lower frequencies and have long been associated with shocks and energetic particles, while Type~III bursts are more frequent radio bursts that also drift from higher to lower frequencies, but much more rapidly, and are associated with electron beams \citep{1950AuSRA...3..387W,1963ARA&A...1..291W,1967SoPh....1..304T,1972ARA&A..10..159W}.

The Sun is a single star, of spectral type~\hbox{G}, observed at a specific time in its evolution.
Modeling and comparisons with solar analogs indicate that the Sun was more active and its space weather was stronger in its past~\citep{2018ApJ...856...53P}; even in the relatively recent past, the Sun has produced Carrington- and Miyake events~\citep{2021ARAandA..59..445H,2022LRSP...19....2C}.
Stars of different spectral types, particularly low-mass stars that support large convective regions within their interiors, have different surface magnetic field strengths and different levels of magnetic activity \citep[e.g.,][]{KochukhovLavail2017}.
Detecting the stellar equivalent of solar Type~II and Type~III bursts would enable comparisons between the properties of the Sun and other stars, placing the Sun in a larger context.

The instantaneous emission frequency for a Type~III burst depends upon the local plasma frequency $f_e \simeq 9\,\mathrm{kHz}\sqrt{n_e}$, where $n_e$ is the local electron density in the solar wind (in units of cm${}^{-3}$).
Thus, for the Sun, Type~III bursts are a well-established means of tracking the electron density of the solar wind \citep{1998SoPh..183..165L}, which, at large distances ($r \gtrsim 10\,R_\sun$), can be converted to an estimate of its mass loss.
By extension, detection of stellar Type~III bursts could provide estimates of their mass loss rates.
There have been numerous efforts to search for stellar Type~III bursts using ground-based radio telescopes.
Due to the Earth's ionospheric cutoff frequency ($\approx 10\,\mathrm{MHz}$), these observations have occurred at frequencies $f \gtrsim 30\,\mathrm{MHz}$, corresponding to electron beams propagating in the lower coronae (distances $r \lesssim 2\,R_*$), insufficient to provide estimates of stellar mass loss rates.

Solar Type~II bursts trace bulk mass motions and shocks, typically resulting from coronal mass ejections (CMEs).
Of particular interest are so-called decametric-hectometric (DH) and ``kilometric Type~II'' bursts, as these occur at frequencies below approximately 14\,MHz.
Like Type~III bursts, Type~II bursts occur at the local plasma frequency, so DH-kilometric Type~II bursts are clear indications of bulk mass motions and shocks (and energetic particles) escaping from the solar corona and into the inner heliosphere and interplanetary medium.
The potential importance of CMEs generated by other stars was recognized within a decade after the first identifications of solar CMEs \citep{1984ARAandA..22..267W}.
Much like the case for stellar Type~III bursts, there have been a number of ground-based efforts to detect transient stellar radio emissions, i.e., Type~II bursts, associated with CMEs, but the observational frequencies have been sufficiently high that the detections likely trace mass motions and shocks only at low altitudes ($r \lesssim 2\,R_*$) in the stellar coronae.
A DH-kilometric stellar Type~II burst would represent a key observational signature for a stellar \hbox{CME}.

The lunar reflection interferometry technique could be used to monitor or survey nearby stars for DH-kilometric stellar radio bursts.
The field of view of a lunar reflection interferometer likely would be quite large (\S\ref{sec:tech.payload}), meaning that multiple stars could be observed simultaneously. 
Being akin to a two-element interferometer, the sensitivity of this technique likely would not be sufficient to detect individual DH-kilometric stellar radio bursts from nearby stars, unless some stars are capable of generating much more intense radio bursts than the Sun does, by factors of $10^3$ or more.
Depending upon the orbit of the spacecraft, observations could be co-added (``stacked'') to obtain higher signal-to-noise ratios and thereby determine an average rate of stellar radio bursts.

\section{Conclusions}
\label{sec:conclude}

We have introduced lunar reflection interferometry (LRI), a single-spacecraft
implementation of adding interferometry in which the lunar surface supplies a
passive, virtual second antenna.  Radiation received directly from the sky
interferes with a delayed and attenuated copy reflected from the lunar surface.
The direct--reflected cross term is a weighted measurement of the mutual
coherence of the sky electric field and can therefore be interpreted within the
standard van Cittert--Zernike framework.  Spacecraft motion changes both the
effective projected baseline and its orientation, allowing repeated
measurements to be combined through orbital synthesis.  For a spacecraft at an
altitude of order 100 km, the maximum direct--reflected delay is approximately
0.67 ms, while sources within about $30^\circ$ of the local zenith provide
projected baselines extending to approximately 100 km.

The viability of LRI depends critically on the coherence of lunar reflections.
Full-wave simulations using representative lunar topography, together with
Kirchhoff integrations over LOLA digital elevation models and comparison with
Kaguya Lunar Radar Sounder results, indicate that substantial portions of the
lunar maria should preserve useful coherent reflection through at least
approximately 5 MHz and probably to approximately 7 MHz for orbits near
100 km altitude.  Operation toward 10 MHz is more restrictive and is expected
to favor lower orbital altitudes and selected exceptionally smooth maria
regions.  The usable reflection area increases rapidly toward lower frequency,
while the lunar highlands are expected to contribute primarily at the lowest
frequencies.  These results identify the lunar maria not merely as an
environment surrounding the instrument, but as a direction-, frequency-, and
position-dependent component of the interferometer response.

End-to-end mapping studies show that the annular delay response of an
individual LRI snapshot can be combined over successive spacecraft positions
to reconstruct both compact and extended emission.  A single orbital passage
has a strongly ring-like point-spread function and may retain an azimuthal
ambiguity, whereas orbital precession supplies additional baseline orientations
and progressively improves the spatial-frequency coverage.  Simulations using
a conservative reflection-loss model recover ensembles of compact sources,
large diffuse structures, and the principal morphology of Centaurus A from
multi-month observing sequences.  Under the adopted sky-noise, reflection, and
radiometer models, a single 10--15 minute passage reaches approximately the
$0.1$--$1$ kJy regime, while an accumulated integration of order 120 hours
reaches several to several tens of Jy.  A representative six-month orbit study
gives sky-coverage fractions ranging from nearly full-sky coverage at
0.3 MHz to approximately two thirds of the sky at 7 MHz, with the effective
exposure determined by the frequency-dependent distribution of coherent maria
reflection regions.

The required flight hardware is compatible with a small-spacecraft
architecture.  A pair of deployable crossed dipoles, high-impedance analog
front ends, a high-dynamic-range ADC, and a polyphase-filter-bank processor can
provide the required polarization and spectral information.  Sub-kilohertz
spectral channels and snapshot integrations of order one second are sufficient
to sample the maximum geometric delay and to limit temporal smearing as the
spacecraft footprint moves across a Fresnel zone.  Preliminary payload studies
indicate instrument resources of order 12 kg and 50 W, although the final
telemetry allocation, electromagnetic-compatibility requirements, and
calibration architecture remain to be established in a complete mission
design.

LRI would provide the first opportunity for resolved, wide-area astronomical
imaging throughout much of the frequency range below the terrestrial
ionospheric cutoff using only one spacecraft.  Potential investigations include
spatially resolved spectral-turnover measurements of Centaurus A, tomography
of nearby Galactic synchrotron and free--free absorbing structures, and
measurements of interstellar and interplanetary angular broadening using bright
compact sources.  Searches for low-frequency stellar radio bursts would provide
an additional exploratory application.  The same data would also constrain the
frequency-dependent dielectric response of the lunar surface and the properties
of the near-lunar plasma environment.

Future work on several fronts will help to verify the LRI methodology. These include more detailed 
derivation of the observables and their noise covariance in spherical geometry, calibration of the
direction-dependent complex reflection kernel, quantitative modeling of the lunar plasma and
terrestrial AKR, and development of wide-field deconvolution methods for the
spatially varying LRI point-spread function.  Subject to these validations, LRI
offers a novel and economical route into the last largely unmapped decade of
the radio spectrum and a practical pathfinder for more extensive future
low-frequency interferometric observatories in lunar space.

\begin{acknowledgments}

We thank I.~Davis and E.~Sholnick for illuminating discussions about exospace weather.
We acknowledge the use of OpenAI's ChatGPT Large Language Model~\cite{chatGPT} for portions of the code development and analysis presented here, and in editing and proofreading the manuscript text.
This research has made use of the Astrophysics Data System, funded by NASA under Cooperative Agreement 80NSSC21M00561.
Part of this research was carried out at the Jet Propulsion Laboratory, California Institute of Technology, under a contract with the National Aeronautics and Space Administration (80NM0018D0004). KP's work is supported by a Jansky Fellowship of the National Radio Astronomy Observatory. The National Radio Astronomy Observatory and Green Bank Observatory are facilities of the U.S. National Science Foundation operated under cooperative agreement by Associated Universities, Inc.

\end{acknowledgments}

\begin{contribution}

P. Gorham, J. Lazio, and K. Plant were primarily responsible for the text and graphics. P. Gorham, R. Burridge, A. Romero-Wolf and E. Oberla provided detailed analysis and simulations to validate performance. P. Allison, M. Anderson, C. Deaconu, J. Lazio, S. Lepri,  E. Oberla, J., and J. Rolla provided details on relevant low-frequency observations and the associated science cases. C. Miki and E. Oberla developed instrument and spacecraft models to test payload viability. A. Connolly and P. Gorham provided resources for concept development and validation. A. Bramson, E. Costello, and M. Siegler provided expertise on lunar regolith modeling and other aspects of relevant lunar science.


\end{contribution}

%


\appendix



\providecommand{\LRIobsbox}[2]{\parbox[t]{#1}{\vspace{0pt}\raggedright #2}}
\providecommand{\LRIobscbox}[2]{\parbox[t]{#1}{\vspace{0pt}\centering #2}}

\section{Prior Astronomical Observations at and Below 30\,MHz}\label{appendix1}

The observational literature below the terrestrial ionospheric cutoff is heterogeneous.  Some experiments produced two-dimensional maps, others sampled a set of drift-scan tracks or selected directions, and many spacecraft instruments measured only an absolute spectrum or the lowest-order anisotropy of the sky.  A single table that treats all of these products as equivalent ``surveys'' therefore obscures both the limited sky coverage of several classic measurements and the distinction between angular resolution and antenna field of view.  We separate diffuse-sky maps, directional or spectral measurements, discrete-source surveys, and recent pathfinder instruments below.

For a nominal rectangular footprint in equatorial coordinates, the gross solid angle quoted in Table~\ref{tab:lf_diffuse_maps} is
\begin{equation}
  \Omega_{\rm foot}=\Delta\alpha\left(\sin\delta_{\max}-\sin\delta_{\min}\right),
  \label{eq:survey_footprint}
\end{equation}
where $\Delta\alpha$ is in radians.  The same expression applies in Galactic coordinates after replacing $(\alpha,\delta)$ by $(\ell,b)$.  We define $f_{\rm foot}=\Omega_{\rm foot}/4\pi$, with $4\pi~\mathrm{sr}=41\,253~\mathrm{deg}^{2}$.  These are \emph{geometric footprints}, not beam solid angles and not necessarily the area of valid calibrated pixels.  A useful distinction is
\begin{equation}
  \Omega_{\rm valid}=C_{\rm fill}\,\Omega_{\rm foot},
  \label{eq:survey_valid_area}
\end{equation}
where $C_{\rm fill}$ represents loss from sparse tracks, gaps, discarded ionospheric intervals, radio-frequency interference (RFI), contamination by the Sun, Earth, or Jupiter, and other editing.  Historical papers usually do not provide $C_{\rm fill}$ in a form that can be reconstructed without digitizing their published maps.  Accordingly, the coverage codes in Table~\ref{tab:lf_diffuse_maps} distinguish published or exactly specified coverage (P), area derived from coordinate limits (D), sparse-track or beam-union estimates (T), irregular masks that require map digitization (F), and composite products assembled from more than one survey (C).

\startlongtable
\begin{deluxetable*}{cclclcc}
\tabletypesize{\scriptsize}
\tablewidth{\textwidth}
\tablecaption{Published Maps of Diffuse Galactic Emission at and Below 30 MHz\label{tab:lf_diffuse_maps}}
\tablehead{\colhead{Year} & \colhead{\shortstack{$\nu$\\(MHz)}} & \colhead{\shortstack{Survey or\\platform}} & \colhead{\shortstack{Angular\\resolution}} & \colhead{\shortstack{Nominal\\footprint}} & \colhead{\shortstack{$\Omega_{\rm foot}$\\{[sr; deg$^{2}$]}}} & \colhead{\shortstack{$f_{\rm foot}$\\{[\%; code]}}}}
\startdata
1954 & \LRIobscbox{0.60in}{18.3} & \LRIobsbox{1.05in}{Shain--Higgins drift survey \citep{shainHiggins1954}} & \LRIobsbox{0.80in}{$\sim17^\circ$} & \LRIobsbox{1.80in}{Five all-RA declination tracks; approximate half-power union $-60^\circ\lesssim\delta\lesssim-4^\circ$} & \LRIobscbox{0.82in}{$\sim5.003$; $\sim16\,424$} & \LRIobscbox{0.55in}{$\sim39.8$; T} \\
1961 & \LRIobscbox{0.60in}{19.7} & \LRIobsbox{1.05in}{Shain Cross \citep{shainEtAl1961}} & \LRIobsbox{0.80in}{$1.4^\circ$} & \LRIobsbox{1.80in}{Extended strip of the southern Milky Way; exact published mask should be digitized} & \LRIobscbox{0.82in}{\ldots} & \LRIobscbox{0.55in}{F} \\
1965 & \LRIobscbox{0.60in}{30.0} & \LRIobsbox{1.05in}{Parkes 64-m telescope \citep{mathewsonEtAl1965}} & \LRIobsbox{0.80in}{$\sim11^\circ$} & \LRIobsbox{1.80in}{All RA, $-90^\circ\leq\delta\leq0^\circ$} & \LRIobscbox{0.82in}{$6.283$; $20\,626$} & \LRIobscbox{0.55in}{$50.0$; D} \\
1966 & \LRIobscbox{0.60in}{4.7} & \LRIobsbox{1.05in}{Penna filled array, Tasmania \citep{ellisHamilton1966}} & \LRIobsbox{0.80in}{$11^\circ\times3^\circ$} & \LRIobsbox{1.80in}{All RA, $-72^\circ\leq\delta\leq-12^\circ$} & \LRIobscbox{0.82in}{$4.669$; $15\,328$} & \LRIobscbox{0.55in}{$37.2$; D} \\
1968 & \LRIobscbox{0.60in}{2.085} & \LRIobsbox{1.05in}{Bothwell array \citep{reber1968}} & \LRIobsbox{0.80in}{$\sim8^\circ$} & \LRIobsbox{1.80in}{Southern-sky map with a nonrectangular published boundary} & \LRIobscbox{0.82in}{\ldots} & \LRIobscbox{0.55in}{F} \\
1968 & \LRIobscbox{0.60in}{10.02} & \LRIobsbox{1.05in}{Penna array, Tasmania \citep{hamiltonHaynes1968}} & \LRIobsbox{0.80in}{$4^\circ\times5^\circ$} & \LRIobsbox{1.80in}{All RA, $-65^\circ\leq\delta\leq-2^\circ$} & \LRIobscbox{0.82in}{$5.475$; $17\,974$} & \LRIobscbox{0.55in}{$43.6$; D} \\
1974 & \LRIobscbox{0.60in}{3.93, 6.55} & \LRIobsbox{1.05in}{RAE-1 \citep{alexanderNovaco1974}} & \LRIobsbox{0.80in}{$\sim60^\circ$} & \LRIobsbox{1.80in}{All RA, $-60^\circ\leq\delta\leq+60^\circ$} & \LRIobscbox{0.82in}{$10.883$; $35\,726$} & \LRIobscbox{0.55in}{$86.6$; D} \\
1974 & \LRIobscbox{0.60in}{29.9} & \LRIobsbox{1.05in}{Fleurs aperture-synthesis survey \citep{jonesFinlay1974}} & \LRIobsbox{0.80in}{$\sim0.8^\circ$ near zenith} & \LRIobsbox{1.80in}{Galactic-plane strip, approximately $225^\circ\lesssim\ell\lesssim360^\circ$ and $0^\circ\lesssim\ell\lesssim30^\circ$; latitude-dependent mask} & \LRIobscbox{0.82in}{\ldots} & \LRIobscbox{0.55in}{F} \\
1976 & \LRIobscbox{0.60in}{10.0} & \LRIobsbox{1.05in}{DRAO northern-sky map \citep{caswell1976}} & \LRIobsbox{0.80in}{$2.6^\circ\times1.9^\circ$} & \LRIobsbox{1.80in}{$0^{\rm h}\leq\alpha\leq16^{\rm h}$, $-6^\circ\leq\delta\leq+74^\circ$; gross footprint includes unusable intervals} & \LRIobscbox{0.82in}{$4.464$; $14\,656$} & \LRIobscbox{0.55in}{$35.5$; D} \\
1977 & \LRIobscbox{0.60in}{3.7, 5.6, 8.3, 13.0, 16.5} & \LRIobsbox{1.05in}{Llanherne multifrequency survey \citep{caneWhitham1977}} & \LRIobsbox{0.80in}{$6.8^\circ$ to $1.5^\circ$} & \LRIobsbox{1.80in}{$320^\circ\leq\ell\leq360^\circ$ and $0^\circ\leq\ell\leq30^\circ$, $-25^\circ\leq b\leq+22^\circ$} & \LRIobscbox{0.82in}{$0.974$; $3\,197$} & \LRIobscbox{0.55in}{$7.75$; D} \\
1978 & \LRIobscbox{0.60in}{1.31, 2.20, 3.93, 4.70, 6.55, 9.18} & \LRIobsbox{1.05in}{RAE-2 \citep{novacoBrown1978}} & \LRIobsbox{0.80in}{Frequency dependent; $70^\circ\times80^\circ$ at 4.70 MHz} & \LRIobsbox{1.80in}{Irregular, frequency-dependent exterior boundary plus a substantial internal unsampled region} & \LRIobscbox{0.82in}{\ldots} & \LRIobscbox{0.55in}{F} \\
1978 & \LRIobscbox{0.60in}{30.0} & \LRIobsbox{1.05in}{Whole-sky composite \citep{cane1978}} & \LRIobsbox{0.80in}{$\sim11^\circ$} & \LRIobsbox{1.80in}{Full sky} & \LRIobscbox{0.82in}{$12.566$; $41\,253$} & \LRIobscbox{0.55in}{$100$; P,C} \\
1982 & \LRIobscbox{0.60in}{2.1--16.5} & \LRIobsbox{1.05in}{Tasmania compilation \citep{ellis1982}} & \LRIobsbox{0.80in}{Frequency dependent; $\sim1.5^\circ$ at 16.5 MHz} & \LRIobsbox{1.80in}{For the 16.5-MHz product: all RA, $-90^\circ\leq\delta\leq0^\circ$; lower-frequency products have more limited sampling} & \LRIobscbox{0.82in}{$6.283$; $20\,626$} & \LRIobscbox{0.55in}{$50.0$; D,C} \\
1987 & \LRIobscbox{0.60in}{1.6} & \LRIobsbox{1.05in}{Llanherne array \citep{ellisMendillo1987}} & \LRIobsbox{0.80in}{$\sim25^\circ$} & \LRIobsbox{1.80in}{Six sparse declination tracks, approximately $10^{\rm h}\rightarrow05^{\rm h}$ in RA and $-12^\circ$ to $-72^\circ$ in declination; value is the track-center envelope} & \LRIobscbox{0.82in}{$\sim3.697$; $\sim12\,135$} & \LRIobscbox{0.55in}{$\sim29.4$; T} \\
1999 & \LRIobscbox{0.60in}{22.0} & \LRIobsbox{1.05in}{DRAO all-sky survey \citep{rogerEtAl1999}} & \LRIobsbox{0.80in}{$1.1^\circ\times1.7^\circ\sec Z$} & \LRIobsbox{1.80in}{All RA, $-28^\circ\leq\delta\leq+80^\circ$} & \LRIobscbox{0.82in}{$9.138$; $29\,997$} & \LRIobscbox{0.55in}{$72.7$; P,D} \\
2001 & \LRIobscbox{0.60in}{10.0} & \LRIobsbox{1.05in}{DRAO--Australia composite \citep{caneErickson2001}} & \LRIobsbox{0.80in}{$\sim5^\circ$} & \LRIobsbox{1.80in}{All RA, $-65^\circ\leq\delta\leq+90^\circ$} & \LRIobscbox{0.82in}{$11.978$; $39\,320$} & \LRIobscbox{0.55in}{$95.3$; D,C} \\
2005--2021 & \LRIobscbox{0.60in}{10, 12.6, 14.7, 16.7, 20, 25} & \LRIobsbox{1.05in}{UTR-2 continuum maps and compilation \citep{vasilenkoEtAl2005,sidorchukEtAl2021}} & \LRIobsbox{0.80in}{Frequency and declination dependent; tens of arcmin in the best products} & \LRIobsbox{1.80in}{All RA, approximately $-15^\circ\leq\delta\leq+85^\circ$ in the compiled northern-sky products} & \LRIobscbox{0.82in}{$7.885$; $25\,887$} & \LRIobscbox{0.55in}{$62.8$; D,C} \\
\enddata
\tablecomments{The date is the publication year.  The area is the gross geometric footprint defined in the text.  Code P: explicitly published or exact coverage; D: derived from coordinate limits; T: sparse-track or half-power-union estimate; F: irregular printed mask requiring digitization; C: composite product.  Ellipses indicate that assigning a numerical area would be less defensible than retaining the published boundary for later digitization.  The RAE-2 receiver covered a wider frequency range than listed here, but the six frequencies shown are those for which Galactic maps were published.}
\end{deluxetable*}

Table~\ref{tab:lf_diffuse_maps} makes clear why statements such as ``all-sky coverage'' must be used cautiously at the lowest frequencies.  In particular, the RAE-2 contour plots have an explicit outer ``limit of the data'' and an internal missing region, and the mask changes with frequency; a single $f_{\rm sky}=1$ entry would therefore overstate the published coverage.  A definitive RAE-2 area should be obtained by tracing each frequency-dependent mask and integrating it on a spherical pixelization.  Likewise, the gross footprint of the 10-MHz Caswell survey does not encode losses to RFI, Jupiter, or rejected observing intervals.  These limitations are distinct from angular resolution: the 4.7-MHz Ellis--Hamilton survey had an $11^\circ\times3^\circ$ beam over roughly 37\% of the sky, whereas the RAE-2 4.70-MHz beam was approximately $70^\circ\times80^\circ$ over a much larger but irregular footprint \citep{ellisHamilton1966,novacoBrown1978}.

\startlongtable
\begin{deluxetable*}{cclllc}
\tabletypesize{\scriptsize}
\tablewidth{\textwidth}
\tablecaption{Directional, Spectral, and Low-Order Sky Measurements at and Below 30 MHz\label{tab:lf_spectral}}
\tablehead{\colhead{Year} & \colhead{\shortstack{$\nu$\\(MHz)}} & \colhead{\shortstack{Instrument or\\platform}} & \colhead{\shortstack{Principal astrophysical\\product}} & \colhead{\shortstack{Angular\\information}} & \colhead{Reference}}
\startdata
1933 & \LRIobscbox{0.62in}{20.5} & \LRIobsbox{1.15in}{Jansky rotating directional antenna} & \LRIobsbox{2.15in}{First repeatable directional detection of Galactic radio emission} & \LRIobsbox{1.15in}{Broad fan beam, of order $30^\circ$} & \LRIobscbox{0.80in}{\citet{jansky1933}} \\
1956 & \LRIobscbox{0.62in}{0.52--2.13} & \LRIobsbox{1.15in}{Reber--Ellis long-wire systems} & \LRIobsbox{2.15in}{Selected-direction low-frequency spectra and the Galactic spectral turnover} & \LRIobsbox{1.15in}{Broad beams, approximately $50^\circ$ in the lowest-frequency work} & \LRIobscbox{0.80in}{\citet{reberEllis1956}} \\
1958 & \LRIobscbox{0.62in}{0.143} & \LRIobsbox{1.15in}{Reber long-wire antenna} & \LRIobsbox{2.15in}{Very-low-frequency background intensity under favorable ionospheric conditions} & \LRIobsbox{1.15in}{Near hemispheric} & \LRIobscbox{0.80in}{\citet{reber1958}} \\
1962 & \LRIobscbox{0.62in}{1.5--10} & \LRIobsbox{1.15in}{Tasmania low-frequency antennas} & \LRIobsbox{2.15in}{Multifrequency spectra in selected Galactic directions} & \LRIobsbox{1.15in}{Broad, approximately $50^\circ$} & \LRIobscbox{0.80in}{\citet{ellisEtAl1962}} \\
1964 & \LRIobscbox{0.62in}{1.5--10} & \LRIobsbox{1.15in}{Alouette I} & \LRIobsbox{2.15in}{Space-based spectra of the cosmic background above the ionosphere} & \LRIobsbox{1.15in}{Broad satellite-antenna response} & \LRIobscbox{0.80in}{\citet{hartz1964}} \\
1965 & \LRIobscbox{0.62in}{$<5$} & \LRIobsbox{1.15in}{Sounding rocket} & \LRIobsbox{2.15in}{Galactic emission measured in selected directions above the ionosphere} & \LRIobsbox{1.15in}{Broad directional sampling} & \LRIobscbox{0.80in}{\citet{alexanderStone1965}} \\
1969 & \LRIobscbox{0.62in}{0.4--6.5} & \LRIobsbox{1.15in}{RAE-1} & \LRIobsbox{2.15in}{Mean Galactic spectrum and large-scale modulation} & \LRIobsbox{1.15in}{Traveling-wave V antennas; broad, frequency-dependent beams} & \LRIobscbox{0.80in}{\citet{alexanderEtAl1969}} \\
1973 & \LRIobscbox{0.62in}{0.13--2.6} & \LRIobsbox{1.15in}{IMP-6} & \LRIobsbox{2.15in}{Absolute mean Galactic background spectrum} & \LRIobsbox{1.15in}{Short-dipole, nearly all-sky weighting} & \LRIobscbox{0.80in}{\citet{brown1973}} \\
1979 & \LRIobscbox{0.62in}{5.2, 9.0, 15.6, 23.0} & \LRIobsbox{1.15in}{Llanherne polar observations} & \LRIobsbox{2.15in}{Spectra toward the north and south Galactic polar regions} & \LRIobsbox{1.15in}{Broad directional beams; not a filled map} & \LRIobscbox{0.80in}{\citet{cane1979}} \\
2001 & \LRIobscbox{0.62in}{0.2--13.8} & \LRIobsbox{1.15in}{Wind/WAVES} & \LRIobsbox{2.15in}{Galactic background spectrum from spin-modulated dipole measurements} & \LRIobsbox{1.15in}{Directionally weighted spectrum, not a conventional image} & \LRIobscbox{0.80in}{\citet{manningDulk2001}} \\
2022 & \LRIobscbox{0.62in}{0.5--7} & \LRIobsbox{1.15in}{Parker Solar Probe/FIELDS} & \LRIobsbox{2.15in}{Monopole and quadrupole ($\ell=0,2$) coefficients at 56 frequencies} & \LRIobsbox{1.15in}{Lowest-order spherical-harmonic reconstruction} & \LRIobscbox{0.80in}{\citet{pageEtAl2022}} \\
2023 & \LRIobscbox{0.62in}{$<6$} & \LRIobsbox{1.15in}{Parker Solar Probe/FIELDS} & \LRIobsbox{2.15in}{Bayesian constraints on a physical low-frequency sky model} & \LRIobsbox{1.15in}{Model-constrained global structure, not a pixelized survey} & \LRIobscbox{0.80in}{\citet{bassettEtAl2023}} \\
2026 & \LRIobscbox{0.62in}{0.06--30} & \LRIobsbox{1.15in}{ROLSES-1 on the lunar surface} & \LRIobsbox{2.15in}{Low-frequency Galactic background spectrum and characterization of terrestrial interference} & \LRIobsbox{1.15in}{Spectral measurement; no two-dimensional sky map} & \LRIobscbox{0.80in}{\citet{hibbardEtAl2026}} \\
\enddata
\tablecomments{This table contains astrophysical sky measurements that should not be assigned a square-degree survey area.  Their relevant sampling descriptors are selected directions, antenna-weighted spectra, spin modulation, or low-order spherical harmonics.  Instrument papers without a published astrophysical background result are instead listed in Table~\ref{tab:lf_pathfinders}.}
\end{deluxetable*}

The historical discrete-source measurements are also worth separating from diffuse-background maps.  Source catalogues can have a large nominal footprint while remaining strongly position dependent in completeness because the diffuse Galactic background, ionospheric scintillation, confusion, and RFI set the limiting flux density.  Table~\ref{tab:lf_sources} lists the principal surveys that established source spectra below 30 MHz.

\startlongtable
\begin{deluxetable*}{cclllc}
\tabletypesize{\scriptsize}
\tablewidth{\textwidth}
\tablecaption{Principal Discrete-Source Surveys at and Below 30 MHz\label{tab:lf_sources}}
\tablehead{\colhead{Year} & \colhead{\shortstack{$\nu$\\(MHz)}} & \colhead{\shortstack{Instrument or\\survey}} & \colhead{\shortstack{Sky\\selection}} & \colhead{\shortstack{Principal product\\or limitation}} & \colhead{Reference}}
\startdata
1954 & \LRIobscbox{0.58in}{18.3} & \LRIobsbox{1.10in}{Shain--Higgins survey} & \LRIobsbox{1.25in}{Broad southern declination strip} & \LRIobsbox{2.08in}{37 discrete sources; $\sim17^\circ$ beam and strong background/confusion dependence} & \LRIobscbox{0.75in}{\citet{shainHiggins1954}} \\
1966 & \LRIobscbox{0.58in}{4.7} & \LRIobsbox{1.10in}{Penna filled array} & \LRIobsbox{1.25in}{$-72^\circ\leq\delta\leq-12^\circ$} & \LRIobsbox{2.08in}{Flux densities for 15 sources within the diffuse-sky survey; $11^\circ\times3^\circ$ beam} & \LRIobscbox{0.75in}{\citet{ellisHamilton1966}} \\
1968 & \LRIobscbox{0.58in}{10.03} & \LRIobsbox{1.10in}{DRAO telescope} & \LRIobsbox{1.25in}{Northern source sample accessible to the transit instrument} & \LRIobsbox{2.08in}{Flux densities for 124 sources with ionospheric and confusion corrections} & \LRIobscbox{0.75in}{\citet{bridlePurton1968}} \\
1969 & \LRIobscbox{0.58in}{22.25} & \LRIobsbox{1.10in}{DRAO telescope} & \LRIobsbox{1.25in}{Large 3C/4C source sample} & \LRIobsbox{2.08in}{Spectral flux densities, with explicit flags for confusing sources and structured backgrounds} & \LRIobscbox{0.75in}{\citet{rogerCostainLacey1969}} \\
1973 & \LRIobscbox{0.58in}{29.9} & \LRIobsbox{1.10in}{Fleurs survey} & \LRIobsbox{1.25in}{Galactic-plane and selected-source fields} & \LRIobsbox{2.08in}{Source strengths accompanying the 29.9-MHz aperture-synthesis program} & \LRIobscbox{0.75in}{\citet{finlayJones1973}} \\
1975 & \LRIobscbox{0.58in}{26.3} & \LRIobsbox{1.10in}{Clark Lake survey} & \LRIobsbox{1.25in}{Broad northern-sky source survey} & \LRIobsbox{2.08in}{471-source survey series; absolute flux scale and source catalogue} & \LRIobscbox{0.75in}{\citet{vinerErickson1975}} \\
1978--2002 & \LRIobscbox{0.58in}{10--25} & \LRIobsbox{1.10in}{UTR-2 northern-sky survey} & \LRIobsbox{1.25in}{Declination-zone series over the northern sky} & \LRIobsbox{2.08in}{Long-running multifrequency catalogue series; the reprocessed compilation contains 483 sources in the final published strip} & \LRIobscbox{0.75in}{\citet{braudeEtAl1978,braudeEtAl2002}} \\
\enddata
\tablecomments{Counts refer to the products stated in the cited papers and should not be interpreted as uniform-completeness source counts over the nominal footprint.  Several surveys deliberately targeted sources known at higher frequencies rather than constructing blind, flux-limited catalogues.}
\end{deluxetable*}

Recent instruments have begun to characterize the observing environment and demonstrate technologies needed for future ultralong-wavelength astronomy.  They are important predecessors to LRI, but their instrument or commissioning publications should not be conflated with a calibrated two-dimensional survey of the diffuse sky.

\startlongtable
\begin{deluxetable*}{cclllc}
\tabletypesize{\scriptsize}
\tablewidth{\textwidth}
\tablecaption{Recent Low-Frequency Pathfinder Instruments Without a Published Filled Diffuse-Sky Map\label{tab:lf_pathfinders}}
\tablehead{\colhead{Year} & \colhead{\shortstack{$\nu$\\(MHz)}} & \colhead{\shortstack{Instrument or\\platform}} & \colhead{Location} & \colhead{\shortstack{Status of the\\cited product}} & \colhead{Reference}}
\startdata
2020 & \LRIobscbox{0.58in}{1.2--125} & \LRIobsbox{1.15in}{ALBATROS} & \LRIobsbox{1.25in}{Sub-Antarctic autonomous stations} & \LRIobsbox{2.30in}{Array and front-end pathfinder for interferometric observations below 30 MHz; the cited paper is an instrument concept and development report} & \LRIobscbox{0.75in}{\citet{chiangEtAl2020}} \\
2023 & \LRIobscbox{0.58in}{1--30} & \LRIobsbox{1.15in}{Longjiang-2 LFIS} & \LRIobsbox{1.25in}{Elliptical lunar orbit} & \LRIobsbox{2.30in}{Published ultralong-wavelength spectra and first observational results from a lunar-orbiting instrument; not a filled calibrated diffuse map} & \LRIobscbox{0.75in}{\citet{yanEtAl2023}} \\
2024 & \LRIobscbox{0.58in}{0.08--80} & \LRIobsbox{1.15in}{NCLE on Queqiao} & \LRIobsbox{1.25in}{Earth--Moon $L_2$ halo orbit} & \LRIobsbox{2.30in}{Architecture and ground-performance characterization of a three-monopole pathfinder; no filled diffuse-sky map in the cited work} & \LRIobscbox{0.75in}{\citet{karapakulaEtAl2024}} \\
\enddata
\tablecomments{The frequency columns give nominal instrumental coverage.  The scientifically useful band can be narrower because of antenna response, spacecraft electromagnetic interference, terrestrial RFI, plasma noise, or calibration limitations.}
\end{deluxetable*}

This organization also clarifies how to compare prior work with LRI.  For imaging experiments, the meaningful comparison is not angular resolution alone but the joint product of frequency, valid solid angle, surface-brightness sensitivity, and repeat sampling.  For broad-beam spectra and low-order reconstructions, the appropriate comparison is instead the number of independent angular modes constrained.  The compiled diffuse-sky models commonly used for foreground prediction inherit these heterogeneous masks and resolutions rather than replacing them with new measurements \citep{deOliveiraCostaEtAl2008}.

\section{Relation of LRI measurements to the van~Cittert-Zernike theorem}\label{appendix2}

\setcounter{equation}{0}

The fundamental observable for \hbox{LRI}, as for any radio interferometer, is the mutual coherence of the electric field, not the image itself. 
Following Thompson, Moran, and Swenson~\citep[hereafter TMS]{TMS2017}, the Fourier transform of the sky brightness distribution is the two-point correlation function of the electric field in the measurement plane, customarily written as the complex visibility.

\begin{figure}[htb!]
\vspace{-2mm}
\centerline{\includegraphics[width=\textwidth]{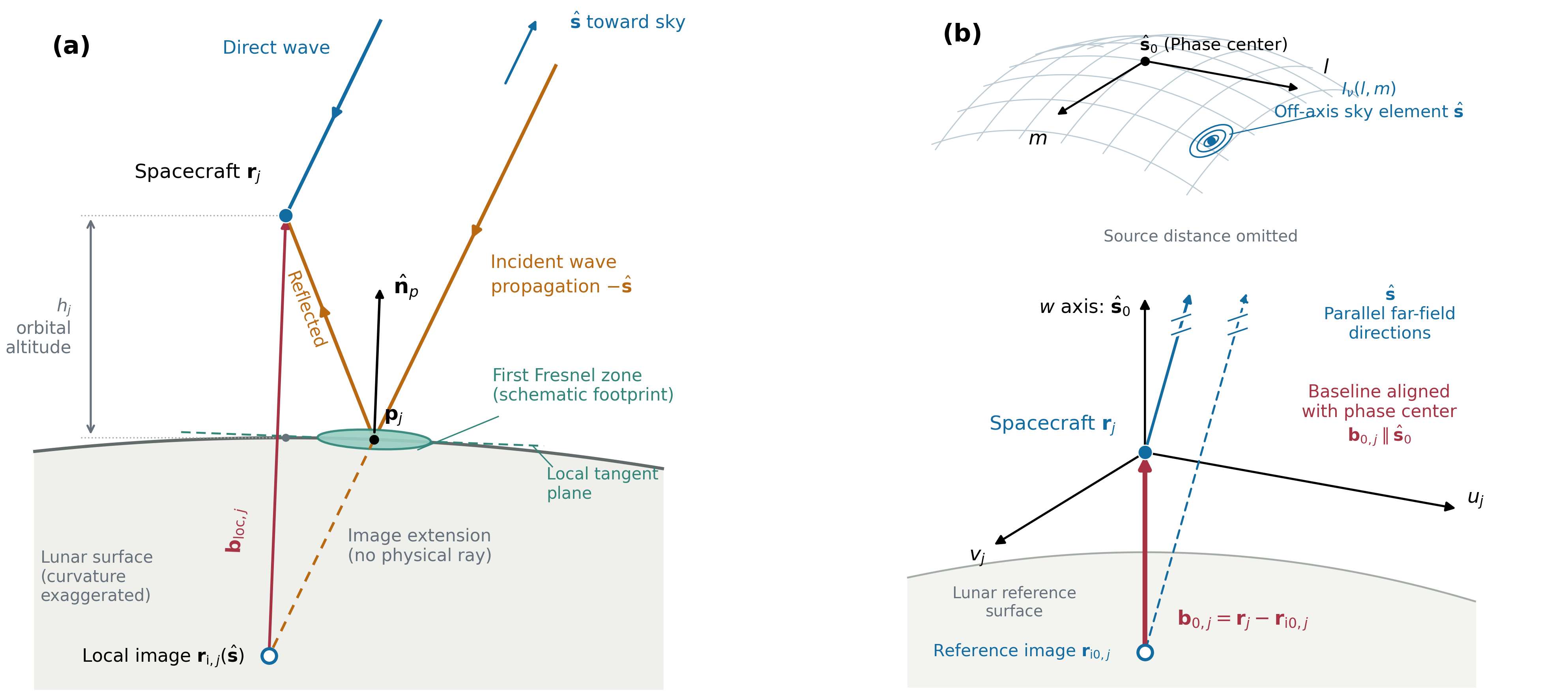}}
\vspace{-3mm}
\caption{ (a) The localized geometry for LRI including the position of a virtual image spacecraft as the second element of a virtual ``end-fire'' interferometer. (b) Similar view now showing the relevant coordinates $u,v,w$, and direction cosines $\ell,m$.
\label{fig:endfire} }
\end{figure}

The van Cittert-Zernike (vCZ) theorem states that, for a spatially incoherent source, the mutual coherence function measured between two sample points in the measurement plane is equivalent to one Fourier component of the
Fourier transform of the intensity distribution in the source plane. For LRI, considered as a spacecraft with its virtual image forming the second element , the interferometer can be described as predominantly an ``end-fire'' geometry, analogous to the case of a two-element ground-based interferometer that tracks a source down near the horizon. This geometry is shown schematically in Figure~\ref{fig:endfire}

\subsection{Two-element virtual inteferometer.}

Under these conditions, the interferometer elements must be described with a full three-dimensional geometry. Initially imagine that the direct and reflected receiving-mode outputs can be accessed separately. Their spectral responses to one angular sky element are

$$
H_{{\rm d},j}=g_{{\rm d},j}\,e^{ik\hat{\mathbf s}\cdot\mathbf r_j},
\qquad
H_{{\rm r},j}=g_{{\rm r},j}\,\Gamma_j\,
e^{ik\hat{\mathbf s}\cdot\mathbf r_{{\rm i},j}}.
$$

Here $g_{{\rm d},j}$ and $g_{{\rm r},j}$ are dimensionless complex antenna voltage responses, defined using the common reference area by

$$
\begin{aligned}
g_{a,j}(\hat{\mathbf s},\nu)
&=\sqrt{\frac{A_{{\rm e},a,j}(\hat{\mathbf s},\nu)}{A_0(\nu)}}
\,e^{i\psi_{a,j}(\hat{\mathbf s},\nu)},
\qquad a\in\{\mathrm d,\mathrm r\},\\
|g_{a,j}|^2&=\frac{A_{{\rm e},a,j}}{A_0}.
\end{aligned}
$$

The area $A_{{\rm e},a,j}$ describes antenna coupling in the selected polarization convention, before applying the lunar transfer. For the reflected mode it is evaluated for the actual arrival from the surface, not automatically for the direct sky direction. The phase $\psi_{a,j}$ describes the antenna response after the explicitly written geometric phase is removed. Lowercase $g$ distinguishes this voltage response from antenna power gain, denoted $G_{\rm p}$ below; the subscript $\rm i$ denotes the image-spacecraft position.

The voltage transfer $\Gamma_j$ may contain a Fresnel sign, absorption, phase, polarization projection, and coherent reflection loss, with geometric free-space phase already extracted as shown. Polarization or loss factors allocated to $\Gamma_j$ must not also be included in $g_{{\rm r},j}$, and the geometric phase must not be counted twice. All responses can depend on $(\nu,\hat{\mathbf s})$. For identical ideal receiving modes and unit reflection this reduces to an ordinary two-antenna problem.

For this case,  the visibility of a snapshot $j$ for baseline coordinates $u_j,v_j,w_j$ and frequency $\nu$ using the TMS formalism is given by
\begin{equation}
\begin{aligned}
{\cal V}_j(u_j,v_j,w_j;\nu)
&=\iint_{\Omega}I_\nu(l,m)\,{\cal W}_j(l,m,\nu)\\
&\quad\times\exp\!\left\{-i2\pi\left[u_jl+v_jm
+w_j\left(\sqrt{1-l^2-m^2}-1\right)\right]\right\}
\frac{dl\,dm}{\sqrt{1-l^2-m^2}} .
\end{aligned}
\end{equation}
where $I_\nu(l,m)$ is the sky brightness for direction cosines $l,m$ relative to the phase center at the zenith, the third direction cosine $n = \sqrt{1 - l^2 - m^2}$, and the factor $\,dl\,dm/n$ is equivalent to the solid angle element $\,d\Omega$. 

As explained in TMS footnote 2 on p. 93, $d\Omega$ is the surface area of an infinitesimal patch on the unit celestial sphere. Its orthogonal projection onto the $(l,m)$ plane has area $dl\,dm=\cos\theta\,d\Omega$, where $\theta$ is angular separation from the phase center. Equivalently, in polar coordinates about $\hat{\mathbf s}_0$, with azimuth $\phi$ measured from $\hat{\mathbf e}_1$ toward $\hat{\mathbf e}_2$,

$$
\begin{aligned}
l&=\sin\theta\cos\phi,\qquad
m=\sin\theta\sin\phi,\qquad
\cos\theta=\sqrt{1-l^2-m^2},\\
d\Omega&=\sin\theta\,d\theta\,d\phi
=\frac{dl\,dm}{\cos\theta}
=\frac{dl\,dm}{\sqrt{1-l^2-m^2}},
\qquad 0\leq\theta<\frac{\pi}{2}.
\end{aligned}
$$

Here the complex weight factor ${\cal W}_j(l,m,\nu) = {g_{{\rm r},j}\Gamma_jg_{{\rm d},j}^*} $ includes the normalized antenna response factors and the transfer function due to the reflection. Following TMS section 3, if $l$ and $m$ are small enough that we can neglect the term
$$ (\sqrt{(1-l^2-m^2)} -1)w \simeq -\frac{1}{2}(l^2 + m^2)w$$
then the visibility above can be approximated as
\begin{equation}
\begin{aligned}
{\cal V}_j(u_j,v_j,w_j) \simeq {\cal V}_j(u_j,v_j,0) &=\iint_{\Omega}I_\nu(l,m)\,{\cal W}_j(l,m,\nu)\\
&\quad\times\exp\!\left\{-i2\pi\left[u_jl+v_jm
\right]\right\}
\frac{dl\,dm}{\sqrt{1-l^2-m^2}} .
\end{aligned}
\end{equation}

The range of $l,m$ over which this is an acceptable approximation for LRI is $\sim 10-15^{\circ}$, and in this case, ${\cal V}_j(u_j,v_j,w_j)$ is approximately independent of $w$, and the inverse transform of the integrand gives
\begin{equation}
 \frac{I_\nu(l,m)\,{\cal W}_j(l,m,\nu)}{{\sqrt{1-l^2-m^2}}} =\iint
{\cal V}_j(u_j,v_j) \exp\!\left\{-i2\pi\left[u_jl+v_jm
\right]\right\}
{du\,dv}
\end{equation}
which is an explicit form of the Van Cittert--Zernicke equation, where the complex weight function, including the transfer function associated with the reflection, replaces a specific normalized antenna response on the left hand side. 

\subsection{Single-element interferometry}

Unlike conventional synthesis imaging arrays, LRI measures the mutual coherence between two receiving modes of a single antenna system: a direct celestial mode and a mode formed by propagation to and reflection from the lunar surface. For a spatially incoherent sky, the direct–reflected cross-spectral density is a linear integral of sky brightness against the product of the two receiving-mode transfer functions as we derived above. This relation does not require the reflected mode to be representable by a point-like virtual antenna. In the limiting case of a planar, perfectly specular reflector, the reflected transfer function becomes that of a fixed image antenna and the LRI measurement reduces to the conventional full-sky interferometer measurement equation. Under a narrow-field approximation it explicitly reduces to the familiar two-dimensional van Cittert–Zernike Fourier relation, but in the general lunar case, imaging is performed by inversion of the exact spherical delay-and-reflection operator.

 A conventional two-element interferometer samples the mutual coherence between two
separate antennas. The signals are then brought together and either summed, in an adding-interferometer, or multiplied, in a correlator.  Adding-interferometers are not often used in modern radio interferometry, as there are clear advantages to the use of correlators.
Lunar reflective interferometry instead sums the signals in a single antenna: one contribution arrives directly from the sky, while a second contribution arrives after reflection from the lunar surface. This is adding interferometry, analogous to how beams are combined from two apertures in an optical interferometer, with intensity then measured after the fields from the two apertures are summed and interference takes place, yielding fringes in the intensity.

The reflected path creates a delayed
copy of the sky field, so that the single-antenna voltage
$$
v_j(t)=v_{{\rm d},j}(t)+v_{{\rm r},j}(t)+n_j(t)
$$
where $n_j(t)$ is a noise term uncorrelated to the sky signal.

The measurable second-order statistic required for interference is therefore the voltage autocorrelation as a function of the lag time $\eta$:
\begin{equation}
\label{acf-eq2}
R_v(\eta)=\left\langle v(t)\,v^*(t-\eta)\right\rangle ,
\end{equation}
or, equivalently, its Fourier transform, the power spectrum
\begin{equation}
S_v(\nu)=\mathcal{F}\{R_v(\eta)\}.
\end{equation}
Thus the autocorrelation function is simply the output of the correlator in the
single-antenna implementation.

Expanding the autocorrelation of the summed voltage gives all nine terms:

$$
\begin{aligned}
R_v(\eta)
&=\left\langle v(t)v^*(t-\eta)\right\rangle\\
&=\left\langle
[v_{\rm d}(t)+v_{\rm r}(t)+n(t)]
[v_{\rm d}(t-\eta)+v_{\rm r}(t-\eta)+n(t-\eta)]^*
\right\rangle\\
&=R_{\rm dd}(\eta)+R_{\rm rr}(\eta)+R_{\rm nn}(\eta)
+R_{\rm rd}(\eta)+R_{\rm dr}(\eta)\\
&\quad+R_{\rm dn}(\eta)+R_{\rm nd}(\eta)
+R_{\rm rn}(\eta)+R_{\rm nr}(\eta).
\end{aligned}
$$

Initially take the additive receiver noise to be zero mean and uncorrelated with both astronomical receiving-mode voltages at every relevant lag. The four signal–noise terms then vanish in the ensemble mean; zero mean alone would not suffice. With $R_n\equiv R_{\rm nn}$, the surviving terms are

$$
R_v(\eta)=R_{\rm dd}(\eta)+R_{\rm rr}(\eta)+R_n(\eta)
+R_{\rm rd}(\eta)+R_{\rm dr}(\eta).
$$

The direct and reflected contributions are driven by the same celestial field, so their cross-correlations remain. This is the autocorrelation of the {\it voltage}, a second-order field statistic, rather than an autocorrelation of detected intensity.

For the stationary snapshot, the Wiener–Khinchin relation now gives the power spectral density as the Fourier transform of this voltage autocorrelation (TMS Section 3.2, pp. 98–99). Define $P_{\rm dd}$, $P_{\rm rr}$, and $P_n$ as the Fourier transforms of the corresponding auto-correlations, and $C_{\rm rd}$ and $C_{\rm dr}$ as those of the cross-correlations. Then

$$
\begin{aligned}
P_v(\nu)
&=\int_{-\infty}^{\infty}R_v(\eta)e^{-i2\pi\nu\eta}\,d\eta\\
&=P_{\rm dd}(\nu)+P_{\rm rr}(\nu)+P_n(\nu)
+C_{\rm rd}(\nu)+C_{\rm dr}(\nu).
\end{aligned}
$$
Stationarity relates the two cross-correlations at opposite lags. With the Fourier convention of the definition of the autocorrelation above, this becomes conjugacy at the same frequency:

$$
\begin{aligned}
R_{\rm dr}(\eta)&=R_{\rm rd}^*(-\eta),\\
C_{\rm dr}(\nu)
&=\int_{-\infty}^{\infty}R_{\rm rd}^*(-\eta)
e^{-i2\pi\nu\eta}\,d\eta
=C_{\rm rd}^*(\nu).
\end{aligned}
$$

Thus the two cross-spectral terms combine as $2\operatorname{Re}C_{\rm rd}(\nu)$; the lag-domain terms cannot in general be replaced by $2\operatorname{Re}R_{\rm rd}(\eta)$ at the same lag. Restoring $j$ and suppressing the common frequency argument, the measured spectrum is

\begin{equation}
{
P_{v,j}=P_{{\rm dd},j}+P_{{\rm rr},j}+P_{n,j}
+C_{{\rm rd},j}+C_{{\rm rd},j}^*
=P_{{\rm dd},j}+P_{{\rm rr},j}+P_{n,j}
+2\operatorname{Re}C_{{\rm rd},j}.}
\label{Fullspectrum}
\end{equation}

A two-element complex-correlation instrument with separate access to these modes would form $C_{{\rm rd},j}=\langle\widetilde v_{{\rm r},j}\widetilde v_{{\rm d},j}^*\rangle$, the explicit cross-spectrum, without the additional total power terms. Equation~\ref{Fullspectrum} shows that the adding measurement contains this same mutual coherence through $2\operatorname{Re}C_{{\rm rd},j}$, together with the individual mode powers and receiver noise. The vCZ relation supplies the sky-to-coherence kernel in either case; the distinction is how the instrument measures that coherence. 



The measurement follows the same field-addition principle as a two-beam optical interferometer: the squared modulus of the summed field amplitude contains an interference term. At optical wavelengths this operation occurs in photodetection. In LRI, the antenna responds linearly to the superposed fields, and the quadratic statistic is formed from the recorded voltage. A single receiver can therefore contain interferometric information from two propagation paths. The optical analogy concerns coherent beam combination followed by intensity detection; it does not require a second physical LRI receiver or a separate square-law detector ahead of voltage recording.

Since the Wiener--Khinchin theorem relates autocorrelation and power spectrum, one may work either in delay space or in frequency space without changing the underlying observable. The spectrum contains the full mutual-coherence information; the ACF is its Fourier-dual correlator output. In practice, LRI sky mapping may be performed in either frequency or delay space because the lunar geometry maps naturally onto differential delays, but the measurement remains fundamentally a spectral measurement of mutual coherence.

\subsection{Lunar surface curvature.}

For each incident sky direction, the smooth reflection region can be represented to leading order by a {\bf tilted and displaced tangent facet at the actual specular point}. The tilt to the local tangent plane simply changes the angles involved, and the displacement to account for the offset of a spherical surface compared to the initial nadir plane just increases the path delay relative to the fixed nadir plane. These terms are separate from the residual phase variation of the true curvature compared to the planar surface of this local facet and do not affect specular coherence. Here we determine that the residual 

To quantify the distinction, let $i$ be the incidence angle relative to the local normal, $\mu=\cos i$, and $L$ the specular-point-to-spacecraft distance. Use local tangent coordinates $x$ in the plane of incidence and $y$ transverse to it. At the specular point the linear path terms cancel. The quadratic optical-path variation on a smooth sphere is

$$
\Delta(x,y)-\Delta(0,0)
\simeq \frac12\left[
\left(\frac{\mu^2}{L}+\frac{2\mu}{R_{\rm M}}\right)x^2
+\left(\frac{1}{L}+\frac{2\mu}{R_{\rm M}}\right)y^2
\right].
$$

The first terms in each coefficient are present for a flat tangent facet; the $2\mu/R_{\rm M}$ terms arise from surface curvature. The flat-facet first Fresnel zone, defined by an excess optical path $\lambda/2$, has semiaxes $a_{x,0}=\sqrt{\lambda L}/\mu$ and $a_{y,0}=\sqrt{\lambda L}$. Across its boundary, the additional curvature phase ranges between

$$
\delta\phi_{y,F}=\frac{2\pi L\mu}{R_{\rm M}},
\qquad
\delta\phi_{x,F}=\frac{2\pi L}{R_{\rm M}\mu}.
$$

The relevant quantity for loss of coherence is not the absolute phase deviation but the RMS deviation $\sigma_{\phi}$ with the mean subtracted. We compute those values and their effect on the loss of coherence via the Ruze equation described in the main text; the results are shown in Table~\ref{tab:curvature} for the exact specular locations on a sphere of radius 1737.4 km.

\begin{table}[htb!]
    \centering
        \caption{Additional curvature phase RMS and expected amplitude losses for a flat-facet Fresnel zone patch approximation.}
    \begin{tabular}{|c|c|c|c|} \hline
        altitude & angular & RMS phase & Ruze-equivalent\\ 
         km & radius  & around mean & power loss\\\hline
        50 & $10^\circ$ & 0.053 rad & 0.28\% \\ \hline
        50 & $20^\circ$ & 0.056 rad & 0.30\%  \\  \hline
        100 & $10^\circ$ & 0.106 rad  & 1.12\% \\ \hline
        100 & $20^\circ$ & 0.111 rad  & 1.22\% \\ \hline
    \end{tabular}

    \label{tab:curvature}
\end{table}

The effect of curvature for the proposed altitude of LRI is thus negligible in its effect on the reflection coherence, which will instead be determined by surface roughness and other related effects. 

These local phase corrections are essentially frequency-independent in the quadratic approximation when each frequency is evaluated over its own Fresnel zone: the zone's area decreases in proportion to $\lambda$. Including curvature makes the first zone slightly smaller. Its actual semiaxes are approximately 3.8–5.6 km across these cases at 1 MHz, and 1.2–1.8 km at 10 MHz. Thus a locally planar coherent facet is a useful approximation at 10–20 degrees, with the remaining smooth phase and geometrical defocusing included in the transfer.

"Coherent" here means that the field contributions have a deterministic hase relationship and a calculable complex sum. It does not mean equal phase everywhere: the first Fresnel zone itself spans $\pi$ of optical-path phase by definition. The above result concerns smooth lunar curvature only; unresolved topography, roughness, dielectric structure, and averaging over changing geometry require their own assessment.

\bibliography{LRIreferences}

@string{aap   	= "Astron.\ \& Astrophys."}

@string{aaps   	= "Astron.\ \& Astrophys.\ Suppl."}

@string{apj   	= "Astrophys.\ J."}

@string{apjl  	= "Astrophys.\ J."}

@string{araa  	= "Ann.\ Rev.\ Astron.\ Astrophys."}

@string{mnras 	= "Mon.\ Not.\ R.\ Astron.\ Soc."}

@string{nar   	= "New Astron.\ Rev."}

@string{nat   	= "Nature"}

@string{pasp  	= "Publ.\ Astron.\ Soc.\ Pacific"}

@string{solphys = "Solar Phys."}

@string{ssr     = "Space Sci.\ Rev."}

@article{thomson1848mathematical,
  author    = {Thomson, William},
  title     = {On the Mathematical Theory of Electricity in Equilibrium},
  journal   = {The Cambridge and Dublin Mathematical Journal
  },
  volume    = {3},
  pages     = {131--148},
  year      = {1848}
}

@article{Alvarez2000,
  author = {Alvarez, H. and Aparici, J. and May, J. and Reich, P.},
  title = {The radio continuum spectrum of {Centaurus A}'s large-scale components},
  journal = {Astronomy and Astrophysics},
  year = {2000},
  volume = {355},
  pages = {863--872},
  url = {https://ui.adsabs.harvard.edu/abs/2000A%26A...355..863A}
}

@article{Bridle1967,
  author = {Bridle, A. H.},
  title = {Flux Densities of {Cassiopeia A} and {Cygnus A} at 10.05 {MHz}},
  journal = {The Observatory},
  year = {1967},
  volume = {87},
  pages = {60--63},
  url = {https://www.nrao.edu/archives/files/original/2235397c42fc1b1a3abe54b0abeea24b.pdf}
}

@article{bridlePurton1968,
  author = {Bridle, A. H. and Purton, C. R.},
  title = {Observations of Radio Sources at 10.03 {MHz}},
  journal = {The Astronomical Journal},
  year = {1968},
  volume = {73},
  pages = {717--726},
  doi = {10.1086/110684}
}

@inproceedings{Baldwin1971,
  author = {Baldwin, J. E.},
  title = {The Electromagnetic Spectrum of the {Crab Nebula}},
  booktitle = {The Crab Nebula},
  series = {IAU Symposium},
  year = {1971},
  volume = {46},
  pages = {22--31},
  doi = {10.1017/S0074180900007178}
}

@article{VinyaikinEtAl1987,
  author = {Vinyaikin, E. N. and Nikonov, V. A. and Tarasov, A. F.
            and Tokarev, Yu. V. and Yurishchev, M. A.},
  title = {Spectrum of low-frequency radio emission of {Cassiopeia A}},
  journal = {Soviet Astronomy},
  year = {1987},
  volume = {31},
  pages = {517},
  url = {https://articles.adsabs.harvard.edu/pdf/1987SvA....31..517V}
}

@article{Tokarev1996,
  author = {Tokarev, Yu. V.},
  title = {Low-frequency radio spectra of {Crab nebula}, {Virgo A}, and {Cygnus A}},
  journal = {Radiophysics and Quantum Electronics},
  year = {1996},
  volume = {39},
  pages = {625--633},
  doi = {10.1007/BF02120983}
}

@article{McKinley2015,
  author = {McKinley, B. and Yang, R. and L{\'o}pez-Caniego, M. and others},
  title = {Modelling of the spectral energy distribution of {Fornax A}:
           leptonic and hadronic production of high-energy emission from the radio lobes},
  journal = {Monthly Notices of the Royal Astronomical Society},
  year = {2015},
  volume = {446},
  pages = {3478--3491},
  doi = {10.1093/mnras/stu2310}
}

@article{Vinyaikin2006,
  author = {Vinyaikin, E. N.},
  title = {Secular decrease of the radio flux density of {Cassiopeia A}
           determined from long-term observations. Observations at 38, 81.5, and 151.5 {MHz}},
  journal = {Astronomy Reports},
  year = {2006},
  volume = {50},
  pages = {143--151},
  doi = {10.1134/S1063772906020077}
}

@article{Vinyaikin2007,
  author = {Vinyaikin, E. N.},
  title = {Evolution of the radio spectrum of {Cassiopeia A} from long-term
           observations. Observations at 290 and 927 {MHz}},
  journal = {Astronomy Reports},
  year = {2007},
  volume = {51},
  pages = {87--96},
  doi = {10.1134/S1063772907020023}
}

@article{Vinyaikin2016,
  author = {Vinyaikin, E. N.},
  title = {The integrated radio spectrum of {Virgo A}},
  journal = {Astronomy Reports},
  year = {2016},
  volume = {60},
  pages = {744--752},
  doi = {10.1134/S1063772916080084}
}

@article{Gaensler2008,
  author = {Gaensler, B. M. and Madsen, G. J. and Chatterjee, S. and Mao, S. A.},
  title = {The Vertical Structure of Warm Ionised Gas in the {Milky Way}},
  journal = {Publications of the Astronomical Society of Australia},
  year = {2008},
  volume = {25},
  pages = {184--200},
  doi = {10.1071/AS08004}
}

@article{Lin2023,
  author = {Lin, Rebecca and van Kerkwijk, Marten H. and Kirsten, Franz
            and Pen, Ue-Li and Deller, Adam T.},
  title = {The Radio Parallax of the {Crab Pulsar}:
           A First {VLBI} Measurement Calibrated with Giant Pulses},
  journal = {The Astrophysical Journal},
  year = {2023},
  volume = {952},
  pages = {161},
  doi = {10.3847/1538-4357/acdc98}
}

@ARTICLE{mcready1947seacliff,
       author = {{McCready}, L.~L. and {Pawsey}, J.~L. and {Payne-Scott}, Ruby},
        title = "{Solar Radiation at Radio Frequencies and Its Relation to Sunspots}",
      journal = {Proceedings of the Royal Society of London Series~A},
         year = 1947,
        month = aug,
       volume = {190},
       number = {1022},
        pages = {357-375},
          doi = {10.1098/rspa.1947.0081},
       adsurl = {https://ui.adsabs.harvard.edu/abs/1947RSPSA.190..357M}
}

@ARTICLE{1950AuSRA...3..387W,
       author = {{Wild}, J.~P. and {McCready}, L.~L.},
        title = "{Observations of the Spectrum of High-Intensity Solar Radiation at Metre Wavelengths. I. The Apparatus and Spectral Types of Solar Burst Observed}",
      journal = {Australian Journal of Scientific Research A Physical Sciences},
         year = 1950,
        month = sep,
       volume = {3},
        pages = {387},
          doi = {10.1071/CH9500387},
       adsurl = {https://ui.adsabs.harvard.edu/abs/1950AuSRA...3..387W}
}

@ARTICLE{1953AuJPh...6..420B,
       author = {{Bolton}, J.~G. and {Slee}, O.~B.},
        title = "{Galactic Radiation at Radio Frequencies. V. The Sea Interferometer}",
      journal = {Australian Journal of Physics},
         year = 1953,
        month = dec,
       volume = {6},
        pages = {420},
          doi = {10.1071/PH530420},
       adsurl = {https://ui.adsabs.harvard.edu/abs/1953AuJPh...6..420B}
}

@ARTICLE{1963ARA&A...1..291W,
       author = {{Wild}, J.~P. and {Smerd}, S.~F. and {Weiss}, A.~A.},
        title = "{Solar Bursts}",
      journal = araa,
         year = 1963,
        month = jan,
       volume = {1},
        pages = {291},
          doi = {10.1146/annurev.aa.01.090163.001451},
       adsurl = {https://ui.adsabs.harvard.edu/abs/1963ARA&A...1..291W}
}

@ARTICLE{1967SoPh....1..304T,
       author = {{Takakura}, Tatsuo},
        title = "{Theory of Solar Bursts (Invited Review Paper)}",
      journal = solphys,
         year = 1967,
        month = jun,
       volume = {1},
       number = {3-4},
        pages = {304-353},
          doi = {10.1007/BF00151359},
       adsurl = {https://ui.adsabs.harvard.edu/abs/1967SoPh....1..304T}
}

@ARTICLE{1972ARA&A..10..159W,
       author = {{Wild}, J.~P. and {Smerd}, S.~F.},
        title = "{Radio Bursts from the Solar Corona}",
      journal = araa,
         year = 1972,
        month = jan,
       volume = {10},
        pages = {159},
          doi = {10.1146/annurev.aa.10.090172.001111},
       adsurl = {https://ui.adsabs.harvard.edu/abs/1972ARA&A..10..159W}
}

@BOOK{TMS2017,
       author = {{Thompson}, A. Richard and {Moran}, James M. and {Swenson}, Jr., George W.},
        title = "{Interferometry and Synthesis in Radio Astronomy, 3rd Edition}",
         year = 2017,
    publisher = "{Springer International Publishing AG}",
          doi = {10.1007/978-3-319-44431-4},
       adsurl = {https://ui.adsabs.harvard.edu/abs/2017isra.book.....T}
}

@article{stefan2013imaging,
  author  = {Stefan, I. I. and Carilli, C. L. and Green, D. A. and Ali, Z. and Aguirre, J. E. and Bradley, R. F. and DeBoer, D. and Dexter, M. and Gugliucci, N. E. and Harris, D. E. and Jacobs, D. C. and Klima, P. and MacMahon, D. and Manley, J. and Moore, D. F. and Parsons, A. R. and Pober, J. C. and Walbrugh, W. P.},
  title   = {Imaging on {PAPER}: {Centaurus A} at 148 {MHz}},
  journal = {Monthly Notices of the Royal Astronomical Society},
  volume  = {432},
  number  = {2},
  pages   = {1285--1293},
  year    = {2013},
  doi     = {10.1093/mnras/stt548}
}

@article{mckinley2013giant,
  author  = {McKinley, B. and Briggs, F. and Gaensler, B. M. and Feain, I. J. and Bernardi, G. and Wayth, R. B. and Johnston-Hollitt, M. and Offringa, A. R. and Arcus, W. and Barnes, D. G. and Bowman, J. D. and Bunton, J. D. and Cappallo, R. J. and Corey, B. E. and Deshpande, A. and deSouza, L. and Emrich, D. and Goeke, R. and Greenhill, L. J. and Hazelton, B. J. and Herne, D. and Hewitt, J. N. and Kaplan, D. L. and Kasper, J. C. and Kincaid, B. B.},
  title   = {The giant lobes of {Centaurus A} observed at 118 {MHz} with the {Murchison Widefield Array}},
  journal = {Monthly Notices of the Royal Astronomical Society},
  volume  = {436},
  number  = {2},
  pages   = {1286--1301},
  year    = {2013},
  doi     = {10.1093/mnras/stt1662}
}

@article{mckinley2021multi,
  author  = {McKinley, B. and Tingay, S. J. and Gaspari, M. and Kraft, R. P. and Matherne, C. and Offringa, A. R. and McDonald, M. and Calzadilla, M. S. and Veilleux, S. and Shabala, S. S. and Gwyn, S. D. J. and Bland-Hawthorn, J. and Crnojevi{\'c}, D. and Gaensler, B. M. and Johnston-Hollitt, M.},
  title   = {Multi-scale feedback and feeding in the closest radio galaxy {Centaurus A}},
  journal = {Nature Astronomy},
  volume  = {6},
  number  = {1},
  pages   = {109--120},
  year    = {2021},
  doi     = {10.1038/s41550-021-01553-3}
}

@article{ellis1966survey,
  author  = {Ellis, G. R. A. and Hamilton, P. A.},
  title   = {Ionospheric Absorption of Cosmic Radio Noise at 2.1 {Mc/s} to 10 {Mc/s}},
  journal = {The Astrophysical Journal},
  volume  = {143},
  pages   = {227--235},
  year    = {1966},
  doi     = {10.1086/148494}
}

@article{ellis1962galactic,
  author  = {Ellis, G. R. A. and Bessell, M. S. and Waterworth, M. D.},
  title   = {Galactic Radio Emission at 4.7 {Mc/s}},
  journal = {Nature},
  volume  = {194},
  number  = {4829},
  pages   = {667--668},
  year    = {1962},
  doi     = {10.1038/194667a0}
}

@ARTICLE{2000GMS...119.....S,
       author = {{Stone}, Robert G. and {Weiler}, Kurt W. and {Goldstein}, Melvn L. and {Bougeret}, Jean-Louis},
        title = "{Radio astronomy at long wavelengths}",
      journal = {Geophysical Monograph Series},
         year = 2000,
        month = jan,
       volume = {119},
          doi = {10.1029/GM119},
       adsurl = {https://ui.adsabs.harvard.edu/abs/2000GMS...119.....S}
}

@article{QTN1989,
  author  = {Meyer-Vernet, Nicole and Perche, Claude},
  title   = {Tool kit for antennae and thermal noise near the plasma frequency},
  journal = {Journal of Geophysical Research: Space Physics},
  volume  = {94},
  number  = {A3},
  pages   = {2405--2415},
  year    = {1989},
  doi     = {10.1029/JA094iA03p02405}
}

@article{QTN2017,
  author  = {Meyer-Vernet, Nicole and Issautier, Karine and Moncuquet, Michel},
  title   = {Quasi-thermal noise spectroscopy: The art and the practice},
  journal = {Journal of Geophysical Research: Space Physics},
  year    = {2017},
  volume  = {122},
  number  = {8},
  pages   = {7925--7945},
  doi     = {10.1002/2017JA024449}
}

@incollection{RickettColes2000,
  title     = {Scattering in the Solar Wind at Long Wavelengths},
  author    = {Rickett, B. J. and Coles, W. A.},
  booktitle = {Radio Astronomy at Long Wavelengths},
  editor    = {Stone, Robert G. and Weiler, Kurt W. and Goldstein, Melvyn L. and Bougeret, Jean-Louis},
  series    = {Geophysical Monograph Series},
  volume    = {119},
  pages     = {97--103},
  year      = {2000},
  publisher = {American Geophysical Union},
  doi       = {10.1029/GM119p0097},
  url       = {https://wiley.com}
}

@article{fogg2022wind,
  title={Wind/WAVES observations of auroral kilometric radiation: Automated burst detection and terrestrial solar wind-magnetosphere coupling effects},
  author={Fogg, A R and Jackman, C M and Waters, J and Bonnin, X and Lamy, L and Cecconi, B and Louis, C},
  journal={Journal of Geophysical Research: Space Physics},
  volume={127},
  number={5},
  pages={e2021JA030209},
  year={2022},
  publisher={Wiley Online Library}
  }

@article{gurnett1974,
  title={The earth as a radio source: Terrestrial kilometric radiation},
  author={Gurnett, Donald A.},
  journal={Journal of Geophysical Research},
  volume={79},
  number={28},
  pages={4227--4238},
  year={1974},
  publisher={Wiley Online Library},
  doi={10.1029/JA079i028p04227}
}

@article{Burns_2021,
	title = {Low Radio Frequency Observations from the Moon Enabled by NASA Landed Payload Missions},
	volume = {2},
	issn = {2632-3338},
	DOI = {10.3847/psj/abdfc3},
	number = {2},
	journal = {The Planetary Science Journal},
	publisher = {American Astronomical Society},
	author = {Burns, Jack O. and Hallinan, Gregg and Lazio, T. Joseph W. and Rawlings, Mark and Spergel, David N. and Kocz, Jon},
	year = {2021},
	month = mar,
	pages = {44}
}

@article{bassett2020,
  title={Low Radio Frequency Observations from the Moon Enabled by NASA Landed Payload Missions},
  author={Bassett, R. and others},
  journal={The Planetary Science Journal},
  volume={2},
  number={1},
  pages={9},
  year={2020},
  publisher={IOP Publishing},
  doi={10.3847/PSJ/abdfc3},
  url={https://iopscience.iop.org/article/10.3847/PSJ/abdfc3}
}

@article{baumjohann2022akr,
  title={Auroral kilometric radiation—The electron cyclotron maser paradigm},
  author={Baumjohann, W. and Treumann, R. A.},
  journal={Frontiers in Astronomy and Space Sciences},
  volume={9},
  pages={1053303},
  year={2022},
  publisher={Frontiers Media SA},
  doi={10.3389/fspas.2022.1053303},
  url={https://www.frontiersin.org/journals/astronomy-and-space-sciences/articles/10.3389/fspas.2022.1053303/full}
}

@article{Wahlund_2025,
  title = {The Radio \& Plasma Wave Investigation (RPWI) for the JUpiter ICy moons Explorer (JUICE)},
  journal = {Space Science Reviews},
  year = {2025},
  volume = {221},
  pages = {1},
  doi = {10.1007/s11214-024-01110-0},
  url = {https://link.springer.com/article/10.1007/s11214-024-01110-0},
  author = {Wahlund, J.-E. and Bergman, J. E. S. and {\AA}hl{\'e}n, L. and Puccio, W. and Cecconi, B. and Kasaba, Y. and others}
}

@article{Goto2011,
  title = {Lunar ionosphere exploration method using auroral kilometric radiation},
  author = {Goto, Yoshitaka and Kasahara, Yoshiya and Fujimoto, Takamasa and Kumamoto, Atsushi and Ono, Takayuki and Hashimoto, Kozo and Omura, Yoshiharu},
  journal = {Earth, Planets and Space},
  volume = {63},
  number = {1},
  pages = {47--53},
  year = {2011},
  doi = {10.5047/eps.2011.01.005},
  url = {https://link.springer.com/article/10.5047/eps.2011.01.005}
}

@article{kaiser1977terrestrial,
  title={Terrestrial kilometric radiation: 3-average spectral properties},
  author={Kaiser, Michael L and Alexander, Joseph K},
  journal={Journal of Geophysical Research},
  volume={82},
  number={22},
  pages={3273--3280},
  year={1977},
  publisher={AGU Publications}
  }

@article{Lane2014VLSSr,
  author = {{Lane}, W.~M. and {Cotton}, W.~D. and {van Velzen}, S. and {Clarke}, T.~E. and {Kassim}, N.~E. and {Helmboldt}, J.~F. and {Lazio}, T.~J.~W. and {Cohen}, A.~S.},
  title = "{The Very Large Array Low-frequency Sky Survey Redux (VLSSr)}",
  journal = {Monthly Notices of the Royal Astronomical Society},
  year = 2014,
  month = mar,
  volume = {440},
  number = {1},
  pages = {327-336},
  doi = {10.1093/mnras/stt2435},
  adsurl = {https://harvard.edu}
}

@article{Zarka2012,
  title   = {Planetary and exoplanetary low frequency radio observations from the Moon},
  author  = {Zarka, P. M. and Bougeret, J. L. and Briand, C. and Aminaei, A.},
  journal = {Planetary and Space Science},
  volume  = {74},
  number  = {1},
  pages   = {156--166},
  year    = {2012},
  issn    = {0032-0633},
  doi     = {10.1016/j.pss.2012.09.010},
  url     = {https://sciencedirect.com}
}

@inbook{cordes2000interstellar,
  author    = {Cordes, J. M.},
  title     = {Interstellar Scattering: Radio Sensing of Deep Space Through the Turbulent Interstellar Medium},
  booktitle = {Radio Astronomy at Long Wavelengths},
  year      = {2000},
  publisher = {American Geophysical Union},
  series    = {Geophysical Monograph Series},
  volume    = {119},
  pages     = {105-117},
  doi       = {10.1029/GM119p0105},
  isbn      = {9780875909776},
  url       = {https://agupubs.onlinelibrary.wiley.com/doi/abs/10.1029/GM119p0105}
}

@article{NPolarSpur2020Gaia,
    author = {Das, Kaustav K and Zucker, Catherine and Speagle, Joshua S and Goo
dman, Alyssa and Green, Gregory M and Alves, João},
    title = {Constraining the distance to the North Polar Spur with Gaia DR2},
    journal = {Monthly Notices of the Royal Astronomical Society},
    volume = {498},
    number = {4},
    pages = {5863-5872},
    year = {2020},
    month = {10},
    issn = {0035-8711},
    doi = {10.1093/mnras/staa2702},
    url = {https://doi.org/10.1093/mnras/staa2702},
    eprint = {https://academic.oup.com/mnras/article-pdf/498/4/5863/33838612/sta
a2702.pdf},
}

@article{CRPHYS_2022__23_S2_1_0,
     author = {Rosine Lallement},
     title = {North {Polar} {Spur/Loop} {I:} gigantic outskirt of the {Northern} {\protect\emph{Fermi} bubble} or
                            nearby hot gas cavity blown by supernovae?},
     journal = {Comptes Rendus. Physique},
     pages = {1--24},
     year = {2022},
     publisher = {Acad\'emie des sciences, Paris},
     volume = {23},
     number = {S2},
     doi = {10.5802/crphys.97},
     language = {en},
}

@ARTICLE{Halekas2011,
       author = {{Halekas}, J.~S. and {Saito}, Y. and {Delory}, G.~T. and {Farrell}, W.~M.},
        title = "{New views of the lunar plasma environment}",
      journal = {Planetary and Space Science},
         year = 2011,
        month = nov,
       volume = {59},
       number = {14},
        pages = {1681-1694},
          doi = {10.1016/j.pss.2010.08.011},
       adsurl = {https://ui.adsabs.harvard.edu/abs/2011P&SS...59.1681H}
}

@article{Lara2009a,
    author = {{Lara}, Martin  and {De Saedeleer}, Bernard and {Ferrer}, Sebasti{\'a}n},
    title={PRELIMINARY DESIGN OF LOW LUNAR ORBITS},
    journal = {Proceedings of the International Symposium on Space Flight Dynamics},
    year={2009},
    booktitle={},
    url={https://issfd.org/ISSFD\_2009/InterMissionDesignII/Lara.pdf}
}

@ARTICLE{Lara2009b,
       author = {{Lara}, Martin and {Ferrer}, Sebasti{\'a}n and {De Saedeleer}, Bernard},
        title = "{Lunar Analytical Theory for Polar Orbits in a 50-Degree Zonal Model Plus Third-Body Effect}",
      journal = {Journal of the Astronautical Sciences},
         year = 2009,
        month = jul,
       volume = {57},
       number = {3},
        pages = {561-577},
          doi = {10.1007/BF03321517},
       adsurl = {https://ui.adsabs.harvard.edu/abs/2009JAnSc..57..561L}
}

@article{ElipeLara2012,
author = {Elipe, Antonio and Lara, Martin},
title = {Frozen Orbits About the Moon},
journal = {Journal of Guidance, Control, and Dynamics},
volume = {26},
number = {2},
pages = {238-243},
year = {2003},
doi = {10.2514/2.5064},
URL = {    
        https://doi.org/10.2514/2.5064
},
eprint = { 
        https://doi.org/10.2514/2.5064 
}
}

@article{Sirwah2020ASO,
  title={A study of the moderate altitude frozen orbits around the Moon},
  author={Magdy Ali Sirwah and Dina Tarek and Mohamed Radwan and Ahmed H. Ibrahim},
  journal={Results in physics},
  year={2020},
  volume={17},
  pages={103148},
  url={https://api.semanticscholar.org/CorpusID:219447932}
}

@article{burns2020transformative,
  title={Transformative science from the lunar farside: observations of the Dark Ages and exoplanetary systems at low radio frequencies},
  author={Burns, Jack O},
  journal={Philosophical Transactions of the Royal Society A},
  volume={378},
  number={2188},
  pages={20190564},
  year={2020},
  publisher={The Royal Society Publishing}
}

@article{burns2019farside,
  title={FARSIDE: a low radio frequency interferometric array on the lunar farside},
  author={Burns, Jack and Hallinan, Gregg and Lux, James and Romero-Wolf, Andrew and Chang, Tzu-Ching and Kocz, Jonathon and Lazio, Joseph and MacDowall, Robert J and Murphy, David and Rapetti, David and others},
  journal={arXiv preprint arXiv:1907.05407},
  year={2019}
}

@article{jones1999alfa,
  title={The Astronomical Low Frequency Array (ALFA)},
  author={Jones, Dayton L and Terrile, Richard J and Kuiper, Thomas BH and Mahoney, Michael J and Marsh, Kenneth A and Murphy, David W and Resch, George L and Unwin, Stephen C and Weinreb, Sander and Woillez, Julien},
  journal={The Westerbork Observatory: Continuing Adventure in Radio Astronomy},
  pages={195--204},
  year={1999},
  publisher={Kluwer Academic Publishers}
}

@article{jones2006low,
  title={Low-frequency radio astronomy from the lunar surface},
  author={Jones, Dayton L and MacDowall, Robert J and Kasper, Justin C and Lazio, Joseph and Burns, Jack and Weiler, Kurt W},
  journal={Moon},
  volume={1},
  pages={2},
  year={2006}
}

@inproceedings{vanvugt2017frequency,
  title={Frequency smearing in full 3D interferometry},
  author={Van Vugt, P and Wijnholds, SJ and Meijerink, A and Bentum, Mark J},
  booktitle={2017 IEEE Aerospace Conference},
  pages={1--9},
  year={2017},
  organization={IEEE}
}

@article{wijnholds2018olfar,
  title={OLFAR: a space-based low-frequency radio telescope},
  author={Wijnholds, Stefan J and others},
  journal={Exoplanet Radio Exploration via the Lunar Farside},
  year={2018}
}

@article{ryle1946solar,
  title={Solar Radiation on 175 Mc./s.},
  author={Ryle, Martin and Vonberg, Derek D.},
  journal={Nature},
  volume={158},
  number={4010},
  pages={339--340},
  year={1946},
  publisher={Nature Publishing Group},
  doi={10.1038/158339b0}
}

@article{ryle1948investigation,
  title={An investigation of radio-frequency radiation from the sun},
  author={Ryle, Martin and Vonberg, Derek D.},
  journal={Proceedings of the Royal Society of London. Series A. Mathematical and Physical Sciences},
  volume={193},
  number={1032},
  pages={98--120},
  year={1948},
  publisher={The Royal Society London},
  doi={10.1098/rspa.1948.0034}
}

@ARTICLE{novacoBrown1978,
       author = {{Novaco}, J.~C. and {Brown}, L.~W.},
        title = "{Nonthermal galactic emission below 10 megahertz.}",
      journal = apj,
         year = 1978,
        month = apr,
       volume = {221},
        pages = {114-123},
          doi = {10.1086/156009},
       adsurl = {https://ui.adsabs.harvard.edu/abs/1978ApJ...221..114N}
}

@article{Parsons_2016,
doi = {10.3847/0004-637X/820/1/51},
url = {https://doi.org/10.3847/0004-637X/820/1/51},
year = {2016},
month = {mar},
publisher = {The American Astronomical Society},
volume = {820},
number = {1},
pages = {51},
author = {Parsons, Aaron R. and Liu, Adrian and Ali, Zaki S. and Cheng, Carina},

title = {OPTIMIZED BEAM SCULPTING WITH GENERALIZED FRINGE-RATE FILTERS},
journal = {The Astrophysical Journal}
}

@ARTICLE{1984ARAandA..22..267W,
       author = {{Wagner}, William J.},
        title = "{Coronal Mass Ejections}",
      journal = araa,
         year = 1984,
        month = jan,
       volume = {22},
        pages = {267-289},
          doi = {10.1146/annurev.aa.22.090184.001411},
       adsurl = {https://ui.adsabs.harvard.edu/abs/1984ARA&A..22..267W}
}

@BOOK{1990LNP...362.....K,
       author = {{Kassim}, Namir E. and {Weiler}, Kurt W.},
        title = "{Low frequency astrophysics from space : proceedings of an international workshop held in Crystal City, Virginia, USA, on~8 and 9~January 1990}",
         year = 1990,
       volume = {362},
    publisher = {Springer-Verlag},
      address = {Berlin}, 
          doi = {10.1007/3-540-52891-1},
       adsurl = {https://ui.adsabs.harvard.edu/abs/1990LNP...362.....K}
}

@ARTICLE{1990ARA&A..28..561R,
       author = {{Rickett}, B.~J.},
        title = "{Radio propagation through the turbulent interstellar plasma.}",
      journal = {\araa},
         year = 1990,
        month = jan,
       volume = {28},
        pages = {561-605},
          doi = {10.1146/annurev.aa.28.090190.003021},
       adsurl = {https://ui.adsabs.harvard.edu/abs/1990ARA&A..28..561R}
}

@INCOLLECTION{1990LNP...362..155S,
       author = {{Spangler}, Steven R. and {Armstrong}, John W.},
        title = "{Low-Frequency Angular Broadening and Diffuse Interstellar Plasma Turbulence}",
    booktitle = {Low Frequency Astrophysics from Space},
         year = 1990,
       editor = {{Kassim}, Namir E. and {Weiler}, Kurt W.},
    publisher = {Springer-Verlag},
       volume = {362},
        pages = {155},
          doi = {10.1007/3-540-52891-110.1007/3-540-52891-1_120},
       adsurl = {https://ui.adsabs.harvard.edu/abs/1990LNP...362..155S}
}

@ARTICLE{1991ARAandA..29..275H,
       author = {{Haisch}, Bernhard and {Strong}, Keith T. and {Rodono}, Marcello},
        title = "{Flares on the Sun and other stars.}",
      journal = araa,
         year = 1991,
        month = jan,
       volume = {29},
        pages = {275-324},
          doi = {10.1146/annurev.aa.29.090191.001423},
       adsurl = {https://ui.adsabs.harvard.edu/abs/1991ARA&A..29..275H}
}

@ARTICLE{1998SoPh..183..165L,
       author = {{Leblanc}, Yolande and {Dulk}, George A. and {Bougeret}, Jean-Louis},
        title = "{Tracing the Electron Density from the Corona to~1 au}",
      journal = solphys,
         year = 1998,
        month = nov,
       volume = {183},
       number = {1},
        pages = {165-180},
          doi = {10.1023/A:1005049730506},
       adsurl = {https://ui.adsabs.harvard.edu/abs/1998SoPh..183..165L}
}

@article{Ruze1966,
  author  = {Ruze, John},
  title   = {Antenna tolerance theory—A review},
  journal = {Proceedings of the IEEE},
  year    = {1966},
  volume  = {54},
  number  = {4},
  pages   = {633--640},
  doi     = {10.1109/PROC.1966.4784}
}

@article{RochblattSeidel1992,
  author  = {Rochblatt, D. J. and Seidel, B. L.},
  title   = {Microwave antenna holography},
  journal = {IEEE Transactions on Microwave Theory and Techniques},
  year    = {1992},
  volume  = {40},
  number  = {6},
  pages   = {1294--1300},
  doi     = {10.1109/22.141363}
}

@article{Goossensetal2020,
  author  = {Goossens, S. and Sabaka, T. J. and Wieczorek, M. A. and Neumann, G. A. and Mazarico, E. and Lemoine, F. G. and et al.},
  title   = {High-resolution gravity field models from GRAIL data and implications for models of the density structure of the Moon's crust},
  journal = {Journal of Geophysical Research: Planets},
  year    = {2020},
  volume  = {125},
  number  = {2},
  doi     = {10.1029/2019je006086}
}

@article{Mazaricoetal2018,
  author  = {Mazarico, Erwan and Neumann, Gregory A. and Barker, Michael K. and Goossens, Sander and Smith, David E. and Zuber, Maria T.},
  title   = {Orbit determination of the {Lunar Reconnaissance Orbiter}: Status after seven years},
  journal = {Planetary and Space Science},
  year    = {2018},
  volume  = {162},
  pages   = {2--19},
  doi     = {10.1016/j.pss.2017.10.004}
}

@article{Kobayashi2010,
  author  = {Kobayashi, T. and Ono, T. and Kumamoto, A.},
  title   = {Identification of subsurface reflectors in the {Kaguya Lunar Radar Sounder} data by a numerical simulation of the surface clutter},
  journal = {IEEE Transactions on Geoscience and Remote Sensing},
  year    = {2010},
  volume  = {48},
  number  = {6},
  pages   = {2639--2650},
  doi     = {10.1109/TGRS.2010.2040479}
}

@article{James2015RRI,
  author  = {James, H. G. and Ware, R. H. and Gillies, R. G. and Yau, A. W. and Hussey, G. C. and Enno, G. and Miles, D. M. and Bernhardt, P. A.},
  title   = {The {e-POP Radio Receiver Instrument} on {CASSIOPE}},
  journal = {Space Science Reviews},
  year    = {2015},
  volume  = {189},
  number  = {1-4},
  pages   = {3--27},
  doi     = {10.1007/s11214-015-0136-1}
}

@article{Ono2009,
  author  = {Ono, T. and Kumamoto, A. and Nakagawa, H. and Yamaguchi, Y. and Oshigami, S. and Yamaji, A. and Kobayashi, T. and Kasahara, Y. and Oya, H.},
  title   = {Lunar Radar Sounder observations of subsurface layers under the mare {Imbrium} of the {Moon}},
  journal = {Science},
  year    = {2009},
  volume  = {323},
  number  = {5916},
  pages   = {909--912},
  doi     = {10.1126/science.1165988}
}

@article{Ono2010LRS,
  author  = {Ono, T. and Kumamoto, A. and Nakagawa, H. and Yamaguchi, Y. and Oshigami, S. and Yamaji, A. and Kobayashi, T. and Kasahara, Y. and Oya, H.},
  title   = {The {Lunar Radar Sounder (LRS)} onboard the {SELENE (Kaguya)} spacecraft},
  journal = {Space Science Reviews},
  year    = {2010},
  volume  = {154},
  number  = {1-4},
  pages   = {145--192},
  doi     = {10.1007/s11214-010-9673-3}
}

@article{Smith2010LOLA,
  author = {Smith, David E. and Zuber, Maria T. and Neumann, Gregory A. and Lemoine, Frank G. and Mazarico, Erwan and Torrence, Mark H. and McGarry, Jan F. and Rowlands, David D. and Head, James W. and Duxbury, Thomas H. and others},
  title = {Initial observations from the Lunar Orbiter Laser Altimeter (LOLA)},
  journal = {Geophysical Research Letters},
  volume = {37},
  number = {18},
  year = {2010},
  doi = {10.1029/2010GL043751}
}

@article{Rosenburgetal2011,
  author  = {Rosenburg, M. A. and Aharonson, O. and Head, J. W. and Kreslavsky, M. A. and Mazarico, E. and Neumann, G. A. and Smith, D. E. and Torrence, M. H. and Zuber, M. T.},
  title   = {Global surface roughness of the Moon as measured by the Lunar Orbiter Laser Altimeter after one year of mission operations},
  journal = {Journal of Geophysical Research: Planets},
  year    = {2011},
  volume  = {116},
  number  = {E2},
  doi     = {10.1029/2010JE003716}
}

@article{Scholten2012LROC,
  author = {Scholten, F. and Gl{\"a}ser, P. and Oberst, J. and Matz, K.-D. and Habermann, M. and Robinson, M. S. and Speyerer, E. J. and Wohlfarth, K.},
  title = {LROC stereo observations: Topography and LROC WAC DTM},
  journal = {Journal of Geophysical Research: Planets},
  volume = {117},
  number = {E12},
  year = {2012},
  doi = {10.1029/2011JE003926}
}

@article{Araki2009Kaguya,
  author = {Araki, H. and Tazawa, S. and Noda, H. and Ishihara, Y. and Goossens, S. and Kawano, N. and Sasaki, S. and Kamiya, I. and Otake, H. and Oberst, J. and Shum, C. K.},
  title = {Lunar Global Shape and Polar Topography Derived from Kaguya-LALT Laser Altimetry},
  journal = {Science},
  volume = {323},
  number = {5916},
  pages = {897--900},
  year = {2009},
  doi = {10.1126/science.1164146}
}

@article{Barker2016SLDEM,
  author = {Barker, M. K. and Mazarico, E. and Neumann, G. A. and Zuber, M. T. and Smith, D. E. and Head, J. W.},
  title = {A new lunar digital elevation model from the Lunar Orbiter Laser Altimeter and SELENE Terrain Camera},
  journal = {Icarus},
  volume = {273},
  pages = {346--355},
  year = {2016},
  doi = {10.1016/j.icarus.2015.07.039}
}

@article{Kaguya2015,
title = {Constraint on subsurface structures beneath Reiner Gamma on the Moon using the Kaguya
 Lunar Radar Sounder},
journal = {Icarus},
volume = {254},
pages = {144-149},
year = {2015},
issn = {0019-1035},
doi = {https://doi.org/10.1016/j.icarus.2015.03.020},
url = {https://www.sciencedirect.com/science/article/pii/S0019103515001207},
author = {Yuichi Bando and Atsushi Kumamoto and Norihiro Nakamura}
}

@inproceedings{kumamoto2021,
  author    = {Kumamoto, A. and Kobayashi, T. and Ono, T. and Yamaji, A. and Oshigami, S. and Ishiyama, K. and Haruyama, J.},
  title     = {Derivation of Lunar Subsurface Loss Tangent from SELENE Lunar Radar Sounder (LRS) Data},
  booktitle = {52nd Lunar and Planetary Science Conference},
  year      = {2021},
  series    = {LPI Contribution No. 2548},
  pages     = {1510},
  address   = {The Woodlands, Texas},
  month     = {March},
  url       = {https://www.hou.usra.edu/meetings/lpsc2021/pdf/1510.pdf}
}

@article{Kaguya2009,
author = {Oshigami, Shoko and Yamaguchi, Yasushi and Yamaji, Atsushi and Ono, Takayuki and Kumamoto, Atsushi and Kobayashi, Takao and Nakagawa, Hiromu},
title = {Distribution of the subsurface reflectors of the western nearside maria observed from Kaguya with Lunar Radar Sounder},
journal = {Geophysical Research Letters},
volume = {36},
number = {18},
pages = {},
doi = {https://doi.org/10.1029/2009GL039835},
url = {https://agupubs.onlinelibrary.wiley.com/doi/abs/10.1029/2009GL039835},
eprint = {https://agupubs.onlinelibrary.wiley.com/doi/pdf/10.1029/2009GL039835},
year = {2009}
}

@misc{RemcomXF7,
  author       = {{Remcom, inc.}},
  title        = {XF7 Finite Difference Time Domain software},
  year         = {2026},
  url          = {https://www.remcom.com/xfdtd-3d-em-simulation-software},
  note         = {Accessed: 2026-03-31},
  howpublished = {\url{https://www.remcom.com/xfdtd-3d-em-simulation-software}}
}

@ARTICLE{2004LRSP....1....2W,
       author = {{Wood}, Brian E.},
        title = "{Astrospheres and Solar-like Stellar Winds}",
      journal = {Living Reviews in Solar Physics},
         year = 2004,
        month = dec,
       volume = {1},
       number = {1},
          eid = {2},
        pages = {2},
          doi = {10.12942/lrsp-2004-2},
       adsurl = {https://ui.adsabs.harvard.edu/abs/2004LRSP....1....2W}
}

@ARTICLE{2006SSRv..126....3W,
       author = {{Wood}, Brian E.},
        title = "{The Solar Wind and the Sun in the Past}",
      journal = ssr,
         year = 2006,
        month = oct,
       volume = {126},
       number = {1-4},
        pages = {3-14},
          doi = {10.1007/s11214-006-9006-0},
       adsurl = {https://ui.adsabs.harvard.edu/abs/2006SSRv..126....3W}
}

@ARTICLE{2008PASP..120.1207P,
       author = {{Parsons}, Aaron and {Backer}, Donald and {Siemion}, Andrew and {Chen}, Henry and {Werthimer}, Dan and {Droz}, Pierre and {Filiba}, Terry and {Manley}, Jason and {McMahon}, Peter and {Parsa}, Arash and et al.},
        title = "{A Scalable Correlator Architecture Based on Modular FPGA Hardware, Reuseable Gateware, and Data Packetization}",
      journal = pasp,
         year = 2008,
        month = nov,
       volume = {120},
       number = {873},
        pages = {1207},
          doi = {10.1086/593053},
archivePrefix = {arXiv},
       eprint = {0809.2266},
 primaryClass = {astro-ph},
       adsurl = {https://ui.adsabs.harvard.edu/abs/2008PASP..120.1207P}
}

@ARTICLE{2008SoPh..247..171H,
       author = {{Harrison}, Richard A. and {Davis}, Christopher J. and {Eyles}, Christopher J. and {Bewsher}, Danielle and {Crothers}, Steve R. and {Davies}, Jackie A. and {Howard}, Russell A. and {Moses}, Daniel J. and {Socker}, Dennis G. and {Newmark}, Jeffrey S. and {Halain}, Jean-Philippe and {Defise}, Jean-Marc and {Mazy}, Emmanuel and {Rochus}, Pierre and {Webb}, David F. and {Simnett}, George M.},
        title = "{First Imaging of Coronal Mass Ejections in the Heliosphere Viewed from Outside the Sun   Earth Line}",
      journal = solphys,
         year = 2008,
        month = jan,
       volume = {247},
       number = {1},
        pages = {171-193},
          doi = {10.1007/s11207-007-9083-6},
       adsurl = {https://ui.adsabs.harvard.edu/abs/2008SoPh..247..171H}
}

@ARTICLE{SWAVES,
       author = {{Bougeret}, J.~L. and {Goetz}, K. and {Kaiser}, M.~L. and {Bale}, S.~D. and {Kellogg}, P.~J. and {Maksimovic}, M. and {Monge}, N. and {Monson}, S.~J. and {Astier}, P.~L. and {Davy}, S. and {Dekkali}, M. and {Hinze}, J.~J. and {Manning}, R.~E. and {Aguilar-Rodriguez}, E. and {Bonnin}, X. and {Briand}, C. and {Cairns}, I.~H. and {Cattell}, C.~A. and {Cecconi}, B. and {Eastwood}, J. and {Ergun}, R.~E. and {Fainberg}, J. and {Hoang}, S. and {Huttunen}, K.~E.~J. and {Krucker}, S. and {Lecacheux}, A. and {MacDowall}, R.~J. and {Macher}, W. and {Mangeney}, A. and {Meetre}, C.~A. and {Moussas}, X. and {Nguyen}, Q.~N. and {Oswald}, T.~H. and {Pulupa}, M. and {Reiner}, M.~J. and {Robinson}, P.~A. and {Rucker}, H. and {Salem}, C. and {Santolik}, O. and {Silvis}, J.~M. and {Ullrich}, R. and {Zarka}, P. and {Zouganelis}, I.},
        title = "{S/WAVES: The Radio and Plasma Wave Investigation on the STEREO Mission}",
      journal = ssr,
         year = 2008,
        month = apr,
       volume = {136},
       number = {1-4},
        pages = {487-528},
          doi = {10.1007/s11214-007-9298-8},
       adsurl = {https://ui.adsabs.harvard.edu/abs/2008SSRv..136..487B}
}

@INPROCEEDINGS{2009SPIE.7436E..0IL,
       author = {{Lazio}, Joseph and {Carilli}, C. and {Hewitt}, J. and {Furlanetto}, S. and {Burns}, Jack},
        title = "{The lunar radio array (LRA)}",
    booktitle = {UV/Optical/IR Space Telescopes: Innovative Technologies and Concepts IV},
         year = 2009,
       editor = {{MacEwen}, Howard A. and {Breckinridge}, James B.},
       series = {Society of Photo-Optical Instrumentation Engineers (SPIE) Conference Series},
       volume = {7436},
        month = aug,
          eid = {74360I},
        pages = {74360I},
          doi = {10.1117/12.827955},
       adsurl = {https://ui.adsabs.harvard.edu/abs/2009SPIE.7436E..0IL}
}

@ARTICLE{2009NewAR..53....1J,
       author = {{Jester}, Sebastian and {Falcke}, Heino},
        title = "{Science with a lunar low-frequency array: From the dark ages of the Universe to nearby exoplanets}",
      journal = {\nar},
         year = 2009,
        month = may,
       volume = {53},
       number = {1-2},
        pages = {1-26},
          doi = {10.1016/j.newar.2009.02.001},
archivePrefix = {arXiv},
       eprint = {0902.0493},
 primaryClass = {astro-ph.CO},
       adsurl = {https://ui.adsabs.harvard.edu/abs/2009NewAR..53....1J}
}

@ARTICLE{2010ARAandA..48..241B,
       author = {{Benz}, Arnold O. and {G{\"u}del}, Manuel},
        title = "{Physical Processes in Magnetically Driven Flares on the Sun, Stars, and Young Stellar Objects}",
      journal = araa,
         year = 2010,
        month = sep,
       volume = {48},
        pages = {241-287},
          doi = {10.1146/annurev-astro-082708-101757},
       adsurl = {https://ui.adsabs.harvard.edu/abs/2010ARA&A..48..241B}
}

@ARTICLE{2014ASTRP...1...43L,
       author = {{Linsky}, J.~L. and {Wood}, B.~E.},
        title = "{Lyman-{\ensuremath{\alpha}} observations of astrospheres}",
      journal = {ASTRA Proceedings},
         year = 2014,
        month = aug,
       volume = {1},
       number = {1},
        pages = {43-49},
          doi = {10.5194/ap-1-43-2014},
archivePrefix = {arXiv},
       eprint = {1408.5934},
 primaryClass = {astro-ph.SR},
       adsurl = {https://ui.adsabs.harvard.edu/abs/2014ASTRP...1...43L}
}

@ARTICLE{jgl+15,
       author = {{Jakosky}, B.~M. and {Grebowsky}, J.~M. and {Luhmann}, J.~G. and {Connerney}, J. and {Eparvier}, F. and {Ergun}, R. and {Halekas}, J. and {Larson}, D. and {Mahaffy}, P. and {McFadden}, J. and {Mitchell}, D.~F. and {Schneider}, N. and {Zurek}, R. and {Bougher}, S. and {Brain}, D. and {Ma}, Y.~J. and {Mazelle}, C. and {Andersson}, L. and {Andrews}, D. and {Baird}, D. and {Baker}, D. and {Bell}, J.~M. and {Benna}, M. and {Chaffin}, M. and {Chamberlin}, P. and {Chaufray}, Y.-Y. and {Clarke}, J. and {Collinson}, G. and {Combi}, M. and {Crary}, F. and {Cravens}, T. and {Crismani}, M. and {Curry}, S. and {Curtis}, D. and {Deighan}, J. and {Delory}, G. and {Dewey}, R. and {DiBraccio}, G. and {Dong}, C. and {Dong}, Y. and {Dunn}, P. and {Elrod}, M. and {England}, S. and {Eriksson}, A. and {Espley}, J. and {Evans}, S. and {Fang}, X. and {Fillingim}, M. and {Fortier}, K. and {Fowler}, C.~M. and {Fox}, J. and {Gr{\"o}ller}, H. and {Guzewich}, S. and {Hara}, T. and {Harada}, Y. and {Holsclaw}, G. and {Jain}, S.~K. and {Jolitz}, R. and {Leblanc}, F. and {Lee}, C.~O. and {Lee}, Y. and {Lefevre}, F. and {Lillis}, R. and {Livi}, R. and {Lo}, D. and {Mayyasi}, M. and {McClintock}, W. and {McEnulty}, T. and {Modolo}, R. and {Montmessin}, F. and {Morooka}, M. and {Nagy}, A. and {Olsen}, K. and {Peterson}, W. and {Rahmati}, A. and {Ruhunusiri}, S. and {Russell}, C.~T. and {Sakai}, S. and {Sauvaud}, J.-A. and {Seki}, K. and {Steckiewicz}, M. and {Stevens}, M. and {Stewart}, A.~I.~F. and {Stiepen}, A. and {Stone}, S. and {Tenishev}, V. and {Thiemann}, E. and {Tolson}, R. and {Toublanc}, D. and {Vogt}, M. and {Weber}, T. and {Withers}, P. and {Woods}, T. and {Yelle}, R.},
        title = "{MAVEN observations of the response of Mars to an interplanetary coronal mass ejection}",
      journal = {Science},
         year = 2015,
        month = nov,
       volume = {350},
       number = {6261},
          eid = {0210},
        pages = {0210},
          doi = {10.1126/science.aad0210},
       adsurl = {https://ui.adsabs.harvard.edu/abs/2015Sci...350.0210J}
}

@article{CassiniJupiter,
author = {Zarka, P. and Cecconi, B. and Kurth, W. S.},
title = {Jupiter's low-frequency radio spectrum from Cassini/Radio and Plasma Wave Science (RPWS) absolute flux density measurements},
journal = {Journal of Geophysical Research: Space Physics},
volume = {109},
number = {A9},
pages = {},
doi = {https://doi.org/10.1029/2003JA010260},
url = {https://agupubs.onlinelibrary.wiley.com/doi/abs/10.1029/2003JA010260},
eprint = {https://agupubs.onlinelibrary.wiley.com/doi/pdf/10.1029/2003JA010260},
year = {2004}
}

@ARTICLE{PassiveSounding,
  author={Peters, Sean T. and Schroeder, Dustin M. and Castelletti, Davide and Haynes, Mark and Romero-Wolf, Andrew},
  journal={IEEE Transactions on Geoscience and Remote Sensing}, 
  title={In Situ Demonstration of a Passive Radio Sounding Approach Using the Sun for Echo Detection}, 
  year={2018},
  volume={56},
  number={12},
  pages={7338-7349},
  doi={10.1109/TGRS.2018.2850662}}

@ARTICLE{2018ApJ...856...53P,
       author = {{Pognan}, Quentin and {Garraffo}, Cecilia and {Cohen}, Ofer and {Drake}, Jeremy J.},
        title = "{The Solar Wind Environment in Time}",
      journal = apj,
         year = 2018,
        month = mar,
       volume = {856},
       number = {1},
          eid = {53},
        pages = {53},
          doi = {10.3847/1538-4357/aaaebb},
archivePrefix = {arXiv},
       eprint = {1802.05153},
 primaryClass = {astro-ph.SR},
       adsurl = {https://ui.adsabs.harvard.edu/abs/2018ApJ...856...53P}
}

@ARTICLE{2021LRSP...18....4T,
       author = {{Temmer}, Manuela},
        title = "{Space weather: the solar perspective: An update to Schwenn~(2006)}",
      journal = {Living Reviews in Solar Physics},
         year = 2021,
        month = dec,
       volume = {18},
       number = {1},
          eid = {4},
        pages = {4},
          doi = {10.1007/s41116-021-00030-3},
archivePrefix = {arXiv},
       eprint = {2104.04261},
 primaryClass = {astro-ph.SR},
       adsurl = {https://ui.adsabs.harvard.edu/abs/2021LRSP...18....4T}
}

@ARTICLE{2021ARAandA..59..445H,
       author = {{Hudson}, Hugh S.},
        title = "{Carrington Events}",
      journal = araa,
         year = 2021,
        month = sep,
       volume = {59},
        pages = {445-477},
          doi = {10.1146/annurev-astro-112420-023324},
       adsurl = {https://ui.adsabs.harvard.edu/abs/2021ARA&A..59..445H}
}

@ARTICLE{2022LRSP...19....2C,
       author = {{Cliver}, Edward W. and {Schrijver}, Carolus J. and {Shibata}, Kazunari and {Usoskin}, Ilya G.},
        title = "{Extreme solar events}",
      journal = {Living Reviews in Solar Physics},
         year = 2022,
        month = dec,
       volume = {19},
       number = {1},
          eid = {2},
        pages = {2},
          doi = {10.1007/s41116-022-00033-8},
archivePrefix = {arXiv},
       eprint = {2205.09265},
 primaryClass = {astro-ph.SR},
       adsurl = {https://ui.adsabs.harvard.edu/abs/2022LRSP...19....2C}
}

@ARTICLE{2022ApJ...926L...1B,
       author = {{Bandyopadhyay}, R. and {Matthaeus}, W.~H. and {McComas}, D.~J. and {Chhiber}, R. and {Usmanov}, A.~V. and {Huang}, J. and {Livi}, R. and {Larson}, D.~E. and {Kasper}, J.~C. and {Case}, A.~W. and {Stevens}, M. and {Whittlesey}, P. and {Romeo}, O.~M. and {Bale}, S.~D. and {Bonnell}, J.~W. and {Dudok de Wit}, T. and {Goetz}, K. and {Harvey}, P.~R. and {MacDowall}, R.~J. and {Malaspina}, D.~M. and {Pulupa}, M.},
        title = "{Sub-Alfv{\'e}nic Solar Wind Observed by the Parker Solar Probe: Characterization of Turbulence, Anisotropy, Intermittency, and Switchback}",
      journal = apjl,
         year = 2022,
        month = feb,
       volume = {926},
       number = {1},
          eid = {L1},
        pages = {L1},
          doi = {10.3847/2041-8213/ac4a5c},
archivePrefix = {arXiv},
       eprint = {2201.10718},
 primaryClass = {physics.space-ph},
       adsurl = {https://ui.adsabs.harvard.edu/abs/2022ApJ...926L...1B}
}

@ARTICLE{2023JGRA..12830884J,
       author = {{Jolitz}, R.~D. and {Rahmati}, A. and {Brain}, D.~A. and {Lee}, C.~O. and {Lillis}, R.~J. and {Thiemann}, E. and {Eparvier}, F. and {Mitchell}, D. and {Halekas}, J. and {Larson}, D. and {Curry}, S.~M. and {Jakosky}, B.~M.},
        title = "{Energy Input of \hbox{EUV}, Solar Wind, and SEPs at Mars: MAVEN Observations During Solar Minimum}",
      journal = {Journal of Geophysical Research (Space Physics)},
         year = 2023,
        month = feb,
       volume = {128},
       number = {2},
          eid = {e2022JA030884},
        pages = {e2022JA030884},
          doi = {10.1029/2022JA030884},
       adsurl = {https://ui.adsabs.harvard.edu/abs/2023JGRA..12830884J}
}

@MISC{KISS,
       author = {{Loyd}, R.~O. Parke and {Shkolnik}, Evgenya L. and {Lazio}, Joseph and {Hallinan}, Gregg W. and {Alvarado-G{\'o}mez}, Juli{\'a}n and {Amaral}, Laura and {Davis}, Ivey and {Farrish}, Alison and {Green}, James and {Brain}, Dave and {Chen}, Bin and {Cohen}, Christina and {Curry}, Shannon and {Dissauer}, Karin and {Egan}, Arika and {Gopalswamy}, Nat and {Gronoff}, Guillaume and {Habbal}, Shadia and {Hu}, Renyu and {Jin}, Meng and {Mason}, James Paul and {Murray-Clay}, Ruth and {Namekata}, Kosuke and {Osten}, Rachel and {Segura}, Ant{\'\i}gona and {Veronig}, Astrid and {Vidotto}, Aline and {Wilson}, Maurice and {Xu}, Yu},
        title = "{The Exospace Weather Frontier}",
 howpublished = {Report prepared for the W. M. Keck Institute for Space Studies (KISS), California Institute of Technology, by R.O.P. Loyd et al, 2025.},
         year = 2025,
        month = oct,
          doi = {10.26206/gmhk5-amp17},
archivePrefix = {arXiv},
       eprint = {2511.02871},
 primaryClass = {astro-ph.IM},
       adsurl = {https://ui.adsabs.harvard.edu/abs/2025kiss.rept.....L}
}

@ARTICLE{polderman2019HII,
       author = {{Polderman}, I.~M. and {Haverkorn}, M. and {Jaffe}, T.~R. and {Alves}, M.~I.~R.},
        title = "{Low-frequency measurements of synchrotron absorbing HII regions and modeling of observed synchrotron emissivity}",
      journal = {\aap},
         year = 2019,
        month = jan,
       volume = {621},
          eid = {A127},
        pages = {A127},
          doi = {10.1051/0004-6361/201834405},
archivePrefix = {arXiv},
       eprint = {1811.10310},
 primaryClass = {astro-ph.GA},
       adsurl = {https://ui.adsabs.harvard.edu/abs/2019A&A...621A.127P}
}

@inproceedings{Kassim1990,
author = {Kassim, Namir E.},
title = {{H II regions in absorption at low frequencies}},
booktitle = {Astrophysics from the Moon},
editor = {Mumma, Michael J. and Smith, Harlan J.},
series = {AIP Conference Proceedings},
volume = {207},
pages = {218--224},
publisher = {American Institute of Physics},
year = {1990},
doi = {10.1063/1.39364}
}

@ARTICLE{2026MNRAS.546ag077M,
       author = {{Manju}, G. and {Pant}, Tarun Kumar and {Mridula}, N. and {Ambili}, K.~M. and {Hossain}, Md Mosarraf and {Kumar}, P. Pradeep and {Sruthi}, T.~V. and {Vishnupriya}, K.~S. and {Tripathi}, Keshav Ram and {Sreelatha}, P. and {John}, Rosmy and {Thampi}, R. Satheesh and {Aneesh}, A.~N.},
        title = "{Langmuir probe-based lunar near-surface plasma measurements associated with the unique hop test of Chandrayaan 3 lander: observations and model simulations}",
      journal = mnras,
         year = 2026,
        month = mar,
       volume = {546},
       number = {4},
          eid = {stag077},
        pages = {stag077},
          doi = {10.1093/mnras/stag077},
       adsurl = {https://ui.adsabs.harvard.edu/abs/2026MNRAS.546ag077M}
}

@TechReport{SunRISEPayload,
    author = {{Romero-Wolf}, Andrew and {Lux}, James and {Robison}, David and {Tarsala}, Jan},
    title = "{SunRISE Project Payload Architecture Design Document D-105635 Rev~C}",
  institution =  {Jet Propulsion Laboratory, California Institute of Technology},
  year = 	 2025,
  OPTtype = 	 {},
  OPTnumber = 	 {},
  OPTaddress = 	 {},
  OPTmonth = 	 {},
  OPTnote = 	 {},
  OPTannote = 	 {},
    doi = {10.48577/jpl.8M0MKX},
    url = {https://doi.org/10.48577/jpl.8M0MKX}
}

@ARTICLE{hogbom1974clean,
       author = {{H{\"o}gbom}, J.~A.},
        title = "{Aperture Synthesis with a Non-Regular Distribution of Interferometer Baselines}",
      journal = {\aaps},
         year = 1974,
        month = jun,
       volume = {15},
        pages = {417},
       adsurl = {https://ui.adsabs.harvard.edu/abs/1974A&AS...15..417H}
}

@ARTICLE{KochukhovLavail2017,
       author = {{Kochukhov}, Oleg and {Lavail}, Alexis},
        title = "{The Global and Small-scale Magnetic Fields of Fully Convective, Rapidly Spinning M Dwarf Pair GJ65 A and B}",
      journal = {\apjl},
         year = 2017,
        month = jan,
       volume = {835},
       number = {1},
          eid = {L4},
        pages = {L4},
          doi = {10.3847/2041-8213/835/1/L4},
archivePrefix = {arXiv},
       eprint = {1702.02946},
 primaryClass = {astro-ph.SR},
       adsurl = {https://ui.adsabs.harvard.edu/abs/2017ApJ...835L...4K}
}

@article{jansky1933,
  author = {Jansky, Karl G.},
  title = {Electrical Disturbances Apparently of Extraterrestrial Origin},
  journal = {Proceedings of the Institute of Radio Engineers},
  year = {1933},
  volume = {21},
  pages = {1387--1398},
  doi = {10.1109/JRPROC.1933.227458}
}

@article{shainHiggins1954,
  author = {Shain, C. A. and Higgins, C. S.},
  title = {Observations of the General Background and Discrete Sources of 18.3 Mc/s Cosmic Noise},
  journal = {Australian Journal of Physics},
  year = {1954},
  volume = {7},
  pages = {130--149},
  doi = {10.1071/PH540130}
}

@article{reberEllis1956,
  author = {Reber, Grote and Ellis, G. R. A.},
  title = {Cosmic Radio-Frequency Radiation Near One Megacycle},
  journal = {Journal of Geophysical Research},
  year = {1956},
  volume = {61},
  pages = {1--10},
  doi = {10.1029/JZ061i001p00001}
}

@article{reber1958,
  author = {Reber, Grote},
  title = {Between the Atmospherics},
  journal = {Journal of Geophysical Research},
  year = {1958},
  volume = {63},
  pages = {109--123},
  doi = {10.1029/JZ063i001p00109}
}

@article{shainEtAl1961,
  author = {Shain, C. A. and Komesaroff, M. M. and Higgins, C. S.},
  title = {A High Resolution Galactic Survey at 19.7 Mc/s},
  journal = {Australian Journal of Physics},
  year = {1961},
  volume = {14},
  pages = {508--514},
  doi = {10.1071/PH610508}
}

@article{ellisEtAl1962,
  author = {Ellis, G. R. A. and Waterworth, M. D. and Bessell, M.},
  title = {Spectrum of the Galactic Radio Emission between 10 Mc/s and 1.5 Mc/s},
  journal = {Nature},
  year = {1962},
  volume = {196},
  pages = {1079},
  doi = {10.1038/1961079a0}
}

@article{hartz1964,
  author = {Hartz, T. R.},
  title = {Observations of the Galactic Radio Emission between 1.5 and 10 Mc/s from the Alouette Satellite},
  journal = {Nature},
  year = {1964},
  volume = {203},
  pages = {173--175},
  doi = {10.1038/203173a0}
}

@article{mathewsonEtAl1965,
  author = {Mathewson, D. S. and Broten, N. W. and Cole, D. J.},
  title = {A Survey of the Southern Sky at 30 Mc/s},
  journal = {Australian Journal of Physics},
  year = {1965},
  volume = {18},
  pages = {665--668},
  doi = {10.1071/PH650665}
}

@article{alexanderStone1965,
  author = {Alexander, J. K. and Stone, R. G.},
  title = {Galactic Radio Emission below 5 Mc/s},
  journal = {The Astrophysical Journal},
  year = {1965},
  volume = {142},
  pages = {1327--1332},
  doi = {10.1086/148418}
}

@article{ellisHamilton1966,
  author = {Ellis, G. R. A. and Hamilton, P. A.},
  title = {Cosmic Radio Noise Survey at 4.7 Mc/s},
  journal = {The Astrophysical Journal},
  year = {1966},
  volume = {143},
  pages = {227--235},
  doi = {10.1086/148493}
}

@article{reber1968,
  author = {Reber, Grote},
  title = {Cosmic Static at 144 Meters Wavelength},
  journal = {Journal of the Franklin Institute},
  year = {1968},
  volume = {285},
  pages = {1--12},
  doi = {10.1016/0016-0032(68)90463-8}
}

@article{hamiltonHaynes1968,
  author = {Hamilton, P. A. and Haynes, R. F.},
  title = {Observations of the Southern Sky at 10.02 MHz},
  journal = {Australian Journal of Physics},
  year = {1968},
  volume = {21},
  pages = {895--902},
  doi = {10.1071/PH680895}
}

@article{alexanderEtAl1969,
  author = {Alexander, J. K. and Brown, L. W. and Clark, T. A. and Stone, R. G. and Weber, R. R.},
  title = {The Spectrum of the Cosmic Radio Background between 0.4 and 6.5 MHz},
  journal = {The Astrophysical Journal},
  year = {1969},
  volume = {157},
  pages = {L163--L168},
  doi = {10.1086/180411}
}

@article{rogerCostainLacey1969,
  author = {Roger, R. S. and Costain, C. H. and Lacey, J. D.},
  title = {Spectral Flux Densities of Radio Sources at 22.25 MHz. I},
  journal = {The Astronomical Journal},
  year = {1969},
  volume = {74},
  pages = {366--372},
  doi = {10.1086/110817}
}

@article{brown1973,
  author = {Brown, L. W.},
  title = {The Galactic Radio Background between 130 and 2600 kHz},
  journal = {The Astrophysical Journal},
  year = {1973},
  volume = {180},
  pages = {359--370},
  doi = {10.1086/151968}
}

@article{finlayJones1973,
  author = {Finlay, E. A. and Jones, B. B.},
  title = {Measurements of Source Strengths at 29.9 MHz},
  journal = {Australian Journal of Physics},
  year = {1973},
  volume = {26},
  pages = {389--402},
  doi = {10.1071/PH730389}
}

@article{alexanderNovaco1974,
  author = {Alexander, J. K. and Novaco, J. C.},
  title = {Survey of the Galactic Background Radiation at 3.93 and 6.55 MHz},
  journal = {The Astronomical Journal},
  year = {1974},
  volume = {79},
  pages = {777--785},
  doi = {10.1086/111608}
}

@article{jonesFinlay1974,
  author = {Jones, B. B. and Finlay, E. A.},
  title = {An Aperture Synthesis Survey of the Galactic Plane},
  journal = {Australian Journal of Physics},
  year = {1974},
  volume = {27},
  pages = {687--711},
  doi = {10.1071/PH740687}
}

@article{vinerErickson1975,
  author = {Viner, M. R. and Erickson, W. C.},
  title = {26.3-MHz Radio Source Survey. I. The Absolute Flux Scale},
  journal = {The Astronomical Journal},
  year = {1975},
  volume = {80},
  pages = {931--954},
  doi = {10.1086/111827}
}

@article{caswell1976,
  author = {Caswell, J. L.},
  title = {A Map of the Northern Sky at 10 MHz},
  journal = {Monthly Notices of the Royal Astronomical Society},
  year = {1976},
  volume = {177},
  pages = {601--616},
  doi = {10.1093/mnras/177.3.601}
}

@article{caneWhitham1977,
  author = {Cane, H. V. and Whitham, P. S.},
  title = {Observations of the Southern Galactic Background at Five Frequencies between 3.7 and 16.5 MHz},
  journal = {Monthly Notices of the Royal Astronomical Society},
  year = {1977},
  volume = {179},
  pages = {21--29},
  doi = {10.1093/mnras/179.1.21}
}

@article{cane1978,
  author = {Cane, H. V.},
  title = {A 30 MHz Map of the Whole Sky},
  journal = {Australian Journal of Physics},
  year = {1978},
  volume = {31},
  pages = {561--565},
  doi = {10.1071/PH780561}
}

@article{braudeEtAl1978,
  author = {Braude, S. Ya. and Megn, A. V. and Ryabov, B. P. and Sharykin, N. K. and Zhouck, I. N.},
  title = {Decametric Survey of Discrete Sources in the Northern Sky. I},
  journal = {Astrophysics and Space Science},
  year = {1978},
  volume = {54},
  pages = {3--36},
  doi = {10.1007/BF00637902}
}

@article{cane1979,
  author = {Cane, H. V.},
  title = {Spectra of the Non-Thermal Radio Radiation from the Galactic Polar Regions},
  journal = {Monthly Notices of the Royal Astronomical Society},
  year = {1979},
  volume = {189},
  pages = {465--478},
  doi = {10.1093/mnras/189.3.465}
}

@article{ellis1982,
  author = {Ellis, G. R. A.},
  title = {Galactic Radio Emission below 16.5 MHz and the Galactic Emission Measure},
  journal = {Australian Journal of Physics},
  year = {1982},
  volume = {35},
  pages = {91--104},
  doi = {10.1071/PH820091}
}

@article{ellisMendillo1987,
  author = {Ellis, G. R. A. and Mendillo, M.},
  title = {A 1.6 MHz Survey of the Galactic Background Radio Emission},
  journal = {Australian Journal of Physics},
  year = {1987},
  volume = {40},
  pages = {705--708},
  doi = {10.1071/PH870705}
}

@article{rogerEtAl1999,
  author = {Roger, R. S. and Costain, C. H. and Landecker, T. L. and Swerdlyk, C. M.},
  title = {The Radio Emission from the Galaxy at 22 MHz},
  journal = {Astronomy and Astrophysics Supplement Series},
  year = {1999},
  volume = {137},
  pages = {7--19},
  doi = {10.1051/aas:1999239}
}

@article{caneErickson2001,
  author = {Cane, H. V. and Erickson, W. C.},
  title = {A 10 MHz Map of the Galaxy},
  journal = {Radio Science},
  year = {2001},
  volume = {36},
  pages = {1765--1768},
  doi = {10.1029/2000RS002463}
}

@article{manningDulk2001,
  author = {Manning, R. and Dulk, G. A.},
  title = {The Galactic Background Radiation from 0.2 to 13.8 MHz},
  journal = {Astronomy and Astrophysics},
  year = {2001},
  volume = {372},
  pages = {663--669},
  doi = {10.1051/0004-6361:20010516}
}

@article{braudeEtAl2002,
  author = {Braude, S. Ya. and Rashkovsky, S. L. and Sidorchuk, K. M. and Sidorchuk, M. A. and Sokolov, K. P. and Sharykin, N. K. and Zakharenko, S. M.},
  title = {Decametric Survey of Discrete Sources in the Northern Sky},
  journal = {Astrophysics and Space Science},
  year = {2002},
  volume = {280},
  pages = {235--299},
  doi = {10.1023/A:1015534108849}
}

@article{vasilenkoEtAl2005,
  author = {Vasilenko, N. M. and Mukha, D. V. and Sidorchuk, M. A. and Sidorchuk, K. M.},
  title = {Making and Processing of the Northern Sky Maps Based on the Continuum Survey with the UTR-2 Radio Telescope},
  journal = {Radio Physics and Radio Astronomy},
  year = {2005},
  volume = {10},
  number = {3},
  pages = {244--253},
  note = {In Russian}
}

@article{deOliveiraCostaEtAl2008,
  author = {{de Oliveira-Costa}, A. and Tegmark, M. and Gaensler, B. M. and Jonas, J. and Landecker, T. L. and Reich, P.},
  title = {A Model of Diffuse Galactic Radio Emission from 10 MHz to 100 GHz},
  journal = {Monthly Notices of the Royal Astronomical Society},
  year = {2008},
  volume = {388},
  pages = {247--260},
  doi = {10.1111/j.1365-2966.2008.13376.x}
}

@article{chiangEtAl2020,
  author = {Chiang, H. C. and others},
  title = {The Array of Long Baseline Antennas for Taking Radio Observations from the Sub-Antarctic (ALBATROS)},
  journal = {Journal of Astronomical Instrumentation},
  year = {2020},
  volume = {9},
  pages = {2050019},
  doi = {10.1142/S2251171720500191}
}

@article{sidorchukEtAl2021,
  author = {Sidorchuk, M. A. and Vasilenko, N. M. and Ulyanov, O. M. and Konovalenko, O. O. and Mukha, D. V. and Abramenkov, E. A. and Sidorchuk, K. M. and Miasoied, A. I.},
  title = {50 Years of Research in Continuum at the UTR-2 Radio Telescope},
  journal = {Radio Physics and Radio Astronomy},
  year = {2021},
  volume = {26},
  number = {4},
  pages = {287--313},
  doi = {10.15407/rpra26.04.287}
}

@article{pageEtAl2022,
  author = {Page, B. and Bassett, N. and Lecacheux, A. and Pulupa, M. and Rapetti, D. and Bale, S. D.},
  title = {The $\ell=2$ Spherical Harmonic Expansion Coefficients of the Sky Brightness Distribution between 0.5 and 7 MHz},
  journal = {Astronomy and Astrophysics},
  year = {2022},
  volume = {668},
  pages = {A127},
  doi = {10.1051/0004-6361/202244621}
}

@article{bassettEtAl2023,
  author = {Bassett, N. and Rapetti, D. and Nhan, B. D. and Page, B. and Burns, J. O. and Pulupa, M. and Bale, S. D.},
  title = {Constraining a Model of the Radio Sky below 6 MHz Using the Parker Solar Probe/FIELDS Instrument in Preparation for Upcoming Lunar-Based Experiments},
  journal = {The Astrophysical Journal},
  year = {2023},
  volume = {945},
  pages = {134},
  doi = {10.3847/1538-4357/acbc76}
}

@article{yanEtAl2023,
  author = {Yan, J. and others},
  title = {Ultra-Low-Frequency Radio Astronomy Observations from a Lunar Orbit},
  journal = {Experimental Astronomy},
  year = {2023},
  volume = {56},
  pages = {333--353},
  doi = {10.1007/s10686-022-09887-0}
}

@article{karapakulaEtAl2024,
  author = {Karapakula, S. and Brinkerink, C. and Vecchio, A. and others},
  title = {Architecture Design and Ground Performance of Netherlands--China Low-Frequency Explorer},
  journal = {Radio Science},
  year = {2024},
  volume = {59},
  pages = {e2023RS007906},
  doi = {10.1029/2023RS007906}
}

@article{hibbardEtAl2026,
  author = {Hibbard, Joshua J. and Burns, Jack O. and MacDowall, Robert and Gopalswamy, Natchimuthuk and Boardsen, Scott A. and Farrell, William and Bradley, Damon and Schulszas, Thomas M. and Dorigo Jones, Johnny and Rapetti, David and Turner, Jake D.},
  title = {Results from NASA's First Radio Telescope on the Moon: Terrestrial Technosignatures and the Low-Frequency Galactic Background Observed by ROLSES-1 Onboard the Odysseus Lander},
  journal = {The Astronomical Journal},
  year = {2026},
  volume = {171},
  pages = {26},
  doi = {10.3847/1538-3881/ae18d8}
}

@ARTICLE{2012RaSc...47.0K04L,
       author = {{Lane}, W.~M. and {Cotton}, W.~D. and {Helmboldt}, J.~F. and {Kassim}, N.~E.},
        title = "{VLSS redux: Software improvements applied to the Very Large Array Low-Frequency Sky Survey}",
      journal = {Radio Science},
         year = 2012,
        month = jan,
       volume = {47},
       number = {3},
          eid = {RS0K04},
        pages = {RS0K04},
          doi = {10.1029/2011RS004941},
archivePrefix = {arXiv},
       eprint = {1205.4695},
 primaryClass = {astro-ph.IM},
       adsurl = {https://ui.adsabs.harvard.edu/abs/2012RaSc...47.0K04L}
}

@ARTICLE{2026ApJ..1002....3O,
       author = {{Ocker}, Stella Koch and {Cordes}, James M.},
        title = "{NE2025: An Updated Electron Density Model for the Galactic Interstellar Medium}",
      journal = {\apj},
         year = 2026,
        month = may,
       volume = {1002},
       number = {1},
          eid = {3},
        pages = {3},
          doi = {10.3847/1538-4357/ae5825},
archivePrefix = {arXiv},
       eprint = {2602.11838},
 primaryClass = {astro-ph.GA},
       adsurl = {https://ui.adsabs.harvard.edu/abs/2026ApJ..1002....3O}
}

@INPROCEEDINGS{SunRISE2024,
  author={Kasper, Justin and Lazio, T. Joseph W. and Romero-Wolf, Andrew and Lux, James P. and Neilsen, Tim},
  booktitle={2024 IEEE Aerospace Conference}, 
  title={The Sun Radio Interferometer Space Experiment (SunRISE)}, 
  year={2024},
  volume={},
  number={},
  pages={1-19},
  doi={10.1109/AERO58975.2024.10521074}}

@manual{chatGPT,
  author = {{OpenAI}},
  title = {{ChatGPT}},
  year = {2026},
  organization = {OpenAI; accessed 2026 September 1},
  url = {https://chatgpt.com/}
}

@ARTICLE{11783057,
  author={Kristinsson, Thorsteinn Hanning and Peters, Sean T. and Voigt, Joana R. C. and Steinbrügge, Gregor and Hamilton, Christopher W. and Diniega, Serina and Williams, Jonathan and Alfonso, Gustavo and Romero-Wolf, Andrew},
  journal={IEEE Transactions on Geoscience and Remote Sensing}, 
  title={In Situ Demonstration of Passive Radar Using HF Jovian Bursts as a Source for Echo Detection}, 
  year={2026},
  volume={},
  number={},
  pages={1-1},
  doi={10.1109/TGRS.2026.3751587}}
\bibliographystyle{aasjournalv7}



\end{document}